\documentclass[12pt]{article}

\usepackage{amssymb, amsmath}
\usepackage{amsfonts}
\usepackage{graphicx}
\usepackage{setspace} 
\usepackage{rotating}
\allowdisplaybreaks
\renewcommand{\theequation}{\thesection .\arabic{equation}}
\usepackage{tikz}
\usepackage{mathrsfs}
\usepackage{url}
\usepackage[authoryear]{natbib}
\newcommand{\be}{\begin{equation}}
	\newcommand{\ee}{\end{equation}}
\newcommand{\bea}{\begin{eqnarray}}
	\newcommand{\eea}{\end{eqnarray}}
\newcommand{\bean}{\begin{eqnarray*}}
	\newcommand{\eean}{\end{eqnarray*}}
\newcommand{\brray}{\begin{array}}
	\newcommand{\erray}{\end{array}}
\newcommand{\ben}{\begin{equation}{nonumber}}
	\newcommand{\een}{\end{equation}{nonumber}}

\newtheorem{dfn}{Definition}[section]
\newtheorem{thm}[dfn]{Theorem}
\newtheorem{lmma}[dfn]{Lemma}
\newtheorem{ppsn}[dfn]{Proposition}
\newtheorem{crlre}[dfn]{Corollary}
\newtheorem{xmpl}[dfn]{Example}
\newtheorem{rmrk}[dfn]{Remark}
\newtheorem{algo}[dfn]{Algorithm}

\newcommand{\bdfn}{\begin{dfn}}
	\newcommand{\bthm}{\begin{thm}}
		\newcommand{\blmma}{\begin{lmma}}
			\newcommand{\bppsn}{\begin{ppsn}}
				\newcommand{\bcrlre}{\begin{crlre}}
					\newcommand{\bxmpl}{\begin{xmpl}}
						\newcommand{\brmrk}{\begin{rmrk}}
							\newcommand{\edfn}{\end{dfn}}
						\newcommand{\ethm}{\end{thm}}
					\newcommand{\elmma}{\end{lmma}}
				\newcommand{\eppsn}{\end{ppsn}}
			\newcommand{\ecrlre}{\end{crlre}}
		\newcommand{\exmpl}{\end{xmpl}}
	\newcommand{\ermrk}{\end{rmrk}}
\newcommand{\p}{\mathbb{P}}

\newcommand{\bs}{\boldsymbol}

\def\argmin{\mathop{\mbox{argmin}}}

\newcommand{\GG}{\mathcal{G}}
\newcommand{\mf}{\mathbf}
\newcommand{\E}{\mathbb{E}}
\newtheorem{theorem}{Theorem}[section]

\newtheorem{remark}{Remark}[section]

\newtheorem{proposition}{Proposition}[section]

\def\a*{{\cal A}_{h,*}}
\def\B{{\cal B}(h)}
\def\B1{{\cal B}_1(h)}
\def\b{{\cal B}^{\rm s.a.}(h)}
\def\b1{{\cal B}^{\rm s.a.}_1(h)}

\newcommand{\FF}{\mathcal{F}}

\newcommand{\lf}{\lfloor}
\newcommand{\rf}{\rfloor}

\newcommand{\Var}{\text{Var}}

\newcommand{\Cov}{\text{Cov}}
\usepackage{geometry}
\usepackage{xr}
\makeatletter
\newcommand*{\addFileDependency}[1]{
  \typeout{(#1)}
  \@addtofilelist{#1}
  \IfFileExists{#1}{}{\typeout{No file #1.}}
}
\makeatother

\newcommand*{\myexternaldocument}[1]{
    \externaldocument{#1}
    \addFileDependency{#1.tex}
    \addFileDependency{#1.aux}
}
\myexternaldocument{Main_Bernoulli}
\begin{document}




  \title{ Comparing time varying regression quantiles under shift invariance}
\author{\small 
	Subhra Sankar Dhar \\
	\small IIT Kanpur\\
	\small Department of Mathematics \& Statistics \\
	\small  Kanpur 208016, India\\
	{\small email: subhra@iitk.ac.in}\\
	\and
	\small Weichi Wu \\
	\small  Center for Statistical Science\\
	\small   Department of Industrial Engineering \\
	\small Tsinghua University
	\small 100084 Beijing, China\\
	{\small email: wuweichi@mail.tsinghua.edu.cn}\\
}
\date{}
\maketitle

\setcounter{equation}{0}

\setcounter{section}{0}\ \\


\begin{abstract}
	This article investigates  whether time-varying quantile regression curves are the same up to the horizontal shift or not.  The errors and the covariates involved in the regression model are allowed to be locally stationary. We formalize this issue in a corresponding  non-parametric hypothesis testing problem, and 
	develop an  integrated-squared-norm based test (SIT) as well as a simultaneous confidence band (SCB) approach. The asymptotic properties of SIT and SCB under null and local alternatives are derived.  Moreover, the asymptotic properties of these tests are also studied when the compared data sets are dependent. We then propose valid wild bootstrap algorithms to implement SIT and SCB. Furthermore, the usefulness of the proposed methodology is illustrated via analysing simulated and  real data related to  COVID-19 outbreak and climate science. 
\end{abstract}
bootstrap, comparison of curves, confidence band, hypothesis testing, locally stationary process, nonparametric quantile regression, COVID-19.

\section{Introduction}\label{Intro}
\cite{koenker1978} proposed the concept of quantile regression as an alternative approach to the least squares estimation (LSE), and it provides us {\color{black} with} conditional quantile surface describing the relation between the univariate response variable and the univariate/multivariate covariates. 
In 1980s and early 1990s, many research articles had been published on parametric quantile regression (see, e.g., \cite{ruppert1980}, \cite{koenker1982}, \cite{efron1991}, \cite{gutenbrunner1992}), and since 1990s-2000s, several attempts {\color{black}were} made for non-parametric quantile regression methods as well (see, e.g., \cite{chaudhuri1991nonparametric}, \cite{koenker1994}, \cite{yu1998local}, \cite{Takeuchi2006}). Once the estimation of the quantile regression developed in both parametric and non-parametric models, comparing the quantile regression curves started to get attention in the literature (see, e.g., \cite{dette2011comparing} and a few references therein) as quantile curves  give us the feature of the conditional distribution of the response variable conditioning on the covariate. Motivated by this powerful 
statistical properties of the quantile curves, we {\color{black}hereby} study the following research problem. 

Consider two regression models with common response variable and the same covariates for two different, \textcolor{black}{possibly dependent} groups. Formally speaking, suppose that $(y_{i,1}, \mf x_{i,1})_{i=1}^{n_1}$ and  $(y_{i,2}, \mf x_{i,2})_{i=1}^{n_2}$ are two sets of data, where the covariates $\mf x_{i,1}=(x_{i,1,1},...x_{i,p_1,1})^\top$ and $\mf x_{i,2}=(x_{i,1,2},...x_{i,p_2,2})^\top$ are $p_1\times 1$ and $p_2\times 1$ vectors, respectively. Now, for $\tau \in (0,1)$ and $s=1$ and $2$, we define the conditional quantiles
\begin{align}\label{Con_quantile}
Q_\tau(y_{i,s}|\mf x_{i,s}):=\inf \{s: F_{y_{i,s}|\mf x_{i,s}}(s|\mf x_{i,s})>\tau\}=\theta_{1,\tau, s}\left(\frac{i}{n_s}\right)x_{i,1,s}+\cdots+\theta_{p_s,\tau, s}\left(\frac{i}{n_s}\right)x_{i,p_s,s}.
\end{align}
Note that model \eqref{Con_quantile} can be written as 
\begin{align}\label{testmodel}
y_{i,s}=\mf x^\top_{i,s}\bs \theta_{\tau,s}\left(\frac{i}{n_s}\right)+e_{i,\tau,s},~~ i=1,\cdots,n_s,~~ s=1~\mbox{and}~2, 
\end{align}
where for $s = 1$ and $2$, $\bs \theta_{\tau,s}(t)=(\theta_{1,\tau,s}(t),...,\theta_{p_s,\tau,s}(t))^\top$ are $p_s\times 1$ vectors with each element being a smooth function on $[0,1]$, and the errors $e_{i,\tau,s}$ satisfy 
$Q_\tau(e_{i,\tau,s}|\mf x_{i,s})=0$. The last condition on the $\tau$-th quantile of the conditional distribution of the errors given the covariate ensures the model \eqref{testmodel} is identifiable. \textcolor{black}{ In particular, we allow $x_{i}$ and $e_{i, \tau}$ to be locally stationary and  correlate with each other, which captures a distinctly complex dependence structure of the covariates and the errors.}  Further technical assumptions on $\mf x_{i,s}$ and $e_{i,\tau,s}$ will explicitly be discussed in Section \ref{Asy-Result}. We are now interested in the following hypothesis problem.  For a pre-specified vectors $\mf c_1\in \mathbb R^{p_1\times 1}$ and $\mf c_2 \in \mathbb R^{p_2\times 1}$,
define $m_s(t)=\mf c_s^\top\bs \theta_{\tau,s}(t)$ for  $s=1,2$, $c_s=(c_{1,s},\cdots,c_{p_s,s})^\top$, and we want to test
\begin{align}\label{hypo1}
H_0: m_1(t)=m_2(t+d)~~~\text{for $0< t< 1-d$  and some unknown constant $d\in [0, 1)$}.
\end{align}

Let us now discuss a special case. Note that when $d = 0$, $H_{0}$ will be equivalent to testing the equivalence of $m_{1}(t)$ and $m_{2} (t)$  for $t\in (0, 1)$. In this case, when $\mf c_1=(1, 0, \ldots, 0)^\top$ and $\mf c_2=(1, 0, \ldots, 0)^\top$, the problem will coincide with comparing the curves $\theta_{1,\tau,1}(t)$ and $\theta_{1,\tau,2}(t)$ for $t\in (0, 1)$, and such comparison can be carried out by an appropriate functional notion of difference between estimated $\theta_{1,\tau,1}(t)$ and $\theta_{1,\tau,2}(t)$. Such types of problems have already been explored in the literature (see \cite{munk1998}). However, our proposed testing of hypothesis problem described in \eqref{hypo1} is fundamentally different from the aforesaid case. Firstly, we are comparing two sets of certain linear combinations of the components of the quantile coefficients of \eqref{Con_quantile}; it is not a direct comparison between particular quantiles of two different distributions. Secondly, note that in \eqref{Con_quantile} and \eqref{testmodel}, the quantiles are time varying, which is entirely different from the usual regression quantiles. We will discuss more about time varying regression models in the next paragraph. Finally, in \eqref{hypo1}, we are checking whether there is any nonnegative shift between two functions $m_1$ and $m_2$ or not.  Here, it should be pointed out that the testing of hypothesis problem described in \eqref{hypo1} can be written for some negative $d$ as well but without loss of generality, we study for $d \geq 0$. 
\textcolor{black}{ 
	Moreover, the model described in \eqref{hypo1} with respect to the time parameter is often applicable to real data as well. For example, the Gross Domestic Product (GDP) curves of two nations over a fixed period of time or the survival rate of women aged more than sixty five of two different nations over a long period of time. For usual mean based time varying models, such type of problem was studied by \cite{gamloumaz2007}, \cite{vimond2010}, \cite{colldala2015} and a few references therein. However, none of them studied such problems in the framework of quantile regression (i.e., \eqref{Con_quantile} or \eqref{hypo1}) for time varying models.}

In this article, we  thoroughly study this problem of testing \eqref{hypo1} \textcolor{black}{assuming the functions $m_1(t)$ and $m_2(t)$ are strictly monotone on $[0,1]$. Then the null hypothesis \eqref{hypo1} holds if and only if $(m_1^{-1}(u))' - (m_2^{-1}(u))' = 0$ when $u$ belongs to a certain interval, a subset of $(0, 1)$ (see Section \ref{ME}). Therefore, testing of \eqref{hypo1} are carried out based on a Bahadur representation of time varying quantile regression coefficients and a Gaussian approximation  to the estimated  difference $\Delta (u):=( m_1^{-1}(u))'-( m_2^{-1}(u))'$.   
	We also discuss the relaxation of the monotone assumption.} Our major
contributions are the following. 

\textcolor{black}{The first major contribution is to develop a formal
	$\mathcal L^2$ test in checking the hypothesis \eqref{hypo1} for dependent and non-stationary data.
	The test, which we denote as squared integrated test (SIT) test has the form of $\int( \hat \Delta^2(t)w(t)dt$, where $w(t)$ is a certain weight function, and $\hat\Delta(t)$ is a suitable {\it smooth} estimate of $\Delta(t)$. Approximating the SIT test statistic in terms of quadratic form and establishing the central limit theorem for the quadratic form  (see \cite{de1987central}), the asymptotic distribution of the SIT statistic is derived under null hypothesis (i.e., the hypothesis described in \eqref{hypo1}) and local alternatives.} 
In this context, we would like to mention that there have been a few research articles on conditional quantiles of independent data, and the readers are referred to  \cite{zheng1998}, \cite{Horowitz2002}, \cite{He2003}, \cite{kim2007quantile} and a few references therein. However, none of the above research articles considered the more widely applicable   \textcolor{black}{hypothesis  testing} problems that we consider here (see \eqref{hypo1}) for time varying quantile regression models (see \eqref{Con_quantile}). 

The second major contribution is to develop the simultaneous confidence band (SCB) for the difference between $(m_1^{-1}(.))'$ and $(m_2^{-1}(.))'$, and an asymptotic property of the SCB is derived,  which asserts the form of the \textcolor{black}{simultaneous} confidence band of the difference between $(m_1^{-1}(.))'$ and $(m_2^{-1}(.))'$ for a preassigned level of significance $\alpha \in (0, 1)$. \textcolor{black}{Specifically, the $100(1-\alpha)\%$  SCB of $\Delta(t)$, $t\in I : = (m_1(0),m_1(1-d))$, is  $(U_{\alpha}(t), L_\alpha(t))$, where
	\begin{align}\label{SCB-def}
	\lim_{n\rightarrow \infty}\p(L_\alpha(t)\leq \Delta(t) \leq U_{\alpha}(t), \text{for all $t\in \hat{\mathcal I}_n$} )= (1-\alpha).
	\end{align} Here $\hat {\mathcal I}_n$ is a consistent estimate of interval $I$ under null hypothesis, and $(U_{\alpha}(t), L_\alpha(t))$ depends on the sample size, which are obtained from the approximate formula for the maximum deviation of Gaussian processes (see \cite{sun1994simultaneous}) based on Weyl's volumes of tube formula. It is clear from \eqref{SCB-def} that one can use SCB as a graphical device like a band, and $\Delta(u)$ will be inside the band with a certain probability. Moreover, one can also estimate the type-I error and the power of the corresponding test associated with the SCB using the one-to-one correspondence between the confidence band and the testing of hypothesis.
	Earlier, \cite{zhou2010nonparametric} derived the limiting correct simultaneous confidence bands of quantile curves for the dependent and locally stationary data when the covariates are fixed. For random design points, \cite{wu2017nonparametric} studied the limiting properties of simultaneous confidence bands of the corresponding functional considered in their article, which is {\color{black} different from} the key term of our work.} 

The third major contribution is to propose a robust Bootstrap procedure to have a good finite sample performance of the SIT and the SCB tests. In principle, one can carry out the test based on SIT and construct the SCB using the results in Theorems \ref{asymptotic} and \ref{SCB}. However, for small or moderate sample size, directly implementing those results may not produce satisfactory performance due to slow convergence rate, and to overcome this problem, the bootstrap method is proposed, and a better rate of convergence of the Bootstrap method is established as well. The readers may also look at \cite{zhou2010simultaneous} and a few references therein.


It is now an appropriate place to mention that  \cite{dette2021identifying} addressed an apparently similar looking hypothesis; however, this article studies an entirely different hypothesis, and the content is far different too. The differences are the following : Firstly, \cite{dette2021identifying} investigates the shift invariance in mean while our article studies shift invariance in quantiles. It should be pointed out that the structure of data in different quantiles can be heterogeneous for non-stationary time series data (see real data analysis), and therefore, our proposed methodology is able to capture richer features than the mean-based method in  \cite{dette2021identifying}. Secondly, \cite{dette2021identifying} is restricted to examining the means of time series, while our method is applicable to {\it regression} quantiles with stochastic and dependent covariates. Finally, this article derives the SCB in addition to the SIT,  which provides a quantitative rule for using the Graphical device in \cite{dette2021identifying} to determine the shift among curves.

The rest of the article is organized as follows. In Section \ref{ME}, we characterize the null
hypothesis stated in \eqref{hypo1}, which is a key observation in subsequent theoretical studies.
Section \ref{llqe} discusses the local linear quantile estimator for time varying regression coefficients, and in Section \ref{BI}, basic ideas related to the estimation of the regression function and its derivative are studied. The two-stage estimator of the shift parameter is also developed in this section along with the formulation of the SIT and the SCB tests. Section \ref{Asy-Result} thoroughly investigates various asymptotic properties and related facts of the SIT and the SCB tests. Section \ref{IT} explores various issues related to implementation of the tests,  
including the bootstrap-based  algorithms.  {\color{black}Section \ref{mutual-dependent} studies the asymptotic properties of the SIT and the SCB tests when the compared data sets are dependent.}
The finite sample performance of the SIT test and the SCB test is investigated via various simulation studies in Section \ref{FSS}, and two benchmark real data sets are analyzed in Section \ref{RDA}. Finally, additional simulation results and all technical details and proofs are moved to the supplemental material.
\section{Methodology}\label{ME}
\setcounter{equation}{0}
\textcolor{black}{We assume that for regression functions $m_1(t)$ and $m_2(t)$, $\inf_{t\in [0,1]} m'_1(t)m'_2(t)>0$ holds. Without loss of generality, we here consider that both functions $m_1 (t)$ and $m_2 (t)$ are monotonically increasing functions, i.e., $m_1'(t)>0$ and $m_2'(t)>0$. We will discuss the relaxation of the monotonicity in Remark \ref{remak1}.} 
Observe that the null hypothesis $H_{0}$ described in \eqref{hypo1} is equivalent to checking $(m_1^{-1}(u))' - (m_2^{-1}(u))' = 0$ when $u$ belongs to a certain interval.  Proposition \ref{prop-monotone} states this result explicitly.  
\begin{proposition} \label{prop-monotone} Let $m_1(t)$ and $m_2(t)$ be strictly increasing functions on $[0,1]$. Then  
	$H_0$ described in \eqref{hypo1} holds if and only if $(m_1^{-1}(u))' - (m_2^{-1}(u))' = 0$ for all  $u\in [m_1(0), m_1(1 -  d)]$ when $d = m_2^{-1}(m_1(0))$.
\end{proposition}	

In practice, under the null hypothesis \eqref{hypo1}, estimating $d$ and deriving its asymptotic properties is a complicated task. Therefore, asymptotic performance of {\color{black}a statistic} based on $\hat m_1(t)-\hat m_2(t+\hat d)$,
where $\hat m_1$, $\hat m_2$ and $\hat d$ are estimators of $m_1$, $m_2$ and $d$, respectively, could be intractable, and moreover, it is likely to have unsatisfactory results as various issues like different rate of convergences are involved. In contrast, the assertion of Proposition \ref{prop-monotone} suggests that we can test \eqref{hypo1} based on the estimate of $(m_1^{-1}(u))' - (m_2^{-1}(u))'$.
To use Proposition \ref{prop-monotone} in the theoretical results, one needs to know about various issues such as the estimation of the time varying quantile regression functions and their derivatives, the formulation of test statistics etc, which are discussed in the following subsections. 
\subsection{Local linear quantile estimate}\label{llqe}
We estimate time varying quantile regression coefficients $\bs \theta_{\tau, s}(t)$ using the concept of local linear quantile estimators. Specifically, for $s=1$ and $2$, the local linear quantile estimate of
$(\bs \theta_{\tau,s}(t),\bs \theta'_{\tau,s}(t))$ is denoted by $(\hat {\bs \theta}_{\tau,s,b_{n, s}}(t), \hat {\bs \theta}_{\tau,s,b_{n, s}}'(t))$, where
\begin{align}\label{Localquantile}
(\hat{\bs \theta}_{\tau,s, b_{n, s}}(t), \hat{\bs \theta}'_{\tau,s, b_{n, s}}(t))=\argmin_{\beta_0\in\mathbb R^{p_{s}}, \beta_1\in\mathbb R^{p_{s}}}\sum_{i=1}^{n_s}\rho_\tau\left(y_{i,s}- {\mf x}^\top_{i,s}\beta_0-\mf x_{i,s}^\top\beta_1  \left(\frac{i}{n_s} - t\right)\right)K_{b_{n, s}}\left(\frac{i}{n_s} - t\right),
\end{align}
where   $\rho_\tau(x)=\tau x 1_{[0, \infty)} (x) - (1-\tau)x 1_{(-\infty, 0)} (x)$, $K(\cdot)$ is a kernel function with  $K_{b_{n, s}}(\cdot)=K\left(\frac{\cdot}{b_{n, s}}\right)$, and $b_{n, s}$ is the sequence of bandwidth  associated with the $s$-th sample ($s = 1$ and 2). 
Note that the
local linear (quantile) estimators have been extensively studied in the literature of non-parametric statistics for both independent and dependent data, see for example,  \cite{yu1998local}, \cite{chaudhuri1991nonparametric}, \cite{dette2008non}, \cite{qu2015nonparametric}, \cite{wu2017nonparametric}, \cite{wu2018simultaneous} among many others. Among them, \cite{wu2017nonparametric} investigated the estimator \eqref{Localquantile} with locally stationary covariates and errors, and this locally stationary processes have been developed in the literature to model the slowly changing stochastic structure, which can be found in many real world time series data; see for instance, \cite{dahlhaus1997fitting}, \cite{zhou2009local}, \cite{dette2020prediction}, \cite{dahlhaus2019towards}. These articles motivated us to work on the hypothesis \eqref{hypo1} assuming local stationarity. 

We now estimate $m_1(t)$ and $m_2(t)$ through a biased-corrected estimate of $\tilde {\bs \theta}_{\tau,s}=(\tilde  \theta_{\tau, s,1},...,\tilde  \theta_{\tau, s,p_s})^\top$ for $s=1$ and 2. That is  
\begin{align}\label{mt}
\hat m_1(t)=\mf c_1^\top\tilde {\bs\theta}_{\tau,1}(t), \ \  \hat m_2(t)=\mf c_2^\top\tilde {\bs\theta}_{\tau,2}(t),
\end{align}
where for $s=1$ and 2, and for $1\leq j\leq p_s$, 
\begin{align}\label{debiased}
\tilde \theta^{(b_{n, s})}_{\tau, s,j}(t)=2\hat \theta_{\tau, s,j}^{(\frac{b_{n, s}}{\sqrt 2})}(t)-\hat \theta^{(b_{n, s})}_{\tau,s,j}(t).
\end{align}
The superscript inside the parentheses denotes the bandwidth used for the corresponding estimator.  Notice that 
$(\tilde \theta^{(b_{n, s})}_{s,j}(t),1\leq j\leq p_s, b_{n, s}\leq t\leq 1-b_{n, s})$ are equivalent to the local linear quantile estimators using the second-order kernel $\bar K(x)=
2\sqrt 2K(\sqrt 2 x)-K(x)$. It can be shown similarly to Section 4.1 of \cite{dette2019detecting} that $\tilde \theta^{(b_{n, s})}_{\tau, s,j}(t)$ has a bias at the order of $b_{n, s}^3$, while the unadjusted estimator $\hat  \theta^{(b_{n, s})}_{\tau, s,j}(t)$ has a bias of the order $O(b_{n, s}^2)$, which is non-negligible and hard to evaluate.  Therefore, the de-biased estimator has been widely applied in non-parametric inference, see for example \cite{schucany1977improvement} and  \cite{wu2007inference}. The superscript will be omitted in the rest of  the article  for the sake of notational simplicity. 

\subsection{Basic ideas and the tests}\label{BI}
Suppose that $H$ is a smooth kernel function, $h_s$ ($s = 1$ and $2$) is a sufficiently small bandwidth, and $N$ is a sufficiently large number. We then estimate $(m^{-1}_1)'(t)$ and $(m^{-1}_2)'(t)$, which are denoted by $\hat g_1(t)$ and $\hat g_2(t)$, respectively :  
\begin{align} \label{g1} 
\hat g_{1}(t) = \frac{1}{Nh_1}\sum\limits_{i = 1}^{N}H\left(\frac{\hat{m}_{1}(\frac{i}{N}) - t}{h_1}\ \right), ~\mbox{and}~~\hat g_{2}(t)  = \frac{1}{Nh_{2}}\sum\limits_{i = 1}^{N}H\left (\frac{\hat{m}_{2}(\frac{i}{N}) - t}{h_{2}}\right).
\end{align} Notice that $N$ is not the sample size; it is used for Riemann approximation. Further, observe that 
\begin{align}\label{hatgst}
\hat g_{s }(t)  &\approx  \frac{1}{Nh_{s}}\sum\limits_{i = 1}^{N}H\left(\frac{{m}_{s}(\frac{i}{N}) - t}{h_s}\right)
\approx  \frac{1}{h_s} \int\limits_{0}^{1}H\left(\frac{{m}_{s}(x) - t}{h_s}\right) dx  \\
&=  \int\limits_{({m}_{s}(0) - t))/{h_s}}^{({m}_{s}(1) - t))/{h_{s}}}  H(u) (({m}_{s})^{-1})'(t + uh_s) du\approx  (({m}_{s})^{-1})'(t)  \mathbf{1} ( {m}_{s}(0) <  t  < {m}_{s}(1) ),
\nonumber   
\end{align}
where $\mf 1(A)$ denotes the indicator function of set $A$. Therefore, the estimator defined in \eqref{g1} is a smooth approximation to the step function $(({m}_{s})^{-1})'(t)  \mathbf{1} ( {m}_{s}(0) <  t  < {m}_{s}(1) )$ and is differentiable with respect to $t$. Such type of estimator was proposed by \cite{dette2006simple} and studied extensively by \cite{dette2019detecting} for locally stationary time series models. 

Now, using $\hat g_s$ ($s = 1$ and 2), one  can estimate $m^{-1}_s(t)$ ($s = 1$ and 2)  by 
\begin{align}
\hat G_s(t)=\int\limits_{m_s(0)}^t \hat g_s(u)du.
\end{align} 
This fact motivates us to estimate the horizontal shift $d$ under null hypothesis as follows. Note that for $0\leq t\leq 1-d$, we have $m_1(t)=m_2(t+d)=u$ when $m_1(0)\leq u\leq m_1(1-d)$, and therefore,  
\begin{align}\label{destimate}d=m_2^{-1}(u)-m_1^{-1}(u), ~~m_1(0)\leq u\leq m_1(1-d).\end{align} This fact drives us to estimate $d$ 
by \begin{align}
\hat d=\frac{\int\limits_{\hat m_1(0)}^{\hat m_1(1-\tilde d)}[\hat G_2(u)-\hat G_1(u)]du}{\hat m_1(1-\tilde d)-\hat m_1(0)},
\end{align}
where $\tilde d=\hat m^{-1}_2(\hat m_1(0))$ is a preliminary estimator of  $d$ by letting $u=m_1(0)$ in \eqref{destimate}. 
With this $\hat d$, one can therefore estimate the endpoints of intervals in \eqref{destimate}, i.e., $a:=m_1(0)$ and   $b:=m_1(1-d)$. Let $\hat{a}$ and $\hat{b}$ be the estimators of $a$ and $b$, respectively, where $\hat a:=\hat m_1(0)$ and $\hat b:=\hat m_1(1-\hat d)$ under null hypothesis, and their properties under null and alternative are discussed in detail in Proposition \ref{propbound} of the supplementary material.

Next, to formulate the test statistic, we use the fact in Proposition \ref{prop-monotone} and 
propose the SIT and the SCB tests to check the hypothesis described in \eqref{hypo1} based on $\hat g_1(t)-\hat g_2(t)$. For the SIT test, the test statistics is  defined as 
\begin{align}\label{Tn1n2}
T_{n_1,n_2}=\int (\hat g_1(t)-\hat g_2(t))^2\hat w(t)dt, ~\text{where} ~~\hat w(t)=\mf 1(\hat a+\eta\leq t\leq \hat b-\eta),
\end{align}
and $\eta=\eta_{n_1,n_2}$ is a positive sequence that diminishes sufficiently slowly as $n_1,n_2\rightarrow \infty$. For instance, one may consider $\eta_{n_1, n_2}$ vanishes at the rate of $ \frac{1}{\log (n_1 + n_2)}$. 
The purpose of introducing $\eta$ here is to avoid the issues related to the boundary points; for details, see remark \ref{remark1} of the supplementary material. Observe that $T_{n_1,n_2}$ is an estimate of distance between $(m^{-1}_1)'(t)$ and $(m^{-1}_2)'(t)$ in $\mathcal L^2$  sense, and we shall reject the null hypothesis when $T_{n_1,n_2}$ is a large enough. The second test is the simultaneous confidence band 
centered around $\hat g_1(t)-\hat g_2(t)$, whose detailed expression is provided in the statement of Theorem \ref{SCB}. Using the relation between the testing of hypothesis and the confidence band, it is easy to see that the SCB test is rejected at significance level $\alpha$ if the curve $(x\in[\hat a+\eta, \hat b-\eta], y=0)$ is not entirely contained by the $100(1-\alpha)\%$ SCB. 
{\color{black}\begin{remark}\label{remak1}
		We now discuss the relaxation of the monotonicity assumption of $m_1(t)$ and $m_2(t)$. Consider $0=a_0<a_1<...<a_{k+1}=1$ for some $k>0$, where on each interval $I_i=(a_i,a_{i+1}]$, $m_1(t)$ is monotone either increasing or decreasing, and $\frac{\partial^{s_i}}{\partial t^{s_i}} m_1(t)|_{t=a_i}=0$, and $\frac{\partial^{s_{i+1}}}{\partial t^{s_{i+1}}} m_1(t)|_{t=a_i}\neq 0$ for $1\leq i\leq k$. Here $\displaystyle\max_{1\leq i\leq k}s_i$ is called the maximal varnishing order of $m_1(t)$ and will be at least $1$. Let $u_i=m_1(a_i)\vee m_1(a_{i+1})$ and $l_i=m_1(a_i)\wedge m_1(a_{i+1})$. Then by the argument of \eqref{hatgst}, $\hat g_1(t)$  approximately {\color{black} equals with} 
		\begin{align}
		g_1(t):=\sum_{i=1}^k|((m_1(t)\mathbf 1(t\in (l_i,u_i]))^{-1})'\mathbf 1(t\in (l_i,u_i])|.
		\end{align}
		Similarly, $\hat g_2(t)$ approximates $g_2(t)$ where $g_2(t)$ can be defined similarly to $g_1(t)$. By the decomposition \eqref{5.33-3-26} and \eqref{5.34-3-26} in the supplemental material and proof of Theorem 4.1 in \cite{dette2019detecting}, we conjecture that our proposed Bootstrap tests {\color{black}(i.e., Algorithms \ref{algo1} and \ref{algo2})} will be consistent for the hypothesis $g_1(t)=g_2(t)$, with the rate of detectable local alternative adjusted by a function of $h_1$, $h_2$ and  the maximal varnishing orders of $m_1(t)$ and $m_2(t)$. {\color{black} Note} that under the null hypothesis of shift invariance \eqref{hypo1}, $g_1(t)=g_2(t)$. Therefore, if we exclude the pairs of curves belonged to the class $\mathcal F:=\{m_1(t)\neq m_2(t), g_1(t)=g_2(t)\}$ from the alternatives, our proposed testing procedure is consistent and asymptotically correct. Notice that by Proposition \ref{prop-monotone}, all pairs of monotone functions $(m_1(t), m_2(t))\not \in \mathcal F$. Simulation studies in {\color{black}Table \ref{tab9}} support this observation, while we leave the theoretical justification as a future work. {\color{black} On the other hand, to the best of our knowledge, there is no test of monotonicity for time series data in the literature.}
\end{remark}}
\section{Asymptotic Results}\label{Asy-Result}
\setcounter{equation}{0} 

In this section, we investigate the  asymptotic properties of $T_{n_1,n_2}$ and the  asymptotic form of the SCB at a presumed significance level $\alpha$. We start from a few concepts and assumptions for the model described in \eqref{testmodel}. Let $(\zeta^{(1)}_i)_{i\in \mathbb Z}$, $(\zeta^{(2)}_i)_{i\in \mathbb Z}$,  $(\eta^{(1)}_i)_{i\in \mathbb Z}$ and $(\eta^{(2)}_i)_{i\in \mathbb Z}$ be i.i.d.\ random vectors, and the filtrations for $s=1$ and $2$ are the following:    $\FF_{i,s}=(\zeta^{(s)}_{-\infty},...,\zeta^{(s)}_{i-1},\zeta^{(s)}_i)$ and $\GG_{i,s}=(\eta^{(s)}_{-\infty},...,\eta^{(s)}_{i-1},\eta^{(s)}_i)$. We assume that the covariates and errors are both locally stationary process in the sense of \cite{zhou2009local},   i.e., $$\mf x_{i,s}=\mf H_s\left(\frac{i}{n_s},\GG_{i,s}\right), e_{i,s}=L_s\left(\frac{i}{n_s},\FF_{i,s},\GG_{i,s}\right)$$ for $s=1$ and 2, where $\mf H_s$ and $L_s$ are the marginal filters. We list some basic assumptions of processes $\mf x_{i,s}$ and $e_{i,s}$ in conditions  (A3) and (A4).
Now, write $\FF^*_{i,s}=(\FF_{-1,s},\zeta^{(s),*}_{0},\zeta^{(s)}_{1}...,\zeta^{(s)}_{i-1},\zeta^{(s)}_i)$, where  $(\zeta^{(s),*}_{i})_{i\in \mathbb Z}$ is an i.i.d.\ copy of  $(\zeta^{(s)}_{i})_{i\in \mathbb Z}$ and define $\GG^*_{i,s}$ also in a similar way. For a $p$ dimensional (random) vector $\mf v : = (v_{1}, \ldots, v_{p})^\top$, let $|\mf v|=\sqrt{(\sum_{i=1}^pv_i^2)}$, and for any random vector $\mf v$, write $\|\mf v\|_q=(\E (|\mf v|^q))^{1/q}$, which is its $\mathcal L^q$ norm for some  $q\geq 1$. Let $\chi\in (0,1)$ be a fixed constant, and suppose that $M$ and $\eta$ are sufficiently large and sufficiently small positive constants, respectively; though it may vary from line to line. For any positive semi-definite matrix $\bs \Sigma$, write $\lambda_{min}(\bs \Sigma)$ as its smallest eigenvalue. We first give out the following set of conditions, which enable us to study the deviation of the nonparametric quantile estimator, $\tilde{\bs \theta}_\tau-\bs \theta_\tau$.
\begin{description}
	\item (A1) Define $\bs \theta''_{\tau,s}(t)=(\theta''_{1,\tau,s}(t),...,\theta''_{p_s,\tau,s}(t))^\top$ for $s=1$ and 2. Assume that $(\theta''_{i,\tau,s}(t),1\leq i\leq p_s)_{s=1,2}$ are Lipschitz continuous on $[0,1]$. 
	\item (A2) Define $Q_\tau(L_s(t,\FF_{i,s},\GG_{i,s})|\GG_{i,s}):=\displaystyle\inf_{x} \{x:\mathbb{P}(L_s(t,\FF_{i,s},\GG_{i,s})\geq x|\GG_{i,s})\geq \tau\}$. Assume that $Q_\tau(L_s(t,\FF_{i,s},\GG_{i,s})|\GG_{i,s})=0$.  
	\item (A3) For the errors processes, we assume that for $s=1$ and 2,
	\begin{align}
	&i)	~\delta(L,i):=\sup_{t\in[0,1]}\|L_s(t,\FF_{i,s}^*,\GG_{i,s}^*)-L_s(t,\FF_{i,s},\GG_{i,s})\|_1=O(\chi^i) ~\text{for $i\geq 0$.}\\
	&ii)	~\sup_{0\leq t_1,t_2\leq 1}\|L_s(t_1,\FF_{0,s},\GG_{0,s})-L_s(t_2,\FF_{0,s},\GG_{0,s}))\|_v\leq M|t_1-t_2|
	\end{align}
	for a constant $v\geq 1$.
	\item (A4) For covariate processes, we assume that   for $s=1$ and $2$, there exists a constant $t_x>0$ such that,
	\begin{align}
	&i)~	\max_{1\leq i\leq n_s}\E(\exp(t_x|\mf x_{i,s}|))\leq M<\infty,\\
	&ii)~\delta(\mf H,i):=	\sup_{t\in [0,1]}\|\mf H_s(t,\GG^*_{i,s})-\mf H_s(t,\GG_{i,s}))\|_1=O(\chi^i) ~~~\text{for $i\geq 0$},\\
	&iii)~	\sup_{0\leq t_1,t_2\leq 1}\|\mf H_s(t_1,\FF_{0,s},\GG_{0,s})-\mf H(t_2,\FF_{0,s},\GG_{0,s}))\|_2\leq M|t_1-t_2|.
	\end{align}
	\item (A5) For  conditional densities, we define for $s=1$ and 2, and for $0\leq q\leq  2p_s+1$,
	\begin{align}
	F_s^{(q)}(t,x|\FF_{i-1,s},\GG_{i,s})=\frac{\partial^q}{\partial x^q}\p(L_s(t,\FF_{i,s},\GG_{i,s})\leq x|\FF_{i-1,s},\GG_{i,s}).
	\end{align}
	In particular, we write $f_s(t,x|\FF_{i-1,s},\GG_{i,s})=F_s^{(1)}(t,x|\FF_{i-1,s},\GG_{i,s})$ for brevity.
	Assume that $\displaystyle\sup_{t\in[0,1],x\in \mathbb R}|F_s^{(q)}(t,x|\FF_{i-1,s},\GG_{i,s})|\leq M$ almost surely. Further, define 
	\begin{align}
	\delta_{l,s}^{(q)}(i-1)=\sup_{t\in [0,1],x\in \mathbb R}\|F_s^{(q)}(t,x|\FF_{i-1,s}^*,\GG_{i,s})-F_s^{(q)}(t,x|\FF_{i-1,s},\GG_{i,s})\|_l, 
	\end{align}
	and assume that $\delta_{1,s}^{(q)}(i-1)=O(\chi^i)$ for $i\geq 0$.
	\item (A6) Define for $s=1$ and 2, conditional on $\GG_{i,s}$, the conditional density and the quantile design matrix as  
	\begin{align}
	f_s(t,x|\GG_{i,s})=\frac{\partial}{\partial x}\p(e_{i,s}(t)\leq x|\GG_{i,s}), \bs \Sigma_s(t)=\E(f_s(t,0|\GG_{i,s})\mf H_s(t,\GG_{i,s})\mf H_s^\top(t,\GG_{i,s})).
	\end{align}
	Assume that \begin{align}
	\sup_{t\in[0,1]}\|f_s(t,0|\GG_{i,s})-f_s(t,0|\GG_{i,s}^*)\|_2=O(\chi^i)~~\text{for $i\geq 0$},\\
	\inf_{t\in [0,1]}\lambda_{min}(\bs \Sigma_s(t))\geq \eta>0 ~\mbox{and} \sup_{t\in [0,1]}|f_s^{(1)}(t,0|\GG_{0,s})|\leq M, a.s..
	\end{align}
	\item (A7) Let $\psi_\tau (x) = \tau -\mf 1 (x \leq  0)$ be the left derivative of $\rho_\tau(x)$. For $s=1$ and 2, define the gradient vector process
	\begin{align}
	\mf U_s(t,\FF_{i,s}, \GG_{i,s})=\psi_\tau(L_s(t,\FF_{i,s},\GG_{i,s}))\mf H_s(t,\GG_{i,s}).
	\end{align}
	Notice that by definition, $\mf U_s(\frac{i}{n_s},\FF_{i,s},\GG_{i,s})=\psi_\tau(e_{i,s})\mf x_{i,s}$, which is the gradient vector. Now define the long run covariance matrices for $\mf U_s$, which is
	\begin{align}\label{lrv}
	\mf V_s^2(t)=\sum_{j=-\infty}^\infty \Cov(\mf U_s(t,\FF_{0,s},\GG_{0,s}),\mf U_s(t,\FF_{j,s},\GG_{j,s}) ).		\end{align}
	Assume that for $s=1$ and 2, there exists an $\tilde \eta>0$, s.t.
	\begin{align}
	\inf_{ t\in (0,1]}\lambda_{min}(\mf V_s^2(t))\geq \tilde \eta>0.
	\end{align}
	\item (A8)  The kernel functions $K(\cdot)$ and $H(\cdot)$ are symmetric and twice differentiable functions with support $[-1,1]$. Also, $\int\limits_{-1}^1 K(x)dx=1$, $\int\limits_{-1}^1 H(x)dx=1$, and $K''$, $H''$ are Lipschitz continuous on $[-1,1]$. 
\end{description}

Conditions (A1)-(A8) are associated with {\color{black}the smoothness of} the quantile regression coefficients, conditional quantiles, errors and covariates.   The quantities $\delta(L,i)$ and $\delta(\mf G,i)$ are called `physical dependence measure' in the literature (see \cite{zhou2009local}), and ii) of conditions (A3) and (A4) postulate stochastic Lipschitz continuity for $L_s$ and ${\bf H}_s$, respectively. In fact, conditions (A3) and (A4) ensure that the errors and covariates are both locally stationary processes with geometrically decaying dependence measure. The verification of these conditions is uncomplicated for a general class of locally stationary processes; we refer to \cite{zhou2009local} for more details.  Condition (A5) is a standard assumption on the dependence measures of the derivatives
of the errors' conditional densities for non-stationary time series quantile regression, see \cite{wu2017nonparametric}, \cite{wu2018gradient} among many others for details and \cite{zhou2009local} for the verification of this condition on representative examples.  Assumption (A6) ensures that the process  $\sum_{i=1}^{n_{s}} f_s (i/n_{s}, 0|\GG_i )\mf x_{i,s}\mf x_{i,s}^\top K_{b_{n, s}} (i/n_{s}-t)/(n_{s}b_{n, s})$ converges to the non-degenerate quantile design matrix. Similar conditions are also assumed in \cite{kim2007quantile},  \cite{qu2008testing} and a few references therein. Condition (A7) means that the long-run covariance matrices of the gradient vectors	$\mf U_s(t,\FF_{i,s}, \GG_{i,s})$ are non-degenerate. Condition (A8) is a mild condition for kernels, and the well known Epanechnikov and many more kernel functions satisfy the assumptions stated in (A8). Notice that conditions (A1)-(A7) generalize the conditions (A1)-(A5) of \cite{wu2017nonparametric} for multiple curves.  We then consider a few more conditions (B) on the bandwidth and the regression function.
\begin{description}
	\item (B1)  For $s=1$ and $2$, $\frac{(\log^{3/2} n_s)}{\sqrt {n_s}b_{n,s}^2}=o(1)$, $ h_s/b_{n,s}=o(1)$,  $\pi_{n,s}=o(\sqrt{\log n_s})$, $\frac{\log^{4/3}n_s}{\sqrt {n_sb_{n,s}}h_s}=o(1)$, and $h_s/b^2_{n,s}\rightarrow \infty$.  Let $\Omega_n=\sum\limits_{s=1}^2 \left(\Theta_{n,s}+h_s+\frac{1}{Nh_s}+\frac{\log^4 n_s}{(nb_{n,s})^{3/2}h_s^2}\right)$ and assume that $\Omega_n=o\left(\sum\limits_{s=1}^2(n_sb_{n,s}^{5/2})^{-1/2}\right)$.
	\item(B2) $(m_2')^{-1}(m_1(0))\in (0,1)$.
	\item (B3) $b_{n,s}\rightarrow 0$, $\frac{n_sb_{n,s}^4}{\log ^8n_s}\rightarrow \infty$, $n_s^\iota b_{n,s}=o(1)$ for some $\iota>0$, and $s=1,2$.
\end{description} 
The condition (B1) implies that in practice we should choose $h_s$ small, which was remarked by \cite{dette2006simple} also. Further, (B1) ensures that our proposed estimators $\hat a$ and $\hat  b$ is well defined under alternative hypothesis, and (B2) means that $d\in(0,1)$ under null. Condition (B3) guarantees that the nonparameteric estimate $\hat m_s$ approximates well $m_s$, $s=1,2$.


Before stating the main results on $T_{n_1,n_2}$ and SCB, we introduce a few more notation. Define for $s=1$ and 2,  $\mathcal T_{n,s}=( b_{n,s}, 1-b_{n,s})$ and $M_{\mf c_s}(t)=((\mf c_s^\top \bs \Sigma_s^{-1}(t))\mf V_{s}^2(\mf c_s^\top \bs \Sigma_s^{-1}(t))^\top)^{1/2}$,
\begin{align}
\check g_s(t)=M_{\mf c_s}(m_s^{-1}(t))((m_s^{-1})'(t))^2\int\limits_{\mathbb R}  H'(y)ydy, \\ \tilde g_s(m_s(u))=\check g_s^2(m_s(u))w(m_s(u))m_s'(u),
\check V_s=\int\limits_\mathbb R\int\limits_\mathbb R\tilde g^2_s(m_s(u)) (\bar K'\star \bar K'(y))^2dudy.
\end{align}
Write $m_{2,1}(u)=m_{2}^{-1}(m_1(u))$, $\tilde g_{1,2}(m_1(u))=\check g_1(m_1(u))\check g_2(m_1(u))w(m_1(u))m_1'(u)$ and $\check V_{12}(r)=\int\limits_{\mathbb R} \int\limits_{\mathbb R}  \tilde g^2_{1,2}(m_1(u))\left[\int\limits_{\mathbb R}  \bar K'(x)\bar K'(rm_{21}'(u)x+y)dx\right]^2dudy$. For $s=1$ and $2$, let $$\check B_s=\left(\int\limits_{\mathbb R}  H'(y)dy\right)^2\int\limits_{\mathbb R}  \bar K^2(x)dx \int\limits_{\mathbb R}  M_{\mf c_s}^2(u)u^4m_s'(u)w(m_s(u))du.$$ Notice that under the null hypothesis \eqref{hypo1}, $m'_{2,1}(u)\equiv 1$.
\begin{theorem}\label{asymptotic}
	Assume the conditions stated in (A1)-(A8) and (B1), (B2), (B3). 
	Now, suppose that $n_1/n_2\rightarrow \gamma_0$, $b_{n,1}/b_{n,2}\rightarrow \gamma_1$,  $\eta^{-1}=O(\log(n_1+n_2))$,  $\eta=o(1)$.
	Further, let $(m_1^{-1})'(t)-(m_2^{-1})'(t)=\rho_n\kappa(t)$ for some bounded function $\kappa(t)$, and $\rho_n:=\rho_{n_1,n_2}=(n_1b^{5/2}_{n_1})^{-1/2}$.  We then have 
	\begin{align}
	n_1b_{n,1}^{5/2}T_{n_1,n_2}-b_{n,1}^{-1/2}(\check B_1+\gamma_0\gamma_1^3\check B_2))-\int_\mathbb R \kappa^2(t)w(t)dt\Rightarrow N(0,V_T),
	\end{align}
	where $V_T=\check V_1+\gamma_0^2\gamma_1^5\check V_2+4\gamma_0\gamma_1^3\check V_{12}(\gamma_1)$.
\end{theorem} 

Under the null hypothesis, $\kappa\equiv 0$. Therefore,  Theorem \ref{asymptotic} suggests to reject null hypothesis of  \eqref{hypo1} whenever 
\begin{align}
T_{n_1,n_2}>\frac{b_{n,1}^{-1/2}\left(\hat{\check B}_1+\gamma_0\gamma_1^3\hat{\check B}_2\right)+z_{1-\alpha}\hat V_T^{1/2}}{n_1b_{n,1}^{5/2}},
\end{align}
where $\alpha$ is the significance level, $z_{1-\alpha}$ is the $(1-\alpha)$-th quantile of a standard normal distribution, $\hat{\check B}_1$,  $\hat{\check B}_2$ and $\hat V_T$ are appropriate estimates of asymptotic bias parameters $B_1$, $B_2$ and the asymptotic variance $V_T$, respectively. Moreover, Theorem \ref{asymptotic} shows that the SIT test is able to detect the alternative which converges to null at a rate of $\sqrt{(n_1b_{n,1}^{5/2})}$, with asymptotic power
\begin{align}
1-\Phi\left(z_{1-\alpha}-\frac{\int\limits_{\mathbb R} \kappa^2(t)w(t)dt}{V^{1/2}_T}\right),
\end{align}
where $\Phi(\cdot)$ denotes the CDF of a standard normal random variable.
\begin{theorem}\label{SCB}
	Assume the conditions stated in (A1)-(A8) and (B1), (B2), (B3) hold. 
	Further, assume that for $s=1$ and 2,  $\frac{b_{n,2}}{b_{n,1}}\rightarrow c_{b,2}\in(0,1)$, $\frac{n_2}{n_1}\rightarrow c_{n,2}\in (0,1)$, $\frac{h_2}{h_1}\rightarrow c_{h,2}\in (0,1)$, $\eta=o(1)$, $\eta^{-1}=O(\log(n_1+n_2))$, and let $c_{b,1}=c_{n,1}=c_{h,1}=1$. Define \begin{align}
	K_1(t)=\sum_{s=1}^2 \frac{M^2_{\mf c_s}(m_s^{-1}(t))(m_s^{-1})'^2(t)}{c_{n,s}c^3_{b,s}(m_s'(m_s^{-1}(t)))^2}\int\limits_{\mathbb R} \bar K'^2(y)dy\left(\int\limits_{\mathbb R} H'(x)xdx\right)^2,\\
	K_2(t)=\sum_{s=1}^2 \frac{M^2_{\mf c_s}(m_s^{-1}(t))}{c_{n,s}c_{b,s}c_{h,s}^4(m_s'(m_s^{-1}(t)))^2}\int\limits_{\mathbb R} \bar K^2(y)dy\left(\int\limits_{\mathbb R} H''(x)dx\right)^2.
	\end{align} Then, if  $(m_1^{-1})^{'}(t)-(m_2^{-1})^{'}(t)=\rho_{n_1,n_2}\kappa(t)$ for some non-zero bounded function $\kappa(t)$ and $\rho_{n_1,n_2}=o(\eta)$, as $\min(n_1, n_2)\rightarrow\infty$, we have
	\begin{align}
	\p\left(\sup_{t\in \mathcal I_{\hat a, \hat b}}\frac{\sqrt{n_1b_{n,1}^3}|(\hat m_1^{-1})'(t)-(\hat m_2^{-1})'(t)-((m_1^{-1})'(t)-(m_2^{-1})'(t))|}{K^{1/2}_1(t)}\geq \sqrt{-2\log(\pi\alpha/\kappa_0)} \right)\rightarrow \alpha,
	\end{align}
	where $\mathcal I_{\hat a, \hat b}=(\hat a+\eta, \hat b-\eta)$ and
	\begin{align}
	\kappa_0=\frac{b_{n,1}}{h_1^2}\int\limits_{m_1(0)}^{m_1(1-m_2^{-1}(m_1(0)))} \left(\frac{K_2(t)}{K_1(t)}\right)^{1/2}dt.\label{Kappa}
	\end{align}
\end{theorem}
Theorem \ref{SCB} gives us the following simultaneous confidence band of $(m_1^{-1})^{'}(t)-(m_2^{-1})^{'}(t)$:
\begin{align}\label{SCB-Aug17}
(\hat m_1^{-1})'(t)-(\hat m_2^{-1})'(t)\pm \frac{\hat K^{1/2}_1(t)}{\sqrt{n_1b_{n,1}^3}}\sqrt{-2\log(\pi\alpha/\hat \kappa_0)},~~ t\in \mathcal I_{\hat a, \hat b},
\end{align}
where $\hat \kappa_0=\frac{b_{n,1}}{h_1^2}\int\limits_{\hat a+\eta}^{\hat b-\eta} \sqrt{\frac{\hat K_2(t)}{\hat K_1(t)}}dt $, and $\hat K_1(t)$ and $\hat K_2(t)$ are appropriate estimates of $K_1(t)$ and $K_2(t)$, respectively. Therefore we can reject the null hypothesis \eqref{hypo1} at significance level $\alpha$. Furthermore, it follows from condition (B) and \eqref{Kappa} that the width of \eqref{SCB-Aug17} is $\frac{\sqrt{\log n-\log \alpha}}{\sqrt{n_1b_{n,1}^3}}$. Consequently, the SCB test is able to detect the alternative converging to null at a rate of $\frac{\sqrt{\log n}}{\sqrt{n_1b_{n,1}^{3}}}$, which indicates the SIT test is asymptotically more powerful than the SCB test when bandwidths are of the same order. However, for moderately large sample size, the SCB test performs
well when  $(m_1^{-1})^{'}(t) - (m_2^{-1})^{'}(t)$ is `bumpy',  or equivalently the majority part of two curves $m_1(t)$ and $m_2(t)$ are same up to the horizontal shift while minor parts of $m_1(t)$ and $m_2(t)$ have notably different shapes so that their differences cannot be eliminated by a horizontal shift.

\section{Implementation of the tests}\label{IT}

\subsection{Estimation of $ M_{\mf c_s}(t)$}\label{estM}
The implementation of the SIT test and the SCB test require the estimation of $M_{\mf c_s}(t)$.  For $s = 1$ and 2, let $\hat e_{i,s}(t)=y_{i,s}-\mf x_{i,s}^\top\hat {\bs\theta}_{\tau,s}(t)$, and
\begin{align}\label{sigmas-estimate}
\hat{\bs \Sigma}_s(t)=\frac{1}{n_sb_{n,s}w_{n,s}}\sum_{i=1}^{n_s}\phi\left(\frac{\hat e_{i,s}(t)}{w_{n,s}}\right)\mf x_{i,s}\mf x_{i,s}^\top K_{b_{n, s}}(i/n_s-t),
\end{align}
where the bandwidth $w_{n,s}$ is such that $w_{n,s}=o(1)$, $n_sw_{n,s}\rightarrow \infty$, and $\phi(\cdot)$ is the probability density function of the standard normal distribution.
It has been shown in Theorem 6 in \cite{wu2017nonparametric} that with appropriate choices of $w_{n, s}$, $\hat{\bs \Sigma}_s(t)$ is a consistent estimator of $\bs \Sigma_s(t)$ uniformly on $[b_{n,s},1-b_{n,s}]$. To estimate $\mf V_s(t)$,
we define \begin{align}
\hat {\bs \Xi}_{i,s}=\sum_{j=-M_s}^{M_s}\left(\tau-\mf 1\left(\hat e_{i+j,s}\left(\frac{i+j}{n_s}\right)\leq 0\right)\right)\mf x_{i+j,s},
\end{align}
where $M_s\rightarrow \infty$, and $M_s=o(n_s)$ is the window size.

Furthermore, let $\hat{\bs \Delta}_i=\frac{\hat {\bs \Xi}_{i,s}\hat {\bs \Xi}_{i,s}^\top}{2M_s+1}$, and
\begin{align}
\hat {\mf V}_s(t)=\frac{1}{n_sb_{n,s}}\sum_{i=1}^{n_s}K_{b_{n,s}}(i/n_s-t)\hat {\bs \Delta}_i.
\end{align}
With appropriate choices of bandwidth, Theorem 5 in \cite{wu2017nonparametric} shows that $\hat {\mf V}_s(t)$ converges to $\mf V(t)$ uniformly on $\left[b_{n,s}+\frac{M_s+1}{n_s},1-b_{n,s}-\frac{M_s+1}{n_s}\right]$.
We then estimate ${ M}_{\mf c_s}(t)$ by
\begin{align}  
\hat { M}_{\mf c_s}(t):=\left(\left(\mf c_s^\top \hat {\bs \Sigma}_s^{-1}(t)\right)\hat {\mf V}_{s}^2\left(\mf c_s^\top \hat{\bs \Sigma}_s^{-1}(t)\right)^\top\right)^{1/2}
\end{align}
for $t\in [b_{n,s}+\frac{M_s+1}{n_s}, 1-b_{n,s}-\frac{M_s+1}{n_s}]$, and $\hat { M}_{\mf c_s}(t)=\hat { M}_{\mf c_s}(b_{n,s}+\frac{M_s+1}{n_s})$ for $0\leq t< b_{n,s}+\frac{M_s+1}{n_s}$, $\hat { M}_{\mf c_s}(t)=\hat { M}_{\mf c_s}(1-b_{n,s}-\frac{M_s+1}{n_s})$ for $t\in (1-b_{n,s}-\frac{M_s+1}{n_s},1]$. Consequently, the estimator $\hat { M}_{\mf c_s}(t)$ is a consistent estimator of $ M_{\mf c_s}(t)$ under appropriate choices of $M_s$ and $w_{n,s}$ which will be discussed in the next section.

\subsection{Bandwidth Selection}\label{BS}
In this section, we first discuss the choices of the smoothing parameters, namely, $b_{n,s}$ and $h_{s}$ ($s = 1$ and 2)  for calculating $T_{n_1,n_2}$. According to Dette et al. (2006), when $h_s$ is sufficiently small, it has a negligible impact on the test, and therefore, by considering bandwidth conditions (B), we recommend choosing $h_s=n_s^{-1/3}$ as a rule of thumb. For $b_{n, s}$, we propose to choose this tuning parameter by a corrected-Generalized Cross Validation (C-GCV) method (see \cite{craven1978smoothing}). Notice that for the local linear regression with bandwidth $b$, the estimator can be written as $\hat {\mf Y}_s=\mf D_s(b) \mf Y_s$ for some $n_s\times n_s$ matrix $\mf D_s(b)$, and then the GCV selects
\begin{align}
\hat b_{n,s,mean}=\arg\min_{b} GCV(b), ~\mbox{where}~~GCV (b) = \frac{n_s^{-1}(\hat {\mf Y}_s(b)-\mf Y_s)^\top(\hat {\mf Y}_s(b)-\mf Y_s)}{(1-trace(\mf D_s(b)/n_s))^2}.
\end{align}
Following the arguments of \cite{yu1998local}, it is appropriate to select $\hat b_{n,s}$ by correcting $\hat b_{n,s,mean}$. First, we define
$\hat b^o_{n,s}=2C_s\hat b_{n,s,mean}$
with
\begin{align}
C_s=\left(\frac{\int\limits_0^1 M_{\mf 1_s}(t)dt}{\int\limits_0^1{trace({\tilde {\mf M}}_s(t)})dt}\right)^{1/5},
\end{align}
where $ M_{\mf 1_s}(t)=((\mf 1_s^\top \bs \Sigma_s^{-1}(t))\mf V_{s}(t)^2(\mf 1_s^\top \bs \Sigma_s^{-1}(t))^\top)^{1/2}$, $\mf 1_s$ is a 
$p_s$-dimensional identity vector, $\mf V_s(t)$ is the same as defined in \eqref{lrv}, $\tilde{\mf M}_s(t)=\tilde{\bs \Sigma}^{-1}_s(t)\tilde {\bs \Lambda}_s(t)\tilde {\bs\Sigma}^{-1}_s(t)$ with
\begin{align}
\tilde{\bs \Lambda}_s(t)=\sum_{j=-\infty}^\infty\Cov(\mf H_s(t,\GG_{0,s})\tilde L(t,\FF_{0,s},\GG_{0,s}),\mf H_s(t,\GG_{j,s})\tilde L(t,\FF_{j,s},\GG_{j,s})),
\end{align}
$(\tilde L(t,\FF_{j,s},\GG_{j,s}),1\leq j\leq n_s)$ is the  errors process of local linear regression for $s$-th sample ($s=1$ and 2), and
$\tilde {\bs \Sigma}_s(t)=\E (\mf H_s(t,\GG_0)\mf H_s(t,\GG_0)^\top)$. We refer to \cite{zhou2010simultaneous} for the estimation of $\tilde {\mf M}_s(t)$. Then as recommended by \cite{zhou2010nonparametric}, for the SIT test, we use $\hat b_{n,s}=\hat b^o_{n,s}\times n_s^{-1/45}$ while 
for the SCB test, we use $\hat b_{n,s}=\hat b^o_{n,s}$.

We now discuss the selection of $w_{n,s}$ and $M_s$ for the estimation of the quantity $\hat { M}_{\mf c_s}(t)$ in Section \ref{IT}. 
As a rule of thumb, we propose to choose $M_s=\lf n_s^{1/3}\rf$ and select $w_{n,s}$ by minimum volatility method. Specifically, consider a grid of possibly $w_{n,s}$: $\{w_{s,1},...,w_{s,k}\}$. Together with $M_s$ and $b_{n,s}$, one can calculate
$\hat { M}_{\mf c_{s,1}},...,\hat { M}_{\mf c_{s,k}}$ using $\{w_{s,1},...,w_{s,k}\}$, respectively.  Then, for a positive integer $u$ ($u=5$ say), define 
\begin{align}
ise(\hat { M}_{\mf c_{s},u}(l))=\frac{1}{u-1}\sum_{v=l}^{l+u-1}\left(\hat { M}_{\mf c_{s,v}}(t)-\frac{1}{u-1}\sum_{v=l}^{l+u-1}\hat { M}_{\mf c_{s,v}}(t)\right)^{2}.
\end{align}
Now, let $l'$ be the minimizer of $ise(\hat { M}_{\mf c_{s},u}(l))$,  and we select 
$w_{s, l'+\lf u/2\rf}$ as $w_{n,s}$.
The validity of these methods for choosing $w_{n,s}$ and $M_s$ are given in \cite{wu2017nonparametric}, which also proposed methods of tuning parameters for refinement. For simplicity, we omit the detailed description of the tuning procedure for refinement in our paper. Our empirical study finds that our choices of tuning parameters $b_{n,s}$, $h_s$, $M_s$, $w_{n,s}$ and the estimate of $ M_{\mf c_s}(t)$ work reasonably well.

\subsection{Boostrap-Based Test}\label{BBT}
Let $\{V_{j,1},j\in \mathbb Z\}$ and  $\{V_{j,2}, j\in \mathbb Z\}$ be i.i.d.\ standard normal random variables. Theorems \ref{asymptotic} and \ref{SCB} are built on the fact that the distribution of $\hat g_1(t)-\hat g_2(t)$ can be well approximated by 
$$(m^{-1}_1(t))'-(m^{-1}_2(t))'+Z(t, \{V_{j,s}\}_{j\in \mathbb Z, s=1,2}),$$
where $Z(t, \{V_{j,s}\}_{j\in \mathbb Z, s=1,2})$ is a Gaussian process defined by $Z(t, \{V_{j,s}\}_{j\in \mathbb Z, s=1,2}):$$=Z_1(t, \{V_{j,1}\}_{j\in \mathbb Z})-Z_2(t,\{V_{j,2}\}_{j\in \mathbb Z})$, and for $s=1$ and 2,
\begin{align*}
& Z_s(t,\{V_{j,s}\}_{j\in \mathbb Z})=\sum_{j=1}^{n_s} W_s(m_s,j,t)V_{j,s}, \\
&W_s(m_s,j,t)=\frac{1}{n_sb_{n,s}Nh_s^2}\sum_{i=1}^NM_{\mf c_s}(i/N)H'\left(\frac{m_s(i/N)-t}{h_s}\right)\bar K_{b_{n,s}}(j/n_s-i/N).\notag
\end{align*}

The limiting distribution is established by the asymptotic limit of quadratic form of the Gaussian process $Z(t, \{V_{j,s}\}_{j\in \mathbb Z, s=1,2})$ for Theorem \ref{asymptotic}, and the convergence of extreme values of $Z(t, \{V_{j,s}\}_{j\in \mathbb Z, s=1,2})$ for Theorem \ref{SCB}. However, the direct implementation of Theorem \ref{asymptotic} and Theorem \ref{SCB} is difficult. The former involves  a complicated bias term of the order $(b_{n, s}^{-1/2})$ to be estimated, and the latter has a slow convergence rate $O\left(\frac{1}{\sqrt {\log n_s}}\right)$, which follows from the proof of Theorem \ref{SCB}. To circumvent this difficulty, we propose the following Bootstrap-assisted algorithm based on $Z(t, \{V_{j,s}\}_{j\in \mathbb Z, s=1,2})$.

\begin{algo}\label{algo1}(Bootstrap-SIT)
	{\rm 
		~~
		
		\noindent
		(a) Estimate $m_1$ and $m_2$, $a$, $b$ and $M_{\mf c_1}(\cdot)$, $M_{\mf c_2}(\cdot)$. 
		\medskip
		
		\noindent  (b) Generate $Q$ copies of i.i.d. standard normal random variables $\{V_{j,1}^{(Q)'}\}_{j=1}^{n_1}$ and  $\{V_{j,2}^{(Q')}\}_{j=1}^{n_2}$ to obtain the statistic
		\begin{align*}
		M_{Q'}=\int\limits_{\mathbb R} 
		\Big(Z^{(Q')}_1(t, \{V_{j,1}\}_{j\in \mathbb Z})-Z^{(Q')}_2(t,\{V_{j,2}\}_{j\in \mathbb Z})\Big)^2\hat w(t)dt, ~~1\leq Q\leq Q'.
		\end{align*}
		\medskip
		\noindent 
		(c) Let $M_{(1)}\leq M_{(2)}\leq\ldots \leq {\color{black}M_{(Q)}}$ be the ordered statistics of $\{M_{s},1\leq s\leq Q\}$. We reject 
		the null hypothesis \eqref{hypo1}
		at level $\alpha$, whenever
		\begin{equation}
		\label{boottest}
		T_{n_1,n_2} > M_{(\lf Q(1-\alpha)\rf)}. 
		\end{equation}
		The $p$-value  of this  test is given by $1-Q^*/Q$,
		where $Q^*=\max\{r:M_{(r)}\leq T_{n_1,n_2}\}$. 
	}
\end{algo}

\begin{algo}\label{algo2}(Bootstrap-SCB)
	{\rm 
		~~
		
		\noindent
		(a)	 Estimate $m_1$ and $m_2$, $a$, $b$ and $M_{\mf c_1}(\cdot)$, $M_{\mf c_2}(\cdot)$. 
		\medskip
		
		\noindent  (b) Generate $Q$ copies of i.i.d.\ standard normal random variables $\{V_{j,1}^{(Q')}\}_{j=1}^{n_1}$ and  $\{V_{j,2}^{(Q')}\}_{j=1}^{n_2}$ to obtain the statistic
		\begin{align*}
		\tilde M_{Q'}=
		\sup_{t\in \mathcal I_{\hat a, \hat b}}\Big|(Z^{(Q')}_1(t, \{V_{j,1}\}_{j\in \mathbb Z})-Z^{(Q')}_2(t,\{V_{j,2}\}_{j\in \mathbb Z}))/\hat K^{1/2}_1(t)\Big|,~~ 1\leq Q'\leq Q. 
		\end{align*}
		\medskip
		\noindent 
		(c) Let $ 	\tilde M_{(1)}\leq 	\tilde M_{(2)}\leq\ldots \leq 	{\color{black}\tilde M_{(Q)}}$ be the ordered statistics of $\{	\tilde M_{s},1\leq s\leq Q\}$. Then, the $(1-\alpha)$- SCB of $(m^{-1}_1)'(t)-(m^{-1}_2)'(t)$ is 
		\begin{equation}
		(\hat m^{-1}_1)'(t)-(\hat m^{-1}_2)'(t)\pm \tilde M_{(\lf Q(1-\alpha)\rf)}\hat K_1^{1/2}(t).
		\end{equation}
	}
\end{algo}

{\color{black}By applying Algorithm \ref{algo1}}, there is no need to estimate the bias term $b_{n,1}^{-1/2}(\check B_1+\gamma_0\gamma_1^3\check B_2))$ as well as the asymptotic variance $V_T$. The validity of these algorithms are based on the approximation of $Z(t, \{V_{j,s}\}_{j\in \mathbb Z, s=1,2})$ to $\hat g_1(t)-\hat g_2(t)$ (see \eqref{g1} for the expressions of $\hat g_1(t)$ and $\hat g_2(t)$), which is discussed in detail in the proof of Theorem \ref{asymptotic}. Notice that the cutoff values $M_{(\lf Q(1-\alpha)\rf)}$ and $\tilde M_{(\lf Q(1-\alpha)\rf)}$ are obtained for fixed $n_1$ and $n_2$, while the critical values in Theorem \ref{asymptotic} and Theorem \ref{SCB} are based on the limiting distribution. Therefore, similar to \cite{zhao2008confidence}, we expect that Algorithms \ref{algo1} and \ref{algo2} will outperform the test using the critical values of Theorems \ref{asymptotic} and \ref{SCB}. Finally, to implement Algorithm \ref{algo2}, we need to estimate $K_1$, which consists of the estimate of $m^{-1}_s$, $(m^{-1}_s)'$ and $m_s'$ for $s=1$ and 2. We suggest to estimate these quantities by $\int\limits_{m_s(0)}^{t}\hat g_s(u)du$, $\hat g_s(u)$, and $\mf c_s^\top\tilde{\bs \theta}'_{\tau,s}(t)$ for $s=1$ and 2, respectively, where   $\tilde{\bs \theta}'_{\tau,s}(t)=(\tilde{ \theta}'_{\tau,s,1}(t), \ldots,\tilde{\theta}'_{\tau,s,p}(t))^\top$ with its $j_{th}$ element $2\hat \theta_{\tau, s,j}^{'(\frac{b_{n, s}}{\sqrt 2})}(t)-\hat \theta^{'(b_{n, s})}_{\tau,s,j}(t)$, and $\hat {\bs \theta}^{'(b_{n, s})}_{\tau,s}(t)=
(\hat \theta^{'(b_{n, s})}_{\tau,s,1}(t), \ldots ,\hat \theta^{'(b_{n, s})}_{\tau,s,p}(t))^\top$ is defined in \eqref{Localquantile} using bandwidth $b_{n, s}$.  
{\color{black}\section{Tests for mutual dependent series}\label{mutual-dependent}
	Algorithms \ref{algo1} and \ref{algo2} are built upon the assumption that two data sets $\tilde{\mathbf  y}_{i,1}:=(y_{i,1},\mf x_{i,1}^\top)^\top$ and $\tilde{\mathbf  y}_{i,2}:=(y_{i,2}, \mf x^\top_{i,2})^\top$ are independent of each other. As pointed out by a referee, it is important to allow the dependence among the two data sets. For this purpose, we should model the two series $\tilde{\mathbf  y}_{i,1}$ and  $\tilde{\mathbf y}_{i,2}$  {\it jointly} assuming that they are generated from certain $p_1+p_2+2$ dimensional vector process $\bs \zeta(t)_{t\in [0,1]}$.  The two vectors $\tilde{\mathbf y}_{i,1}$ and $\tilde{ \mathbf y}_{i,2}$ correspond to the first $p_1+1$ and the next $p_2+1$ components of the $\bs \zeta(t)$, respectively, but at possibly different time points. For instance, if $\tilde {\mathbf y}_{i,2}$ is collected at a subsequent period of when $\tilde {\mathbf y}_{i,1}$, then one could assume that  $\tilde {\mathbf y}_{i,1}$ is realized from $\bs \zeta(t)_{t\in \mathcal T_1}$ and  $\tilde {\mathbf y}_{i,2}$ is realized from $\bs \zeta(t)_{t\in \mathcal T_2}$ where $\mathcal T_1\in [0,1/2)$ and $\mathcal T_2\in (1/2, 1]$. On the other hand, if the two series are both collected in the same period, when the realization time of $\{\tilde{\mf y}_{i,1}\}$  and $\{\tilde {\mf y}_{i,2}\}$ are distinct, the test and the asymptotic properties will be different from when most of the observation of the two series are generated at the same time points. Therefore, an exhaustive discussion of our tests for the two mutual dependent series is prohibitive due to the page limit. In this section, we focus on the following scenario. Let $n_1\geq n_2$, and $\tilde{\mf y}_{i,1}$ are realized at time $\frac{i}{n_1}$ for $1\leq i\leq n_1$ and   $\tilde{\mf y}_{i,2}$ are realized at time $\frac{\lf in_1/n_2\rf}{n_1}$ for $1\leq i\leq n_2$. Thus the realization time of $\tilde{\mf y}_{i,2}$ is a subset of that of $\tilde{\mf y}_{i,1}$ if $n_1/n_2\in \mathbb Z^+$. When $n_1=n_2=n$, the scenario reduces to that $(\tilde{\mf y}^\top_{i,1},\tilde{\mf y}_{i,2}^\top )^\top$ is generated at time $i/n$, $1\leq i\leq n$. Recall the condition  (A7) and the definition of $\mf U_s(t,\FF_{i,s}, \GG_{i,s})$, $s=1,2$ therein. Now, define 
	$\mf U(t,\FF_{i}, \GG_{i})=(\mf U^\top_1(t,\FF_{i,1},\GG_{i,1}), \mf U^\top_2(t,\FF_{i,2},\GG_{i,2}))^\top$ and $\FF_i=((\zeta^{(1)}_{-\infty},\zeta^{(2)}_{-\infty})^\top,...,(\zeta^{(1)}_{i},\zeta^{(2)}_{i})^\top)$, $\GG_i=((\eta^{(1)}_{-\infty},\eta^{(2)}_{-\infty})^\top,...,(\eta^{(1)}_{i},\eta^{(2)}_{i})^\top)$. The modified condition (A7'):
	\begin{align}
	\inf_{ t\in (0,1]}\lambda_{min}(\mf V^2(t))\geq \tilde \eta>0,
	\end{align}
	where 
	\begin{align}\label{lrvb}
	\mf V^2(t)=\sum_{j=-\infty}^\infty \Cov(\mf U(t,\FF_{0},\GG_{0}),\mf U(t,\FF_{j},\GG_{j}) ).		\end{align}
	Let $\bs \Sigma_{\mf c}(t)=\begin{pmatrix}
	\mf c^\top \bs \Sigma_1(t) &\\& \mf c^\top \bs \Sigma_2(t)
	\end{pmatrix}$ and the $2\times 2$ matrix
	\begin{align}  \label{eq536-new}
	\mf{ M}_{\mf c}(t):=\left(\bs\Sigma_{\mf c}(t) {\mf V}_{}^2(t)\bs\Sigma_{\mf c}(t)^\top\right)^{1/2}:=\begin{pmatrix}
	M^c_{11}(t), M^c_{12}(t)\\ M^c_{12}(t),  M^c_{22}(t).
	\end{pmatrix}.
	\end{align}
	Suppose that $M^c_{21}(t)=M^c_{12}(t)$ and redefine two Gaussian processes $Z_{s}(t,\{V_{j,s}\}_{j\in \mathbb Z})$, $s=1,2$ in Section \ref{BBT} with respect to  $i.i.d.$ standard normals $\{V_{j,s}\}_{j\in\mathbb Z,s=1,2}$ as 
	\begin{align}
	Z_1(t, \{V_{j,1}\}_{j\in \mathbb Z}  )=Z_{11}(t)-Z_{21}(t), ~~~
	Z_2(t, \{V_{j,2}\}_{j\in \mathbb Z} )=Z_{12}(t)-Z_{22}(t), \label{newZ1}
	\end{align}
	where
	\begin{align*}
	Z_{11}(t)=\sum_{j=1}^{n_1}W_{11}(j,t)V_{j,1},~W_{11}(j,t)=\sum_{i=1}^N\frac{M_{11}^c(i/N)}{n_1b_{n,1}Nh_1^2}H'\left(\frac{m_1(i/N)-t}{h_1}\right)\bar K_{b_{n,1}}(j/n_1-i/N),\\
	Z_{12}(t)=\sum_{j=1}^{n_1}W_{12}(j,t)V_{j,2},~W_{12}(j,t)=\sum_{i=1}^N\frac{M_{12}^c(i/N)}{n_1b_{n,1}Nh_1^2}H'\left(\frac{m_1(i/N)-t}{h_1}\right)\bar K_{b_{n,1}}(j/n_1-i/N),\\
	Z_{21}(t)=\sum_{j=1}^{n_2}W_{21}(j,t)V_{\lf \frac{n_1j}{n_2}\rf,1},~W_{21}(j,t)=\sum_{i=1}^N\frac{M_{12}^c(i/N)}{n_2b_{n,2}Nh_2^2}H'\left(\frac{m_2(i/N)-t}{h_2}\right)\bar K_{b_{n,2}}(j/n_2-i/N),\\
	Z_{22}(t)=\sum_{j=1}^{n_2}W_{22}(j,t)V_{\lf \frac{n_1j}{n_2}\rf,2},~W_{22}(j,t)=\sum_{i=1}^N\frac{M_{22}^c(i/N)}{n_2b_{n,2}Nh_2^2}H'\left(\frac{m_2(i/N)-t}{h_2}\right)\bar K_{b_{n,2}}(j/n_2-i/N).
	\end{align*}
	
	\noindent The following theorem describes the asymptotic properties.
	\begin{theorem}\label{core-dependent}
		Consider the two possibly correlated time series $\tilde{\mf y}_{i,1}$ and $\tilde {\mf y}_{i,2}$ with sample sizes $n_1\geq n_2$, where $\tilde {\mf y}_{i,1}$ is realized at time $i/n_1$ for $1\leq i\leq n_1$  and $\tilde {\mf y}_{i,2}$ is realized at $\frac{\lf in_1/n_2\rf}{n_1}$ for $1\leq i\leq n_2$. Then we have:
		\begin{description}
			\item (a)	Under the conditions of Theorem \eqref{asymptotic} with $(A7)$ replaced by $(A7')$, the test of Algorithm \ref{algo1} is still asymptotically correct   if $Z_{s}(t,\{V_{j,s}\}_{j\in \mathbb Z})$, $s=1,2$ in Section \ref{BBT} are replaced by \eqref{newZ1}.
			\item (b)	Under the conditions of Theorem \ref{SCB} with $(A7)$ replaced by $(A7')$, the SCB given by Algorithm \ref{algo2} is still asymptotically level $\alpha$  if $Z_{s}(t,\{V_{j,s}\}_{j\in \mathbb Z})$, $s=1,2$ in Section \ref{BBT} are replaced by \eqref{newZ1}.
		\end{description}
		
	\end{theorem}
	To implement Theorem \ref{core-dependent}, we shall estimate  the quantity $\mf M_{\mf c}(t)$ in \eqref{eq536-new}, which consists of $\bs\Sigma_s(t)$ for $s=1,2$ and $\mf V(t)$ of \eqref{lrvb}. The former can be estimated via \eqref{sigmas-estimate}, and the latter can be estimated by (say $n_1\geq n_2$)
	\begin{align}
	\hat {\mf V}(t)=\frac{1}{n_sb_{n,s}}\sum_{i=1}^{n_2}K_{b_{n,2}}(i/n_2-t)\hat {\bs \Delta}_i.
	\end{align}
	Here  $\hat{\bs \Delta}_i=\frac{\tilde {\bs \Xi}_{i}\tilde {\bs \Xi}_{i}^\top}{2M+1}$, $\tilde {\bs \Xi}_{i}=(\tilde {\bs \Xi}_{\lf in_1/n_2\rf,1},\tilde {\bs \Xi}_{i,2})^\top$, and for $s=1,2$
	\begin{align}
	\tilde {\bs \Xi}_{i,s}=\sum_{j=-M}^{M}\left(\tau-\mf 1\left(\hat e_{i+\lf jn_s/n_2\rf,s}\left(\frac{i+\lf jn_s/n_2\rf}{n_s}\right)\leq 0\right)\right)\mf x_{i+\lf j n_s/n_2\rf,s},
	\end{align}
	where $M$ is the window size such that $M\rightarrow \infty$ and $M=o(n_2)$. 
	In the supplemental material, we proves Theorem \ref{core-dependent} via carefully showing the asymptotic convergence and calculating the asymptotic formula of $T_{n_1,n_2}$ and  $(\hat m_1^{-1})'(t)-(\hat m_2^{-1})'(t)-((m_1^{-1})'(t)-(m_2^{-1})'(t))$ under the new conditions of Theorem \ref{core-dependent}. In particular, we show that when the two series are correlated, under null hypothesis of shift invariance, the asymptotic results of the above two statistics will be very different between the scenarios of $d=0$ and $d\neq 0$, while the new algorithms in Theorem \ref{core-dependent} are adaptive to the two scenarios, i.e., the asymptomatic correctness of the algorithms hold under  both scenarios.}

\section{Finite Sample Simulation Studies}\label{FSS}
This section studies the finite sample performance of the SIT test and the SCB test. The performance of the tests is carried out for $n_1 = n_2 = n = 50, 100, 200$ and $500$, the number of repetitions $= 1000$ and the number of Bootstrap replication (i.e., B) $= 500$. In this study, we consider the Epanechnikov kernel (e.g., see Silverman (1998)) unless mentioned otherwise, and the upper limit of the Riemann sum is the same as the sample size, i.e., $N = n$. Apart from these choices, we set $h_{s} = n_{s}^{-1/3}$ and choose $b_{n, 1}$ and $b_{n, 2}$ as described in Section \ref{BS}.




The covariate random variables and the error random variables are generated as follows. Let  $Q_\tau(\cdot)$ be the $\tau_{th}$ quantile of a random variable. For $s = 1$ and $2$, consider the $p_s$ dimensional covariate $\mf x_{i,s}$  and the error $e_{i,s}$:  $$\mf x_{i,s}=\mf H_s\left(\frac{i}{n},\GG_{i,s}\right), e_{i,s}=L_s\left(\frac{i}{n},\FF_{i,s}\right), e_{i,\tau, s}=e_{i,s}-Q_\tau(e_{i,s}),$$ where $\GG_{i,s}=(\bs \zeta_{-\infty,s},...,\bs \zeta_{i,s})$  and $\bs \zeta_{i,s}=(\zeta_{i,1,s},...,\zeta_{i,p_s,s})^\top$ are independent random vectors, and $\FF_{i,s}=(\varepsilon_{-\infty,s},...,\varepsilon_{i,s})$.
Furthermore, $(\{\zeta_{i,j,s}\}_{i\in \mathbb Z,1\leq j\leq p_s,s=1,2}, \{\varepsilon_{k,s}\}_{k\in \mathbb Z, s=1,2})$ are jointly independent random variables with  $\{\zeta_{i,j,s}\}_{i\in \mathbb Z,s=1,2}$ following $\chi^2_j/j$, where $\chi^2_j$ denotes a $\chi^2$ distribution with $j$ degrees of freedom. The innovations $\{\varepsilon_{k,1}\}_{k\in \mathbb Z}$ follow standard normal distribution, and $\{\varepsilon_{k, 2}\}_{k\in \mathbb Z}$ follow $t_{5}/\sqrt{5/3}$, where $t_5$ denotes the standardized student $t$ distribution with 5 degrees of freedom. 

The nonlinear filters $L$ and $\mf H$ are defined as follows. For $e_i$, 
$$L_{s} (t, \FF_{i, s}) = 0.6 (t - 0.3)^{2} L_{s} (t, \FF_{i - 1, s}) + \varepsilon_{i, s}.$$ 
Let us denote $\mf H_s=(H_{1,s},...,H_{p_s,s})^\top$. For $s = 1$ and $2$ and $j$-th covariate, $1\leq j\leq p_s$, $$H_{j, s} \left(t, \GG_{i, s}\right) = 0.2 \left(t- 0.3\right)^{2} H_{j, s} \left(t, \GG_{i-1, s}\right) + \zeta_{i, j, s}.$$


\noindent{\bf Example 1:} Let $$y_{1, i} = \theta_{1, \tau, 1}\left(\frac{i}{n}\right) x_{1, i, 1} + \theta_{2, \tau, 1}\left(\frac{i}{n}\right) x_{2, i, 1} + e_{i,\tau, 1} $$ and $$y_{2, i} = \theta_{1, \tau, 2}\left(\frac{i}{n}\right) x_{1, i, 2} + \theta_{2, \tau, 2}\left(\frac{i}{n}\right)x_{2, i, 2} + e_{i, \tau, 2}.$$ Suppose that $\theta_{1, \tau, 1} (t) = t$,  $\theta_{2, \tau, 1} (t) = \log t$, and $\theta_{1, \tau, 2} (t) = t - 0.1$, $\theta_{2, \tau, 2} (t) = \log (t - 0.1)$. Further, consider $c_{1, 1} = c_{2, 1} = c_{1, 2} = c_{2, 2}= 1$. In the numerical studies, we consider $\tau = 0.5$, $0.7$ and $0.8$.  

\noindent{\bf Example 2:} Let 
$$y_{1, i} = \theta_{1, \tau, 1}\left(\frac{i}{n}\right) x_{1, i, 1} + \theta_{2, \tau, 1}\left(\frac{i}{n}\right) x_{2, i, 1} + \theta_{3, \tau, 1}\left(\frac{i}{n}\right) x_{3, i, 1} + e_{i, \tau,1} $$ and $$y_{2, i} = \theta_{1, \tau, 2}\left(\frac{i}{n}\right) x_{1, i, 2} + \theta_{2, \tau, 2}\left(\frac{i}{n}\right) x_{2, i, 2} + \theta_{3, \tau, 2}\left(\frac{i}{n}\right) x_{3, i, 2} + e_{i, \tau,2} .$$ 
Suppose that $\theta_{1, \tau, 1}(t) = t^{2}$, 
$\theta_{2, \tau, 1}(t) = \sin\frac{\pi t}{2}$, $\theta_{3, \tau, 1}(t) = e^{t}$, and $\theta_{1, \tau, 2}(t) = (t - 0.1)^{2}$, 
$\theta_{2, \tau, 2}(t) = \sin\frac{\pi (t - 0.1)}{2}$ and $\theta_{3, \tau, 2}(t) = e^{t - 0.1}$. Further, consider  $c_{1, 1} = c_{2, 1} = c_{3, 1} = c_{1, 2} = c_{2, 2} = c_{3, 2} =1$, and here also, $\tau = 0.5$, $0.7$ and $0.8$ are considered in the numerical study. 

Note that for both Examples 1 and 2, the choices of time varying coefficients (i.e, $\theta(t)$'s) satisfy the null hypothesis described in \eqref{hypo1}. Tables \ref{tab1} and \ref{tab3} show the rejection probabilities of the SIT test and the SCB test for Examples 1 and 2, respectively when the level of significance is 5\% and 10\%. 

\begin{table}[h!]
	
	\begin{center}		
		
		\begin{tabular}{ccccc}\hline
			model & $n = 50$ & $n = 100$ & $n = 200$ & $n = 500$\\ \hline 
			Example 1 ($\alpha = 5\%$ , $\tau = 0.5$) & {$0.063$} & {$0.061$} & {$0.058$} & {$0.051$}\\ \hline
			Example 1 ($\alpha = 10\%$, $\tau = 0.5$) & {$0.119$} & {$0.116$} & {$0.110$} & {$0.103$}\\ \hline
			Example 1 ($\alpha = 5\%$, $\tau = 0.7$) & {$0.062$} & {$0.061$} & 
			{$0.057$} & {$0.052$} \\ \hline
			Example 1 ($\alpha = 10\%$, $\tau = 0.7$) & {$0.118$} & {$0.117$} & 
			{$0.113$} & {$0.104$} \\ \hline
			Example 1 ($\alpha = 5\%$, $\tau = 0.8$) & {$0.063$} & {$0.060$} & 
			{$0.057$} & {$0.052$} \\ \hline
			Example 1 ($\alpha = 10\%$, $\tau = 0.8$) & {$0.115$} & {$0.114$} & 
			{$0.109$} & {$0.106$} \\ \hline
			Example 2 ($\alpha = 5\%$, $\tau = 0.5$) & {$0.067$} & {$0.060$} & {$0.059$} & {$0.052$}\\ \hline
			Example 2 ($\alpha = 10\%$, $\tau = 0.5$) & {$0.122$} & {$0.120$} & {$0.114$} & {$0.107$}\\ \hline
			Example 2 ($\alpha = 5\%$, $\tau = 0.7$) & {$0.065$} & {$0.059$} & {$0.056$} & {$0.051$}\\ \hline
			Example 2 ($\alpha = 10\%$, $\tau = 0.7$) & {$0.120$} & {$0.119$} & {$0.111$} & {$0.104$}\\ \hline
			Example 2 ($\alpha = 5\%$, $\tau = 0.8$) & {$0.061$} & {$0.059$} & {$0.056$} & {$0.053$}\\ \hline
			Example 2 ($\alpha = 10\%$, $\tau = 0.8$) & {$0.117$} & {$0.115$} & {$0.109$} & {$0.105$}\\ \hline
		\end{tabular}
	\end{center}
	\caption{\it The estimated size of the SIT test for different sample sizes $n_1= n_2=n$. The levels of significance (denoted as $\alpha$) are $5\%$ and $10\%$.}
	
	\label{tab1}
	
\end{table}

\begin{table}[h!]
	
	\begin{center}		
		
		\begin{tabular}{ccccc}\hline
			model & $n = 50$ & $n = 100$ & $n = 200$ & $n = 500$\\ \hline 
			Example 1 ($\alpha  = 5\%$, $\tau = 0.5$) & {$0.067$} & {$0.063$} & {$0.057$} & {$0.053$}\\ \hline
			Example 1 ($\alpha  = 10\%$, $\tau = 0.5$) & {$0.122$} & {$0.111$} & {$0.107$} & {$0.102$}\\ \hline
			Example 1 ($\alpha  = 5\%$, $\tau = 0.7$) & {$0.064$} & {$0.060$} & 
			{$0.056$} & {$0.051$} \\ \hline
			Example 1 ($\alpha  = 10\%$, $\tau = 0.7$) & {$0.125$} & {$0.119$} & 
			{$0.111$} & {$0.105$} \\ \hline
			Example 1 ($\alpha  = 5\%$, $\tau = 0.8$) & {$0.062$} & {$0.061$} & 
			{$0.055$} & {$0.050$} \\ \hline
			Example 1 ($\alpha  = 10\%$, $\tau = 0.8$) & {$0.124$} & {$0.117$} & 
			{$0.106$} & {$0.101$} \\ \hline
			Example 2 ($\alpha  = 5\%$, $\tau = 0.5$) & {$0.065$} & {$0.062$} & {$0.057$} & {$0.049$}\\ \hline
			Example 2 ($\alpha  = 10\%$, $\tau = 0.5$) & {$0.129$} & {$0.118$} & {$0.111$} & {$0.104$}\\ \hline
			Example 2 ($\alpha  = 5\%$, $\tau = 0.7$) & {$0.067$} & {$0.063$} & {$0.054$} & {$0.050$}\\ \hline
			Example 2 ($\alpha  = 10\%$, $\tau = 0.7$) & {$0.126$} & {$0.118$} & {$0.110$} & {$0.102$}\\ \hline
			Example 2 ($\alpha  = 5\%$, $\tau = 0.8$) & {$0.065$} & {$0.057$} & {$0.053$} & {$0.050$}\\ \hline
			Example 2 ($\alpha  = 10\%$, $\tau = 0.8$) & {$0.123$} & {$0.116$} & {$0.107$} & {$0.101$}\\ \hline
		\end{tabular}
	\end{center}
	\caption{\it The estimated size of the SCB test for different sample sizes $n_1= n_2=n$. The levels of significance (denoted as $\alpha$) are $5\%$ and $10\%$.}
	
	\label{tab3}
	
\end{table}

For power study, we consider the the same error and covariate processes and used the following examples : 

\noindent {\bf Example 3:} Let $$y_{1, i} = \theta_{1, \tau, 1}\left(\frac{i}{n}\right) x_{1, i, 1} + \theta_{2, \tau, 1}\left(\frac{i}{n}\right) x_{2, i, 1} + e_{i,\tau, 1} $$ and $$y_{2, i} = \theta_{1, \tau, 2}\left(\frac{i}{n}\right) x_{1, i, 2} + \theta_{2, \tau, 2}\left(\frac{i}{n}\right) x_{2, i, 2} + e_{i,\tau,2} .$$  Suppose that $\theta_{1, \tau, 1} (t) = t$,  $\theta_{2, \tau, 1} (t) = \log t$, and $\theta_{1, \tau, 2} (t) = t^{2}$, $\theta_{2, \tau, 2} (t) = (\log t)^{2}$. Further, consider $c_{1, 1} = c_{2, 1} = c_{1, 2} = c_{2, 2}= 1$. In the numerical studies, we consider $\tau = 0.5$, $0.7$ and $0.8$.

\noindent{\bf Example 4:} Let 
$$y_{1, i} = \theta_{1, \tau, 1}\left(\frac{i}{n}\right) x_{1, i, 1} + \theta_{2, \tau, 1}\left(\frac{i}{n}\right) x_{2, i, 1} + \theta_{3, \tau, 1}\left(\frac{i}{n}\right) x_{3, i, 1} + e_{i,\tau, 1} $$ and $$y_{2, i} = \theta_{1, \tau, 2}\left(\frac{i}{n}\right) x_{1, i, 2} + \theta_{2, \tau, 2}\left(\frac{i}{n}\right) x_{2, i, 2} + \theta_{3, \tau, 2}\left(\frac{i}{n}\right) x_{3, i, 2} + e_{i,\tau, 2} .$$ 
Suppose that $\theta_{1, \tau, 1}(t) = t^{2}$, 
$\theta_{2, \tau, 1}(t) = \sin\frac{\pi t}{2}$, $\theta_{3, \tau, 1}(t) = e^{t}$, and $\theta_{1, \tau, 2} (t)= t^{3}$, 
$\theta_{2, \tau, 2}(t) = \cos\frac{\pi t}{2}$ and $\theta_{3, \tau, 2}(t) = \log t$. Further, consider  $c_{1, 1} = c_{2, 1} = c_{3, 1} = c_{1, 2} = c_{2, 2} = c_{3, 2} =1$, and here also, $\tau = 0.5$, $0.7$ and $0.8$ are considered in the numerical study.

Note that in Examples 3 and 4, the choices of the time varying regression coefficients do not satisfy the assertion of null hypothesis described in \eqref{hypo1}. Tables \ref{tab2-1} and  \ref{tab2-2} show the rejection probabilities of the SIT and SCB tests based on $T_{n_1, n_2}$ when data follow the models described in Examples 3 and 4.

\begin{table}[h!]
	
	\begin{center}

		\begin{tabular}{ccccc}\hline
			model & $n = 50$ & $n = 100$ & $n = 200$ & $n = 500$\\ \hline 
			Example 3 ($\alpha  = 5\%$, $\tau = 0.5$) & {$0.367$} & {$0.401$} & {$0.546$} & {$0.777$}\\ \hline
			Example 3 ($\alpha  = 10\%$, $\tau = 0.5$) & {$0.445$} & {$0.517$} & {$0.699$} & {$0.901$}\\ \hline
			Example 3 ($\alpha  = 5\%$, $\tau = 0.7$) & {$0.422$} & {$0.499$} & {$0.627$} & {$0.818$} \\ \hline
			Example 3 ($\alpha  = 10\%$, $\tau = 0.7$) & {$0.497$} & {$0.563$} & 
			{$0.776$} & {$0.923$} \\ \hline
			Example 3 ($\alpha  = 5\%$, $\tau = 0.8$) & {$0.378$} & {$0.444$} & 
			{$0.622$} & {$0.816$} \\ \hline
			Example 3 ($\alpha  = 10\%$, $\tau = 0.8$) & {$0.412$} & {$0.535$} & {$0.701$} & {$0.888$} \\ \hline
			Example 4 ($\alpha  = 5\%$, $\tau = 0.5$) & {$0.422$} & {$0.477$} & {$0.661$} & {$0.825$}\\ \hline
			Example 4 ($\alpha  = 10\%$, $\tau = 0.5$) & {$0.447$} & {$0.500$} & {$0.708$} & {$0.917$}\\ \hline
			Example 4 ($\alpha  = 5\%$, $\tau = 0.7$) & {$0.475$} & {$0.503$} & {$0.688$} & {$0.848$}\\ \hline
			Example 4 ($\alpha  = 10\%$, $\tau = 0.7$) & {$0.510$} & {$0.582$} & {$0.727$} & {$0.949$}\\ \hline
			Example 4 ($\alpha  = 5\%$, $\tau = 0.8$) & {$0.398$} & {$0.419$} & {$0.589$} & {$0.801$}\\ \hline
			Example 4 ($\alpha  = 10\%$, $\tau = 0.8$) & {$0.419$} & {$0.475$} & {$0.623$} & {$0.878$}\\ \hline
		\end{tabular}
	\end{center}
	\caption{\it The estimated power of the test of the test based on $T_{n_1, n_2}$, i.e., the SIT test for different sample sizes $n_1= n_2=n$. The levels of significance (denoted as $\alpha$) are $5\%$ and $10\%$.}

	\label{tab2-1}
	
\end{table}

\begin{table}[h!]
	
	\begin{center}

		\begin{tabular}{ccccc}\hline
			model & $n = 50$ & $n = 100$ & $n = 200$ & $n = 500$\\ \hline 
			Example 3 ($\alpha  = 5\%$, $\tau = 0.5$) & {$0.343$} & {$0.376$} & {$0.519$} & {$0.743$}\\ \hline
			Example 3 ($\alpha  = 10\%$, $\tau = 0.5$) & {$0.421$} & {$0.487$} & {$0.665$} & {$0.874$}\\ \hline
			Example 3 ($\alpha  = 5\%$, $\tau = 0.7$) & {$0.395$} & {$0.470$} & {$0.592$} & {$0.789$} \\ \hline
			Example 3 ($\alpha  = 10\%$, $\tau = 0.7$) & {$0.466$} & {$0.525$} & 
			{$0.739$} & {$0.888$} \\ \hline
			Example 3 ($\alpha  = 5\%$, $\tau = 0.8$) & {$0.344$} & {$0.417$} & 
			{$0.589$} & {$0.786$} \\ \hline
			Example 3 ($\alpha  = 10\%$, $\tau = 0.8$) & {$0.387$} & {$0.509$} & {$0.668$} & {$0.849$} \\ \hline
			Example 4 ($\alpha  = 5\%$, $\tau = 0.5$) & {$0.434$} & {$0.496$} & {$0.675$} & {$0.842$}\\ \hline
			Example 4 ($\alpha  = 10\%$, $\tau = 0.5$) & {$0.472$} & {$0.521$} & {$0.735$} & {$0.946$}\\ \hline
			Example 4 ($\alpha  = 5\%$, $\tau = 0.7$) & {$0.499$} & {$0.530$} & {$0.723$} & {$0.880$}\\ \hline
			Example 4 ($\alpha  = 10\%$, $\tau = 0.7$) & {$0.538$} & {$0.611$} & {$0.752$} & {$0.981$}\\ \hline
			Example 4 ($\alpha  = 5\%$, $\tau = 0.8$) & {$0.421$} & {$0.443$} & {$0.614$} & {$0.824$}\\ \hline
			Example 4 ($\alpha  = 10\%$, $\tau = 0.8$) & {$0.445$} & {$0.498$} & {$0.651$} & {$0.901$}\\ \hline
		\end{tabular}
	\end{center}
	\caption{\it The estimated power of the SCB test for different sample sizes $n_1= n_2=n$. The levels of significance (denoted as $\alpha$) are $5\%$ and $10\%$.}

	\label{tab2-2}
	
\end{table}

It follows from the results of Examples 1 and 2 that the test based on $T_{n_1, n_2}$, i.e., the SIT test and the SCB test can achieve the nominal level of significance when $\tau = 0.5$, $0.7$ and $0.8$. In terms of estimated power, the results of Examples 3 and 4 indicate that the SIT and the SCB tests can achieve the maximum power as the sample size increases. Precisely speaking, for Example 3, the SIT test is marginally more powerful than the SCB test whereas for Example 4, the SCB test is faintly more powerful than the SIT test. We also observe the same phenomena for unequal $n_1$ and $n_2$ but for the sake of concise presentation, we have not here reported the values of the estimated size and power.  

{\color{black} At the end, as one reviewer pointed out the issue, we want to discuss the performance of the SIT and the SCB tests when $m_{1}(t)$ and $m_{2}(t)$ are non-monotone. Let us consider the following two examples, and the results are summarized in Table \ref{tab9}.
	
	\noindent{\bf Example NM1:} Let $$y_{1, i} = \theta_{1, \tau, 1}\left(\frac{i}{n}\right) x_{1, i, 1} + e_{i,\tau, 1} $$ and $$y_{2, i} = \theta_{1, \tau, 2}\left(\frac{i}{n}\right) x_{1, i, 2} + e_{i, \tau, 2}.$$ Suppose that $\theta_{1, \tau, 1} (t) = t(0.5 - t)(1 - t)$, and $\theta_{1, \tau, 2} (t) = (t - 0.1)(0.6 - t)(1.1 - t)$. Further, consider $c_{1, 1} = c_{1, 2} = 1$. In the numerical studies, we consider $\tau = 0.5$.
	
	\noindent{\bf Example NM2:} Let $$y_{1, i} = \theta_{1, \tau, 1}\left(\frac{i}{n}\right) x_{1, i, 1} + e_{i,\tau, 1} $$ and $$y_{2, i} = \theta_{1, \tau, 2}\left(\frac{i}{n}\right) x_{1, i, 2} + e_{i, \tau, 2}.$$ Suppose that $\theta_{1, \tau, 1} (t) = t(0.5 - t)(1 - t)$, and $\theta_{1, \tau, 2} (t) = t(0.5 - t)^{2}(1 - t)$. Further, consider $c_{1, 1} = c_{1, 2} = 1$. In the numerical studies, we consider $\tau = 0.5$.
	
	\begin{table}[h!]
		
		\begin{center}		
			
			\begin{tabular}{ccccc}\hline
				model & $n = 50$ & $n = 100$ & $n = 200$ & $n = 500$\\ \hline 
				SIT test : Example NM1 ($\alpha = 5\%$ , $\tau = 0.5$) & {$0.061$} & {$0.059$} & {$0.060$} & {$0.055$}\\ \hline
				SIT test : Example NM1 ($\alpha = 10\%$, $\tau = 0.5$) & {$0.117$} & {$0.111$} & {$0.108$} & {$0.102$}\\ \hline
				SCB test : Example NM1 ($\alpha = 5\%$, $\tau = 0.5$) & {$0.062$} & {$0.059$} & 
				{$0.058$} & {$0.054$} \\ \hline
				SCB test : Example NM1 ($\alpha = 10\%$, $\tau = 0.5$) & {$0.119$} & {$0.114$} & 
				{$0.109$} & {$0.106$} \\ \hline
				SIT test : Example NM2 ($\alpha = 5\%$, $\tau = 0.5$) & {$0.167$} & {$0.221$} & 
				{$0.364$} & {$0.501$} \\ \hline
				SIT test : Example NM2 ($\alpha = 10\%$, $\tau = 0.5$) & {$0.205$} & {$0.256$} & 
				{$0.422$} & {$0.563$} \\ \hline
				SCB test : Example NM2 ($\alpha = 5\%$, $\tau = 0.5$) & {$0.134$} & {$0.185$} & {$0.244$} & {$0.314$}\\ \hline
				SCB test : Example NM2 ($\alpha = 10\%$, $\tau = 0.5$) & {$0.151$} & {$0.199$} & {$0.261$} & {$0.332$}\\ \hline
			\end{tabular}
		\end{center}
		\caption{\it The values in first, second, third and fourth rows are the estimated sizes of the SIT and the SCB tests. The values in fifth, sixth, seventh and eighth rows are the estimated powers of the SIT and the SCB tests.
			The levels of significance (denoted as $\alpha$) are $5\%$ and $10\%$, and $\tau = 05$.}
		
		\label{tab9}
		
	\end{table}
	
	Note that in both {\bf Example NM1} and {\bf Example NM2}, $\theta_{1, \tau, 1} (t)$ is a non-monotone function of $t\in (0, 1)$. Further, in {\bf Example NM1}, the choices of time varying coefficients (i.e, $\theta(t)$'s) satisfy the null hypothesis described in \eqref{hypo1}, and both the SIT and the SCB tests can achieve the nominal level of significance when $\tau = 0.5$. Furthermore, in {\bf Example NM2}, where the choices of time varying coefficients (i.e, $\theta(t)$'s) does not satisfy the null hypothesis, the power of the SIT and SCB tests increase as the sample sizes increase.}

{\color{black}\subsection{Sensitivity Analysis via Various Tuning parameters}\label{CTP}
	The study in Section \ref{BS} indicates that the performance of the SIT and the SCB tests depends on various tuning parameters, namely, $b_{n, s}$, $h_{s}$, $w_{n, s}$ and $M_{s}$. In this section, we carry out the power and the level study for various choices of $b_{n, s}$ (i.e., bandwidth) and $w_{n, s}$ for the same examples considered in Section \ref{FSS}. As it is mentioned in Section \ref{BS}, we choose $M_s=\lf n_s^{1/3}\rf$ and $h_{s} = n_{s}^{-1/3}$, where $\lf x \rf$ denotes the largest integer less then or equal to $x$. In order to satisfy the condition described on $w_{n, s}$ in Section \ref{estM}, we choose $w_{n, s} = n_{s}^{-1/3}$, $n_{s}^{-1/5}$ and $n_{s}^{-1/7}$, and in order to satisfy the conditions on $b_{n, s}$ in Section \ref{Asy-Result}, we choose $b_{n, s} = n_{s}^{-1/4}$ and $n_{s}^{-1/5}$. Tables \ref{tab5} and \ref{tab6} show the rejection probabilities (i.e., estimated size) of the SIT test and the SCB test for Examples 1 and 2 (considered in Section \ref{FSS}), respectively when the level of significance is 5\% for the aforementioned choices of $b_{n, s}$, $h_{s}$, $w_{n, s}$ and $M_{s}$, and $n_{s} = 100$ and 500.

	\begin{table}[h!]
		
		\begin{center}		
			
			\begin{tabular}{ccc}\hline
				model & $n = 100$ & $n = 500$\\ \hline 
				Example 1 ($\alpha = 5\%$ , $\tau = 0.5$, $b_{n, s} = n_{s}^{-1/4}$)  & {$(0.059, 0.057, 0.061)$} & {$(0.049, 0.052, 0.048)$}\\ \hline
				Example 1 ($\alpha = 5\%$, $\tau = 0.5$, $b_{n, s} = n_{s}^{-1/5}$)  & {$(0.057, 0.056, 0.060)$} & {$(0.050, 0.053, 0.049)$}\\ \hline
				Example 1 ($\alpha = 5\%$, $\tau = 0.7$, $b_{n, s} = n_{s}^{-1/4}$)  & {$(0.059, 0.062, 0.061)$}  & {$(0.051, 0.052, 0.054)$} \\ \hline
				Example 1 ($\alpha = 5\%$, $\tau = 0.7$, $b_{n, s} = n_{s}^{-1/5}$)  & {$(0.056, 0.060, 0.059)$} & {$(0.050, 0.051, 0.050)$}
				\\ \hline
				Example 1 ($\alpha = 5\%$, $\tau = 0.8$, $b_{n, s} = n_{s}^{-1/4}$)  & {$(0.057, 0.053, 0.055)$} & 
				{$(0.053, 0.051, 0.050)$} \\ \hline
				Example 1 ($\alpha = 5\%$, $\tau = 0.8$, $b_{n, s} = n_{s}^{-1/5}$)  & {$(0.056, 0.054, 0.056)$} & 
				{$(0.052, 0.050, 0.053)$} \\ \hline
				Example 2 ($\alpha = 5\%$, $\tau = 0.5$, $b_{n, s} = n_{s}^{-1/4}$)  & {$(0.055, 0.057, 0.054)$} &  {$(0.051, 0.054, 0.052)$}\\ \hline
				Example 2 ($\alpha = 5\%$, $\tau = 0.5$, $b_{n, s} = n_{s}^{-1/4}$)  & {$(0.056, 0.052, 0.055)$} &  {$(0.054, 0.052, 0.053)$}\\ \hline
				Example 2 ($\alpha = 5\%$, $\tau = 0.7$, $b_{n, s} = n_{s}^{-1/4}$)  & {$(0.057, 0.059, 0.059)$} &  {$(0.050, 0.051, 0.049)$}\\ \hline
				Example 2 ($\alpha = 5\%$, $\tau = 0.7$, $b_{n, s} = n_{s}^{-1/5}$)  & {$(0.056, 0.060, 0.058)$} &  {$(0.053, 0.047, 0.050)$}\\ \hline
				Example 2 ($\alpha = 5\%$, $\tau = 0.8$, $b_{n, s} = n_{s}^{-1/4}$)  & {$(0.055, 0.061, 0.056)$} &  {$(0.053, 0.052, 0.055)$}\\ \hline
				Example 2 ($\alpha = 5\%$, $\tau = 0.8$, $b_{n, s} = n_{s}^{-1/5}$)  & {$(0.054, 0.057, 0.055)$} &  {$(0.054, 0.051, 0.056)$}\\ \hline
			\end{tabular}
		\end{center}
		\caption{\it The estimated size of the SIT test for different sample sizes $n_1= n_2=n$. The levels of significance (denoted as $\alpha$) is $5\%$. In each cell, from the left, the first, the second and the third values are corresponding to $w_{n, s} = n_{s}^{-1/3}$, $n_{s}^{-1/5}$ and $n_{s}^{-1/7}$, respectively.}
		
		\label{tab5}
		
	\end{table}
	
	\begin{table}[h!]
		
		\begin{center}		
			
			\begin{tabular}{ccc}\hline
				model & $n = 100$ & $n = 500$\\ \hline 
				Example 1 ($\alpha = 5\%$ , $\tau = 0.5$, $b_{n, s} = n_{s}^{-1/4}$)  & {$(0.060, 0.057, 0.062)$} & {$(0.052, 0.053, 0.050)$}\\ \hline
				Example 1 ($\alpha = 5\%$, $\tau = 0.5$, $b_{n, s} = n_{s}^{-1/5}$)  & {$(0.061, 0.057, 0.063)$} & {$(0.052, 0.055, 0.051)$}\\ \hline
				Example 1 ($\alpha = 5\%$, $\tau = 0.7$, $b_{n, s} = n_{s}^{-1/4}$)  & {$(0.061, 0.064, 0.062)$}  & {$(0.053, 0.054, 0.051)$} \\ \hline
				Example 1 ($\alpha = 5\%$, $\tau = 0.7$, $b_{n, s} = n_{s}^{-1/5}$)  & {$(0.054, 0.063, 0.060)$} & {$(0.051, 0.055, 0.052)$}
				\\ \hline
				Example 1 ($\alpha = 5\%$, $\tau = 0.8$, $b_{n, s} = n_{s}^{-1/4}$)  & {$(0.059, 0.055, 0.056)$} & 
				{$(0.055, 0.052, 0.052)$} \\ \hline
				Example 1 ($\alpha = 5\%$, $\tau = 0.8$, $b_{n, s} = n_{s}^{-1/5}$)  & {$(0.059, 0.057, 0.060)$} & 
				{$(0.057, 0.052, 0.056)$} \\ \hline
				Example 2 ($\alpha = 5\%$, $\tau = 0.5$, $b_{n, s} = n_{s}^{-1/4}$)  & {$(0.052, 0.058, 0.055)$} &  {$(0.052, 0.056, 0.055)$}\\ \hline
				Example 2 ($\alpha = 5\%$, $\tau = 0.5$, $b_{n, s} = n_{s}^{-1/4}$)  & {$(0.055, 0.053, 0.058)$} &  {$(0.053, 0.054, 0.057)$}\\ \hline
				Example 2 ($\alpha = 5\%$, $\tau = 0.7$, $b_{n, s} = n_{s}^{-1/4}$)  & {$(0.055, 0.061, 0.060)$} &  {$(0.053, 0.057, 0.053)$}\\ \hline
				Example 2 ($\alpha = 5\%$, $\tau = 0.7$, $b_{n, s} = n_{s}^{-1/5}$)  & {$(0.058, 0.061, 0.059)$} &  {$(0.057, 0.050, 0.053)$}\\ \hline
				Example 2 ($\alpha = 5\%$, $\tau = 0.8$, $b_{n, s} = n_{s}^{-1/4}$)  & {$(0.056, 0.063, 0.058)$} &  {$(0.054, 0.055, 0.057)$}\\ \hline
				Example 2 ($\alpha = 5\%$, $\tau = 0.8$, $b_{n, s} = n_{s}^{-1/5}$)  & {$(0.056, 0.056, 0.059)$} &  {$(0.058, 0.056, 0.061)$}\\ \hline
			\end{tabular}
		\end{center}
		\caption{\it The estimated size of the SCB test for different sample sizes $n_1= n_2=n$. The level of significance (denoted as $\alpha$) is  $5\%$. In each cell, from the left, the first, the second and the third values are corresponding to $w_{n, s} = n_{s}^{-1/3}$, $n_{s}^{-1/5}$ and $n_{s}^{-1/7}$, respectively.}
		
		\label{tab6}
		
	\end{table}
	
	Recall that in Examples 3 and 4 in Section \ref{FSS}, the time varying regression coefficients do not satisfy the assertion of null hypothesis described in \eqref{hypo1}. Tables \ref{tab7} and \ref{tab8} show the rejection probabilities (i.e., estimated power) of the SIT and the SCB tests when the data follow the models described in Examples 3 and 4. It is clear from the values in Tables \ref{tab5} and \ref{tab6} that both the SIT and the SCB tests can approximately attain the exact size of the tests for various choices of $b_{n, s}$ and $w_{n, s}$, and in the case of power study, the values in Tables \ref{tab7} and \ref{tab8} indicate that the power does not vary more than 4\% for various choices of $b_{n, s}$ and $w_{n, s}$.

	\begin{table}[h!]
		
		\begin{center}		
			
			\begin{tabular}{ccc}\hline
				model & $n = 100$ & $n = 500$\\ \hline 
				Example 3 ($\alpha = 5\%$ , $\tau = 0.5$, $b_{n, s} = n_{s}^{-1/4}$)  & {$(0.456, 0.433, 0.479)$} & {$(0.722, 0.755, 0.744)$}\\ \hline
				Example 3 ($\alpha = 5\%$, $\tau = 0.5$, $b_{n, s} = n_{s}^{-1/5}$)  & {$(0.448, 0.459, 0.488)$} & {$(0.747, 0.781, 0.777)$}\\ \hline
				Example 3 ($\alpha = 5\%$, $\tau = 0.7$, $b_{n, s} = n_{s}^{-1/4}$)  & {$(0.486, 0.503, 0.499)$}  & {$(0.799, 0.832, 0.818)$} \\ \hline
				Example 3 ($\alpha = 5\%$, $\tau = 0.7$, $b_{n, s} = n_{s}^{-1/5}$)  & {$(0.501, 0.552, 0.527)$} & {$(0.834, 0.811, 0.857)$}
				\\ \hline
				Example 3 ($\alpha = 5\%$, $\tau = 0.8$, $b_{n, s} = n_{s}^{-1/4}$)  & {$(0.492, 0.481, 0.505)$} & 
				{$(0.850, 0.848, 0.888)$} \\ \hline
				Example 3 ($\alpha = 5\%$, $\tau = 0.8$, $b_{n, s} = n_{s}^{-1/5}$)  & {$(0.503, 0.511, 0.495)$} & 
				{$(0.872, 0.865, 0.889)$} \\ \hline
				Example 4 ($\alpha = 5\%$, $\tau = 0.5$, $b_{n, s} = n_{s}^{-1/4}$)  & {$(0.485, 0.469, 0.500)$} &  {$(0.836, 0.854, 0.847)$}\\ \hline
				Example 4 ($\alpha = 5\%$, $\tau = 0.5$, $b_{n, s} = n_{s}^{-1/5}$)  & {$(0.502, 0.487, 0.499)$} &  {$(0.851, 0.866, 0.879)$}\\ \hline
				Example 4 ($\alpha = 5\%$, $\tau = 0.7$, $b_{n, s} = n_{s}^{-1/4}$)  & {$(0.511, 0.526, 0.515)$} &  {$(0.858, 0.871, 0.880)$}\\ \hline
				Example 4 ($\alpha = 5\%$, $\tau = 0.7$, $b_{n, s} = n_{s}^{-1/5}$)  & {$(0.518, 0.538, 0.542)$} &  {$(0.864, 0.881, 0.885)$}\\ \hline
				Example 4 ($\alpha = 5\%$, $\tau = 0.8$, $b_{n, s} = n_{s}^{-1/4}$)  & {$(0.449, 0.467, 0.480)$} &  {$(0.822, 0.845, 0.848)$}\\ \hline
				Example 4 ($\alpha = 5\%$, $\tau = 0.8$, $b_{n, s} = n_{s}^{-1/5}$)  & {$(0.449, 0.466, 0.478)$} &  {$(0.883, 0.901, 0.892)$}\\ \hline
			\end{tabular}
		\end{center}
		\caption{\it The estimated power of the SIT test for different sample sizes $n_1= n_2=n$ . The levels of significance (denoted as $\alpha$) is $5\%$. In each cell, from the left, the first, the second and the third values are corresponding to $w_{n, s} = n_{s}^{-1/3}$, $n_{s}^{-1/5}$ and $n_{s}^{-1/7}$, respectively.}
		
		\label{tab7}
		
	\end{table}

	\begin{table}[h!]
		
		\begin{center}		
			
			\begin{tabular}{ccc}\hline
				model & $n = 100$ & $n = 500$\\ \hline 
				Example 3 ($\alpha = 5\%$ , $\tau = 0.5$, $b_{n, s} = n_{s}^{-1/4}$)  & {$(0.436, 0.411, 0.460)$} & {$(0.701, 0.738, 0.725)$}\\ \hline
				Example 3 ($\alpha = 5\%$, $\tau = 0.5$, $b_{n, s} = n_{s}^{-1/5}$)  & {$(0.429, 0.433, 0.465)$} & {$(0.725, 0.769, 0.770)$}\\ \hline
				Example 3 ($\alpha = 5\%$, $\tau = 0.7$, $b_{n, s} = n_{s}^{-1/4}$)  & {$(0.461, 0.480, 0.468)$}  & {$(0.762, 0.800, 0.792)$} \\ \hline
				Example 3 ($\alpha = 5\%$, $\tau = 0.7$, $b_{n, s} = n_{s}^{-1/5}$)  & {$(0.487, 0.517, 0.511)$} & {$(0.806, 0.784, 0.833)$}
				\\ \hline
				Example 3 ($\alpha = 5\%$, $\tau = 0.8$, $b_{n, s} = n_{s}^{-1/4}$)  & {$(0.466, 0.459, 0.480)$} & 
				{$(0.829, 0.816, 0.865)$} \\ \hline
				Example 3 ($\alpha = 5\%$, $\tau = 0.8$, $b_{n, s} = n_{s}^{-1/5}$)  & {$(0.480, 0.483, 0.469)$} & 
				{$(0.839, 0.844, 0.866)$} \\ \hline
				Example 4 ($\alpha = 5\%$, $\tau = 0.5$, $b_{n, s} = n_{s}^{-1/4}$)  & {$(0.466, 0.448, 0.581)$} &  {$(0.817, 0.833, 0.828)$}\\ \hline
				Example 4 ($\alpha = 5\%$, $\tau = 0.5$, $b_{n, s} = n_{s}^{-1/5}$)  & {$(0.485, 0.466, 0.476)$} &  {$(0.822, 0.841, 0.855)$}\\ \hline
				Example 4 ($\alpha = 5\%$, $\tau = 0.7$, $b_{n, s} = n_{s}^{-1/4}$)  & {$(0.477, 0.495, 0.503)$} &  {$(0.837, 0.844, 0.862)$}\\ \hline
				Example 4 ($\alpha = 5\%$, $\tau = 0.7$, $b_{n, s} = n_{s}^{-1/5}$)  & {$(0.496, 0.511, 0.523)$} &  {$(0.840, 0.853, 0.864)$}\\ \hline
				Example 4 ($\alpha = 5\%$, $\tau = 0.8$, $b_{n, s} = n_{s}^{-1/4}$)  & {$(0.428, 0.441, 0.457)$} &  {$(0.789, 0.816, 0.823)$}\\ \hline
				Example 4 ($\alpha = 5\%$, $\tau = 0.8$, $b_{n, s} = n_{s}^{-1/5}$)  & {$(0.434, 0.447, 0.451)$} &  {$(0.846, 0.870, 0.855)$}\\ \hline
			\end{tabular}
		\end{center}
		\caption{\it The estimated power of the SCB test for different sample sizes $n_1= n_2=n$. The level of significance (denoted as $\alpha$) is $5\%$. In each cell, from the left, the first, the second and the third values are corresponding to $w_{n, s} = n_{s}^{-1/3}$, $n_{s}^{-1/5}$ and $n_{s}^{-1/7}$, respectively.}
		
		\label{tab8}
		
	\end{table}

	\subsection{Simulation Studies : Dependent Data Set} It is already observed that the SIT and the SCB tests perform well for two independent data sets. In this section, we investigate the finite sample of performance of the SIT and the SCB tests when the two compared data sets are dependent as it often happens in practice. Here we consider the same models as in Examples 1, 2, 3 and 4  considered in Section \ref{FSS} but we generated the data in a different way so that data sets become dependent. Strictly speaking, we generate dependent errors in the following way: ${\bf e}_{i} = (e_{i, 1}, e_{i, 2}) = \left(0.8 L_1\left(\frac{i}{n},\FF_{i,1}\right) + 0.2 L_2\left(\frac{i}{n},\FF_{i,2}\right), 0.2 L_1\left(\frac{i}{n},\FF_{i,1}\right) + 0.8 L_2\left(\frac{i}{n},\FF_{i,2}\right) \right)$, where all notation are the same as defined at the beginning of Section \ref{FSS}, and the covariates are generated in the same way as we described in Section \ref{FSS}. 
	
	In this study, we choose the same set of the tuning parameters and the sample sizes (i.e., $n_1$ and $n_2$) as they are considered in Section \ref{CTP}. All results are reported in Section \ref{Sec_F} the supplementary materials. It follows from those results that when the models are the same as the models in Examples 1 and 2, i.e., the null hypothesis is true, the estimated sizes of the SIT test and the SCB test based on Theorem \ref{core-dependent} are not deviated more than 1\% from the estimated sizes when the data sets are mutually independent. Next, when the models are the same as the models in Examples 3 and 4, i.e., the alternative hypothesis is true, the estimated powers of the SIT test and the SCB test based on Theorem \ref{core-dependent} are not deviated more than 6\% from the estimated powers when the data sets are mutually independent.}

\vspace{0.1in}
\section{Real Data Analysis}\label{RDA}

\vspace{0.1in}

\subsection{ Cumulative infected cases and deaths due to COVID-19}
This data set consists of two variables, namely, the cumulative number of infected cases and the cumulative number of deaths due to COVID-19 outbreak in a particular country for the period from December 31, 2019 to October 7, 2020, i.e., $n = 282$ days. We here consider two countries, namely, France and Germany as they are from the same continent. Our analysis is based on the log transformed data since the data is varying from small values to quite large values. The data set is available at \url{https://ourworldindata.org/coronavirus-source-data}. The analysis has three parts, namely, (A) Analysis of cumulative infected cases and deaths in France due to COVID-19 outbreak, (B) Analysis of cumulative infected cases in France and Germany due to COVID-19 outbreak and (C) Analysis of cumulative deaths in France and Germany due to COVID-19 outbreak. All three analyses are done for $\tau$ {\color{black}equals to} 0.8, 0.5 and 0.2 . In order to implement our proposed methodology, we consider $n = 282$ equally spaced points on $[0, 1]$, and the plots are prepared on the time interval $[0, 1]$. 
\subsubsection{Cumulative infected cases and deaths in France : COVID-19 outbreak} 
\begin{figure}
	\includegraphics[width= 2.5in, height = 2.5in]{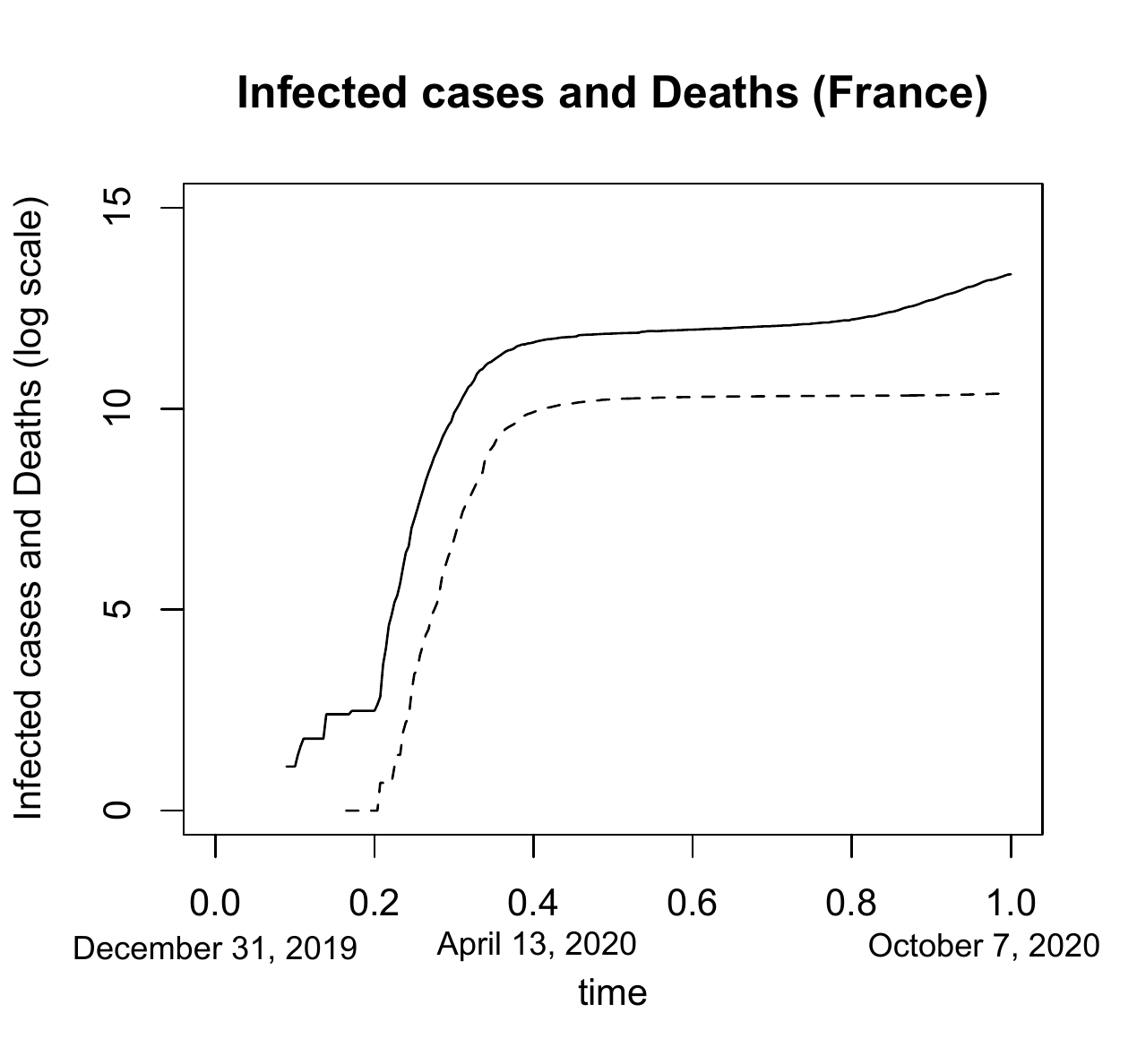}
	\includegraphics[width= 2.5in, height = 2.5in]{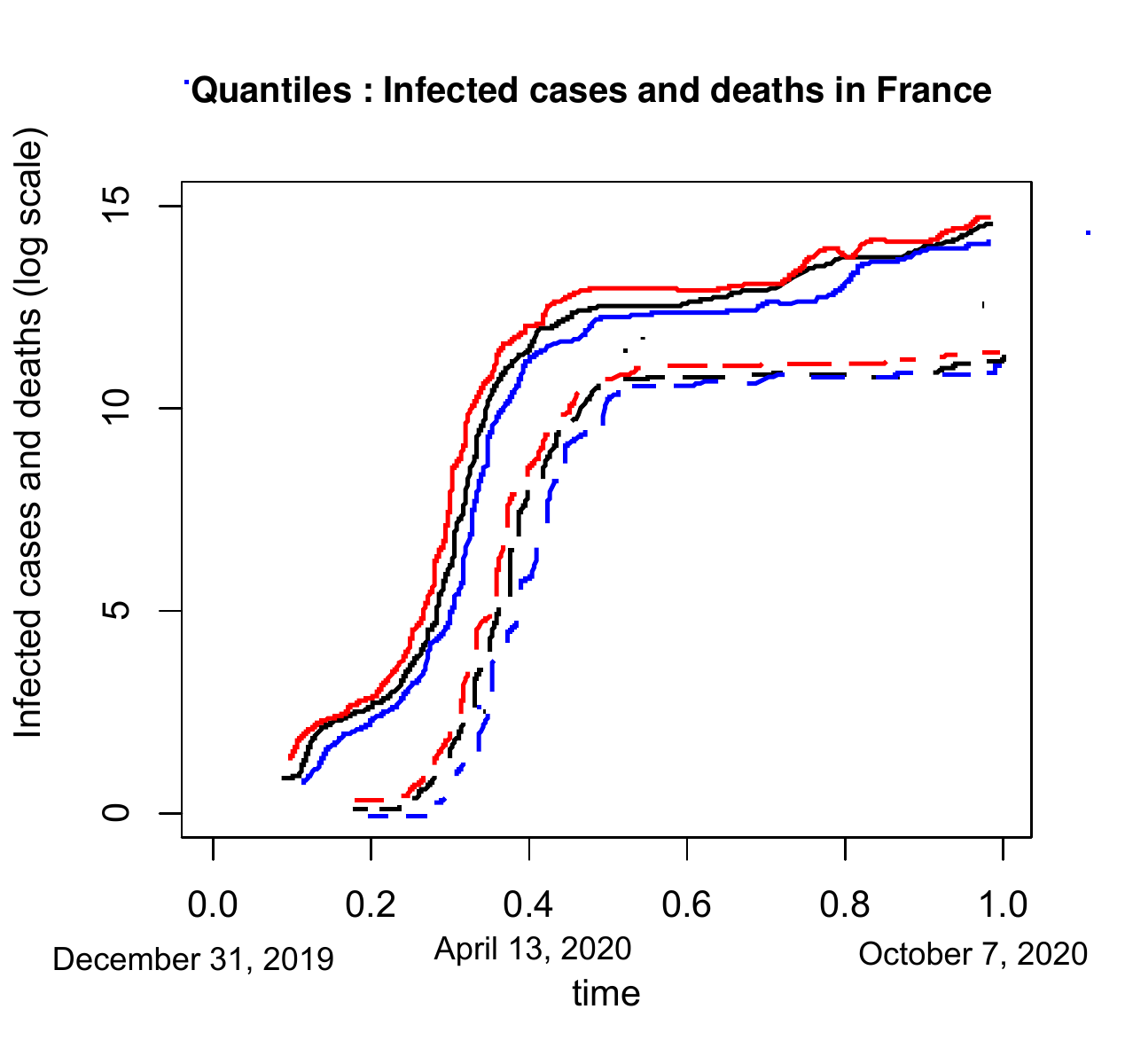}
	\caption{\it  
		\label{Fig_France}
		The left diagram plots the cumulative infected cases (solid line) and deaths (dashed line) in France. The right diagram plots the quantile curves of cumulative infected cases (solid line) and deaths (dashed line) in France. The red  curve {\color{black}corresponds to} $\tau = 0.8$, the black curve {\color{black}corresponds to} $\tau = 0.5$, and the blue curve is {\color{black}corresponds to} $\tau = 0.2$.
	} 
\end{figure}

Let us first discuss a few observations. The left diagram in Figure \ref{Fig_France} indicates that both cumulative infected cases and deaths are increasing over time in France, which is also expected as the new cases are added to the data {\color{black}everyday}. In fact, it is observed in the right diagram in Figure \ref{Fig_France} that the quantile curves of the cumulative infected cases and the deaths have an increasing trend over time. First, we now implement the SIT and the SCB tests on the full data, and the tests are carried out using the procedure explained in Section \ref{BBT}. In this context, we would like to mention that there is no co-variate here, and hence, the choice of ${\bf c}$ does not have any role. For $B = 1000$, the $p$-values of the SIT and the SCB tests are computed when $\tau$ {\color{black}equals to} 0.8, 05 and 02.  For $\tau = (0.8, 0.5, 0.2)$, the $p$-values are $(0.053, 0.067, 0.055)$ and $(0.066, 0.072, 0.062)$ for the SIT and the SCB tests, respectively. These $p$-values of both tests indicate the rejection of the null hypothesis at $8\%$ level of significance, i.e., in other words, the cumulative infected cases and deaths in France due to COVID-19 do not follow the model described in \eqref{hypo1}. {\color{black} We also implement  the test for dependent data (see Theorem \ref{core-dependent}, and we obtain the $p$-values as (0.066, 0.057, 0.058) and (0.069, 0.062, 0.066) for the SIT and the SCB tests, respectively. Hence, the conclusion remains the same.}

However, for $\tau = (0.8, 0.5, 0.2)$, we obtain the large $p$-values as $(0.326$, $0.473$, $0.375)$ and $(0.299$, $0.446$, $0.311)$ for the SIT and the SCB tests, respectively when the tests are implements on the data corresponds to time on $[0, 0.4]$, i.e., the data for the period from December 31, 2019 to April 13, 2020, i.e., altogether for the period of $105$ days. {\color{black} For the same $\tau$, the $p$-values of the SIT and the SCB tests based on Theorem \ref{core-dependent} are (0.377, 0.456, 0.403) and (0.324, 0.478, 0.336), respectively.} These $p$-values indicate that the cumulative infected cases and deaths in France have pattern like the model described in $\eqref{hypo1}$. In this study,  for $\tau = (0.8, 0.5, 0.2)$, we obtain $\hat{d} = (0.0499, 0.0564, 0.0482)$, i.e., in other words, till April 13, 2020, in France, the curves of cumulative infected cases and deaths followed the same pattern but the cumulative infected cases was ahead about five to six days (as $0.0499*105 = 5.239$, $0.0564\times 105 = 5.922$ and $0.0482*105 = 5.061$) compared to the cumulative deaths. Afterwards, as the death rate went down, the same shift difference was not observed till October 7, 2020.

\subsubsection{Cumulative infected cases in France and Germany : COVID-19 outbreak}  
In this case, we implement the SIT and the SCB tests on the cumulative infected cases of France and Germany. For $B = 1000$ and $\tau = (0.8, 0.5, 02)$, we obtain the $p$-values as $(0.376, 0.523, 0.401)$ and $(0.353, 0.458$, 0.366) of the SIT test and the SCB test, respectively, which indicates that the cumulative infected cases in France and Germany follow the model described in \eqref{hypo1}. {\color{black}In fact, the large $p$-values are obtained for the SIT and the SCB tests based on Theorem \ref{core-dependent} as well.} Moreover, for {\color{black}$\tau = (0.8, 0.5, 0.2)$}, we obtain $\hat{d} = (0.004, 0.003, 0.004)$, i.e., in other words, one can conclude that the cumulative infected cases in France have the same pattern as that of Germany, but they are approximately ahead of a day (as $0.003\times 282 = 0.846$ and $0.004\times282 = 1.128$) compared to Germany's number for the period from December 31, 2019 to October 7, 2020, i.e., altogether the period of $282$ days.  

\begin{figure}
	\includegraphics[width= 2.5in, height = 2.5in]{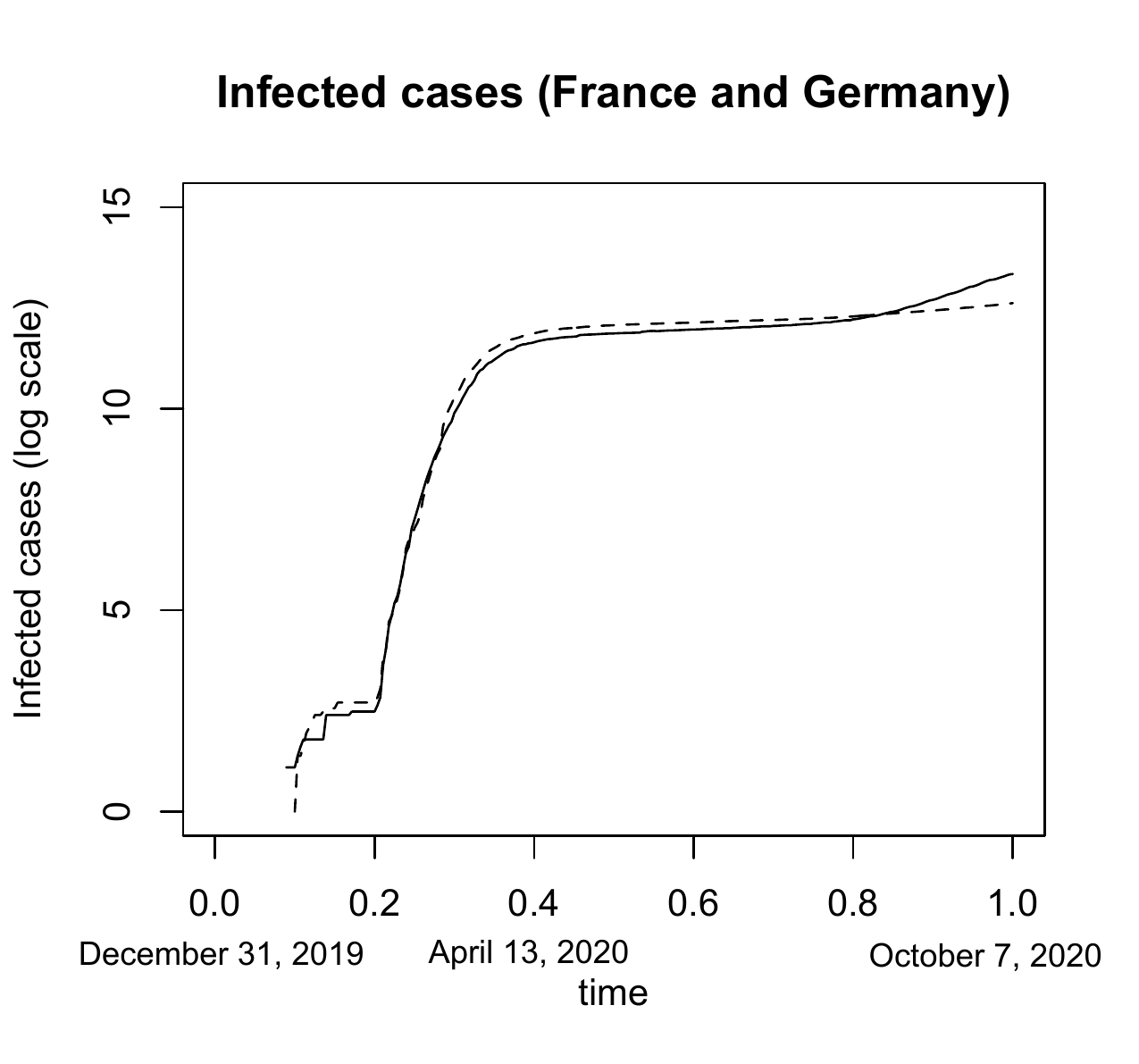}
	\includegraphics[width= 2.5in, height = 2.5in]{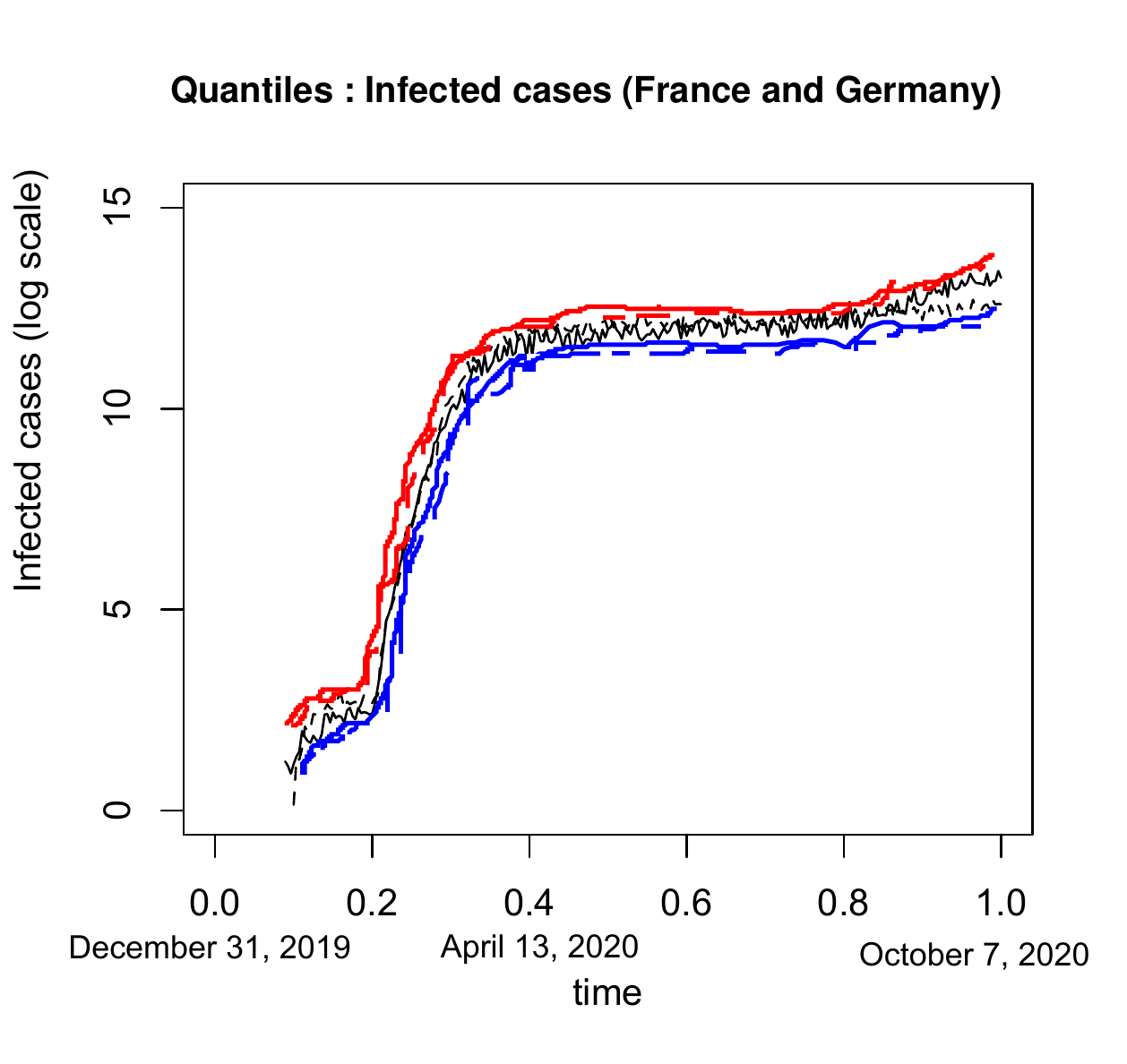}
	\caption{\it  
		\label{Fig_France_Germany_Infected}
		The left diagram  plots the cumulative infected cases of France (solid line) and that of Germany (dashed line). The right diagram plots the quantile curves of cumulative infected cases of France (solid line) and that of Germany (dashed line). The red curve {\color{black}corresponds to} $\tau = 0.8$, the black curve {\color{black}corresponds to} $\tau = 0.5$, and the blue curve is {\color{black}corresponds to} $\tau = 0.2$.
	} 
\end{figure}

\subsubsection{Cumulative deaths in France and Germany : COVID-19 outbreak}

\begin{figure}
	\includegraphics[width= 2.5in, height = 2.5in]{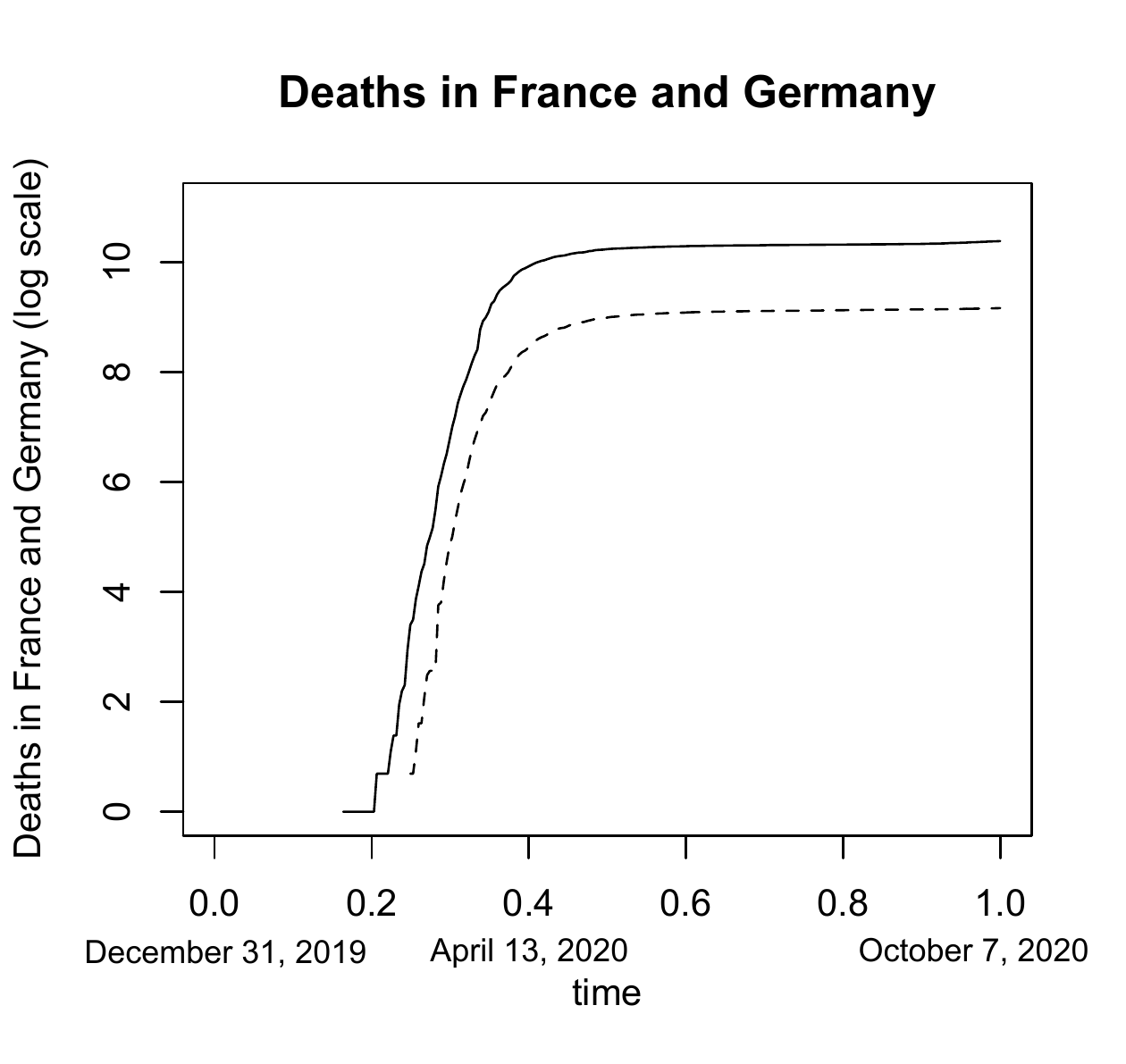}
	\includegraphics[width= 2.5in, height = 2.5in]{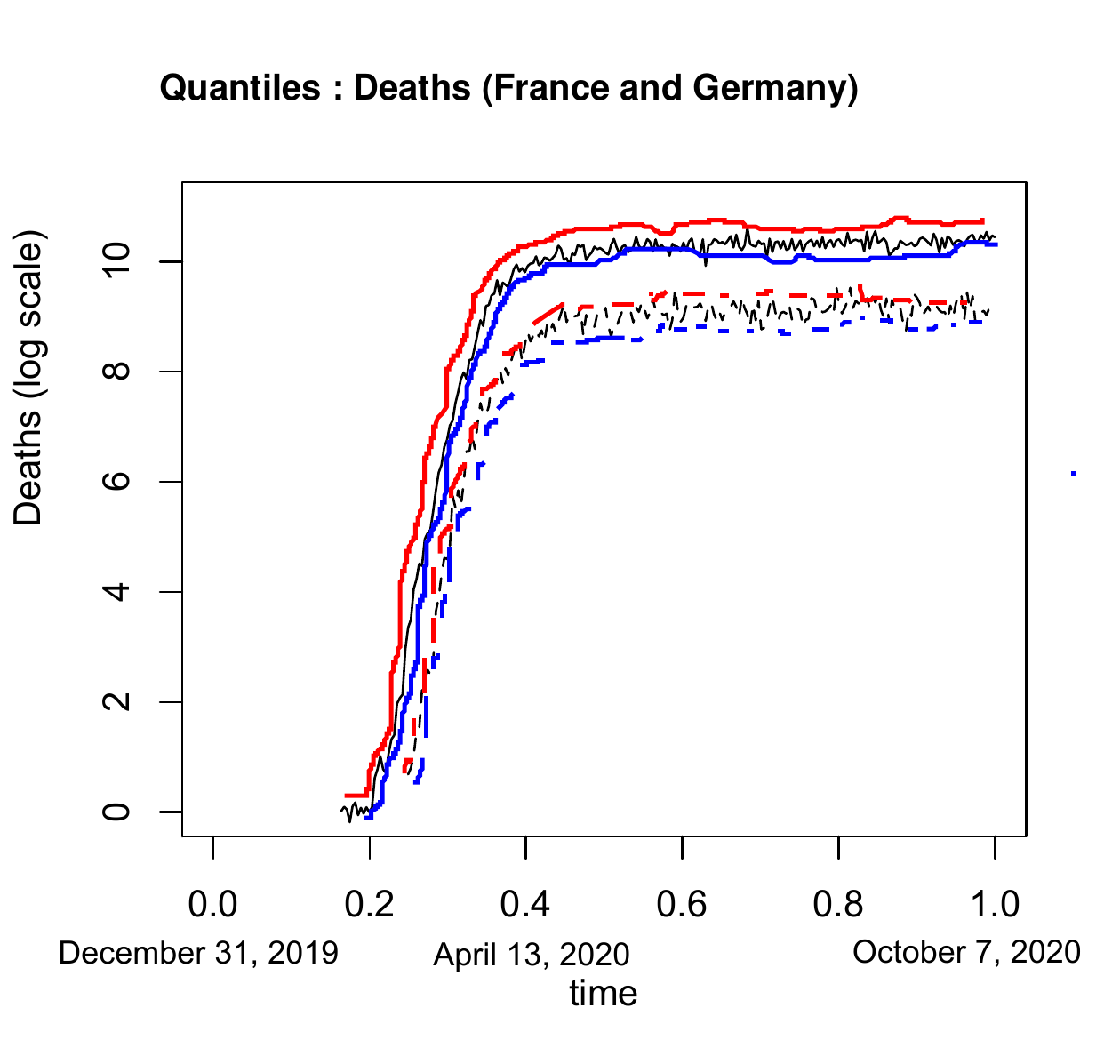}
	\caption{\it  
		\label{Fig_France_Germany_Deaths}
		The left diagram plots the cumulative death of France (solid line) and that of Germany (dashed line). The right diagram plots the quantile curves of cumulative death of France (solid line) and that of Germany (dashed line). The red curve {\color{black}corresponds to} $\tau = 0.8$, the black curve {\color{black}corresponds to} $\tau = 0.5$, and the blue curve {\color{black}corresponds to} $\tau = 0.2$.
	} 
\end{figure}

First observe that the left diagram in Figure \ref{Fig_France_Germany_Deaths} indicates that the cumulative death cases in both France and Germany are increasing over time, and it is also expected as new cases are added to the data every day. In addition, it is observed in the right diagram in Figure \ref{Fig_France_Germany_Deaths} that the quantile curves of the cumulative deaths in France and Germany have an increasing trend over time. We now implement the SIT and the SCB tests on the full data, and the tests are carried out as the earlier cases. For $B = 1000$ and $\tau = (0.8, 0.5, 0.2)$, the $p$-values are obtained as $(0.062, 0.084, 0.065)$ and $(0.057, 0.081, 0.059)$ of the SIT and the SCB tests, respectively. {\color{black} Similarly, the small $p$-values are obtained for the SIT and the SCB tests based on Theorem \ref{core-dependent}}. These $p$-values of both tests indicate the rejection of the null hypothesis at $9\%$ level of significance, i.e., in other words, the cumulative deaths due to COVID-19 in France and Germany do not follow the model described in \eqref{hypo1}.

However, we see the opposite scenario when the SIT and the SCB tests are implemented on the data corresponds to time on $[0, 0.4]$, i.e., the data for the period from December 31, 2019 to April 13, 2020, i.e., altogether the period of $105$ days. During this period of time, for $\tau = (0.8, 0.5, 0.2)$, the $p$-values are $(0.577, 0.633, 0.601)$ and $(0.501, 0.596, 0.552)$ for the SIT test and the SCB test, respectively. {\color{black} We also obtain the large $p$-values for the SIT and the SCB tests based on Theorem \ref{core-dependent}}. These large $p$-values indicate that the cumulative deaths in France and Germany have patterns like the model described in $\eqref{hypo1}$. In this study, for $\tau = (0.8, 0.5, 0.2)$, we obtain $\hat{d} = (0.079, 0.075, 0.080)$, i.e., in other words, till April 13, 2020, the cumulative deaths in France have the same pattern as that of Germany, but they were approximately ahead of eight days (as $0.079\times 105 = 8.295$, $0.075\times 105 = 7.875$ and $0.080\times 105 = 8.4$) compared to Germany's number. Afterwards, as the death rate went down, the same shift difference was not observed till October 7, 2020.

\vspace{0.1in}

\subsection{Average Temperature Anomaly}\label{Tem}
This data set consists of four variables: average temperature anomaly, the carbon emission in the form of gas, solid and liquid. We consider two regions, namely, the northern hemisphere and the southern hemisphere since the feature of average temperature anomaly and the carbon emission in the form of gas, solid and liquid are different in two hemispheres, and they are monotonically increasing over time which causes interest of study in climate science (see, e.g., \cite{Raupach2014}).  The data set for these two regions of the aforementioned four variables are available in \url{https://ourworldindata.org/co2-and-other-greenhouse-gas-emissions} and \url{https://cdiac.ess-dive.lbl.gov/trends/emis/glo_2014.html}. These yearly data sets reported the values of the variables for the period from 1850 to 2018, i.e., $n = 169$. In this study, the average temperature anomaly is considered as the response variable (denoted as $y$), and the carbon emission in the form of gas (denoted as $x_1$), liquid (denoted as $x_2$) and solid (denoted as $x_3$) are the covariates. 

\begin{figure}
	\includegraphics[width= 2.7in, height = 2.7in]{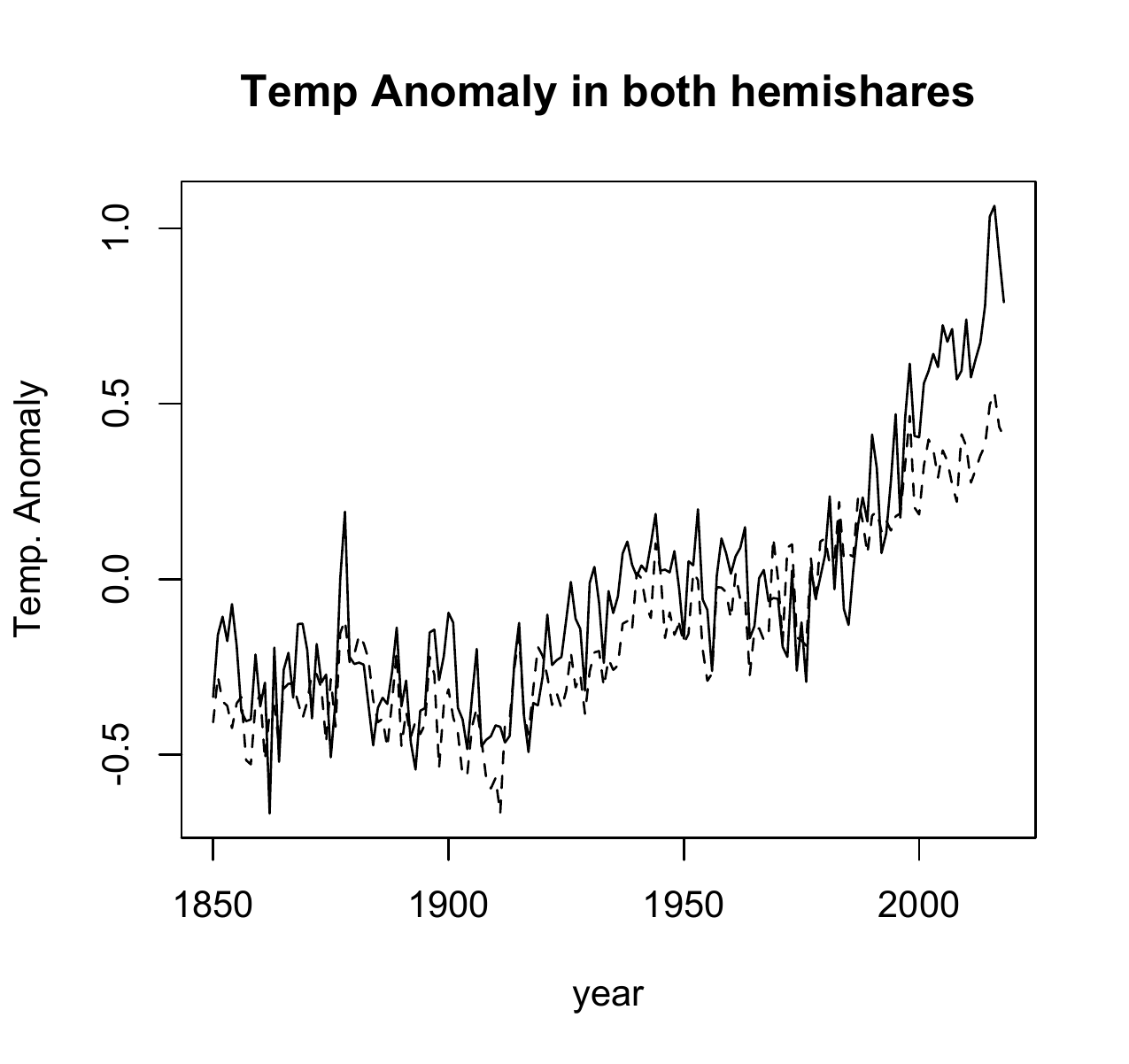}
	\includegraphics[width= 2.7in, height = 2.7in]{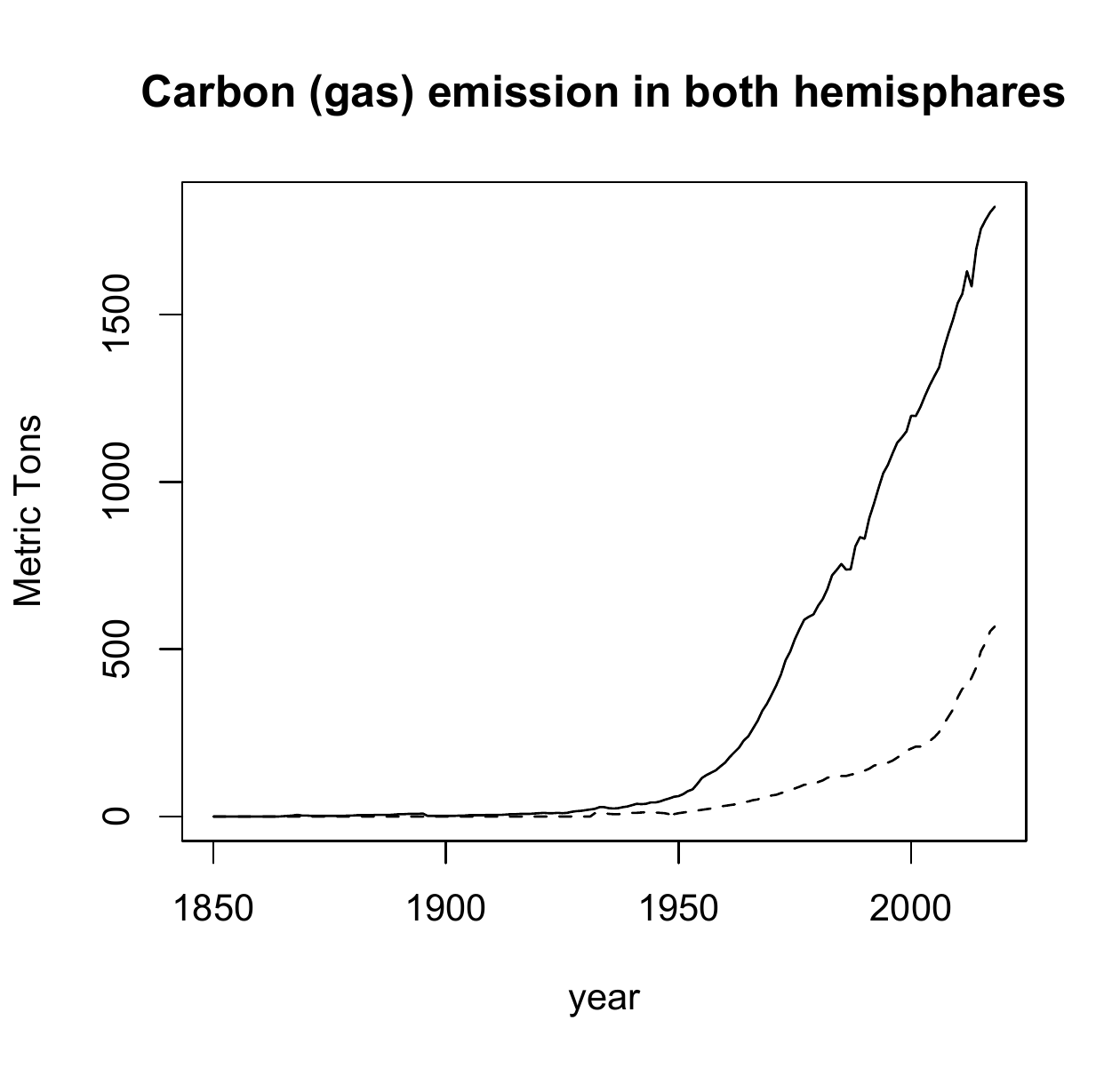}\\
	\includegraphics[width= 2.7in, height = 2.7in]{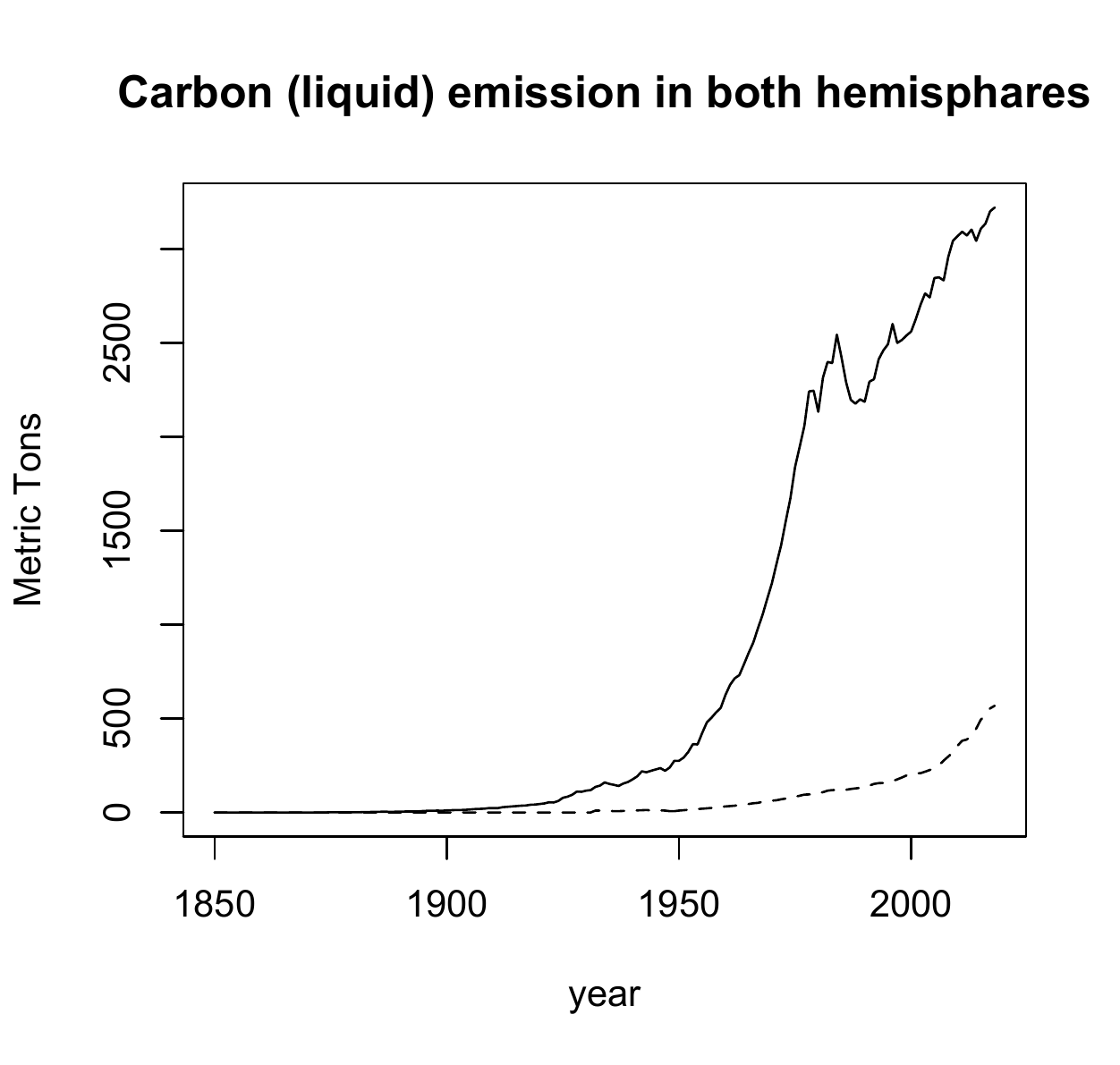}
	\includegraphics[width= 2.7in, height = 2.7in]{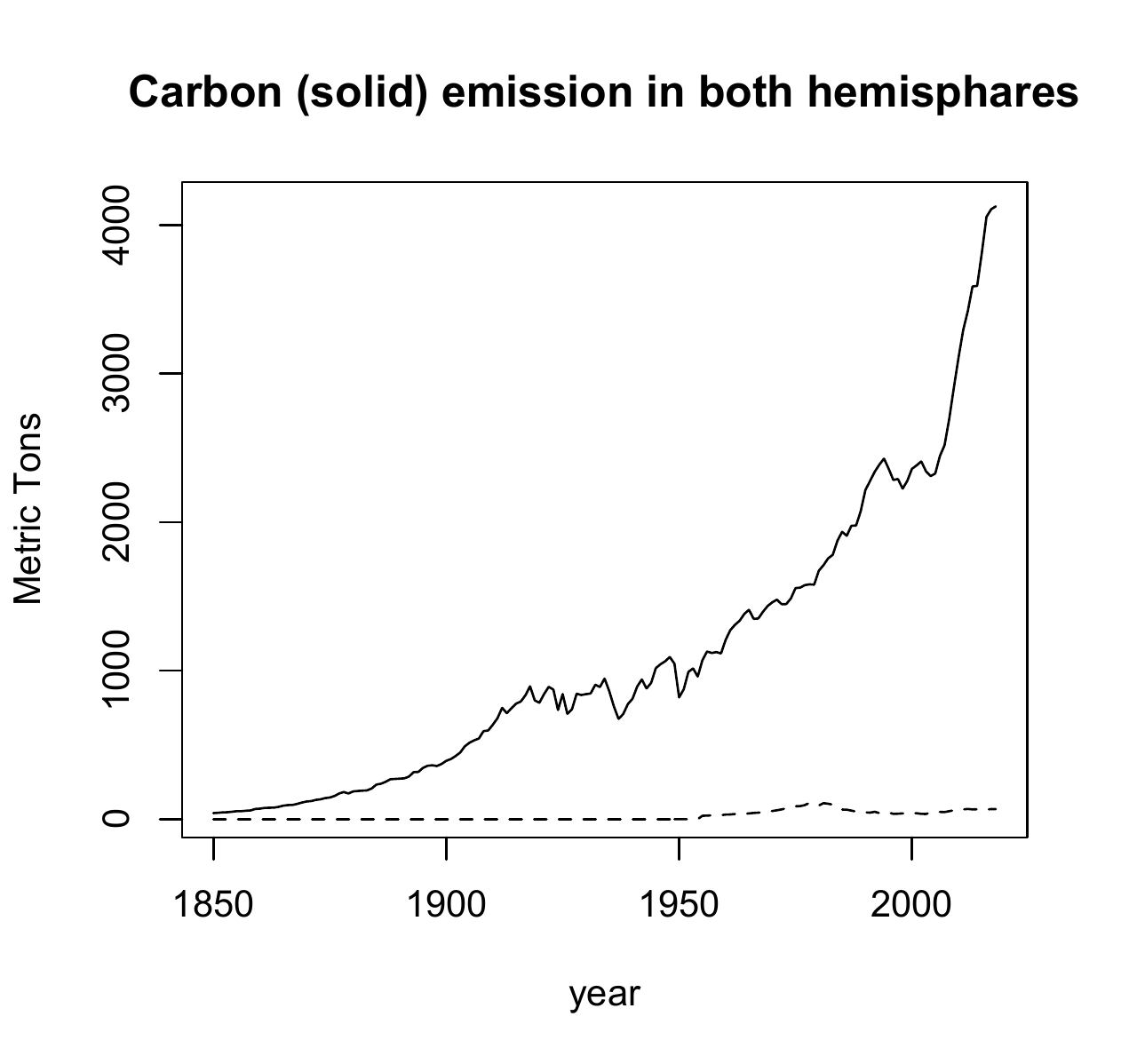}
	\caption{\it  
		\label{Fig3}
		Plots of temperature anomaly, carbon (gas, liquid and solid) emission in northern and southern hemispheres. In each diagram, the line curve represents the northern hemisphere, and the dashed curve represents the southern hemisphere. 
	} 
\end{figure}

\begin{figure}
	\includegraphics[width= 1.8in, height = 2in]{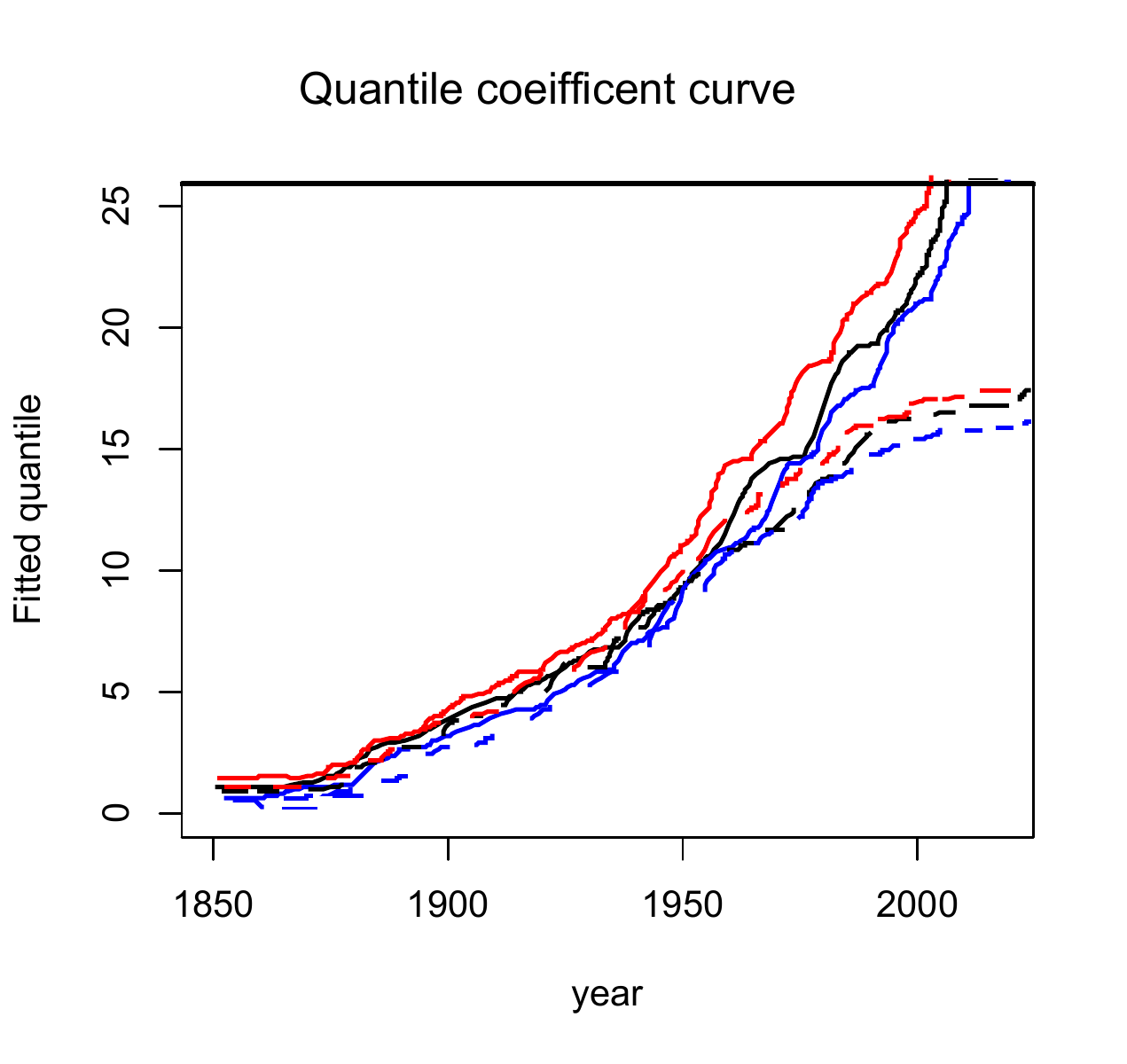}
	\includegraphics[width= 1.8in, height = 2in]{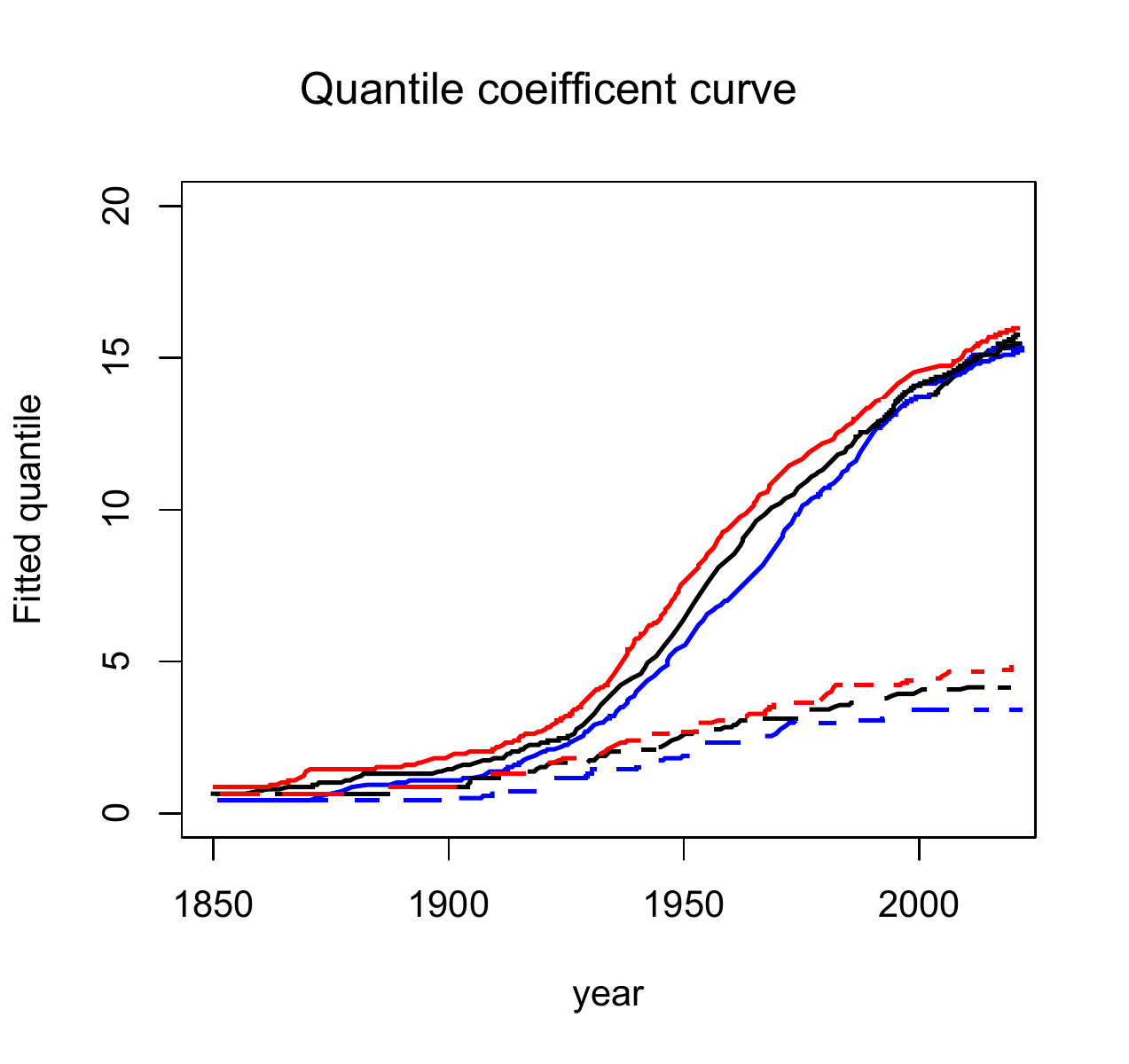}
	\includegraphics[width= 1.8in, height = 2in]{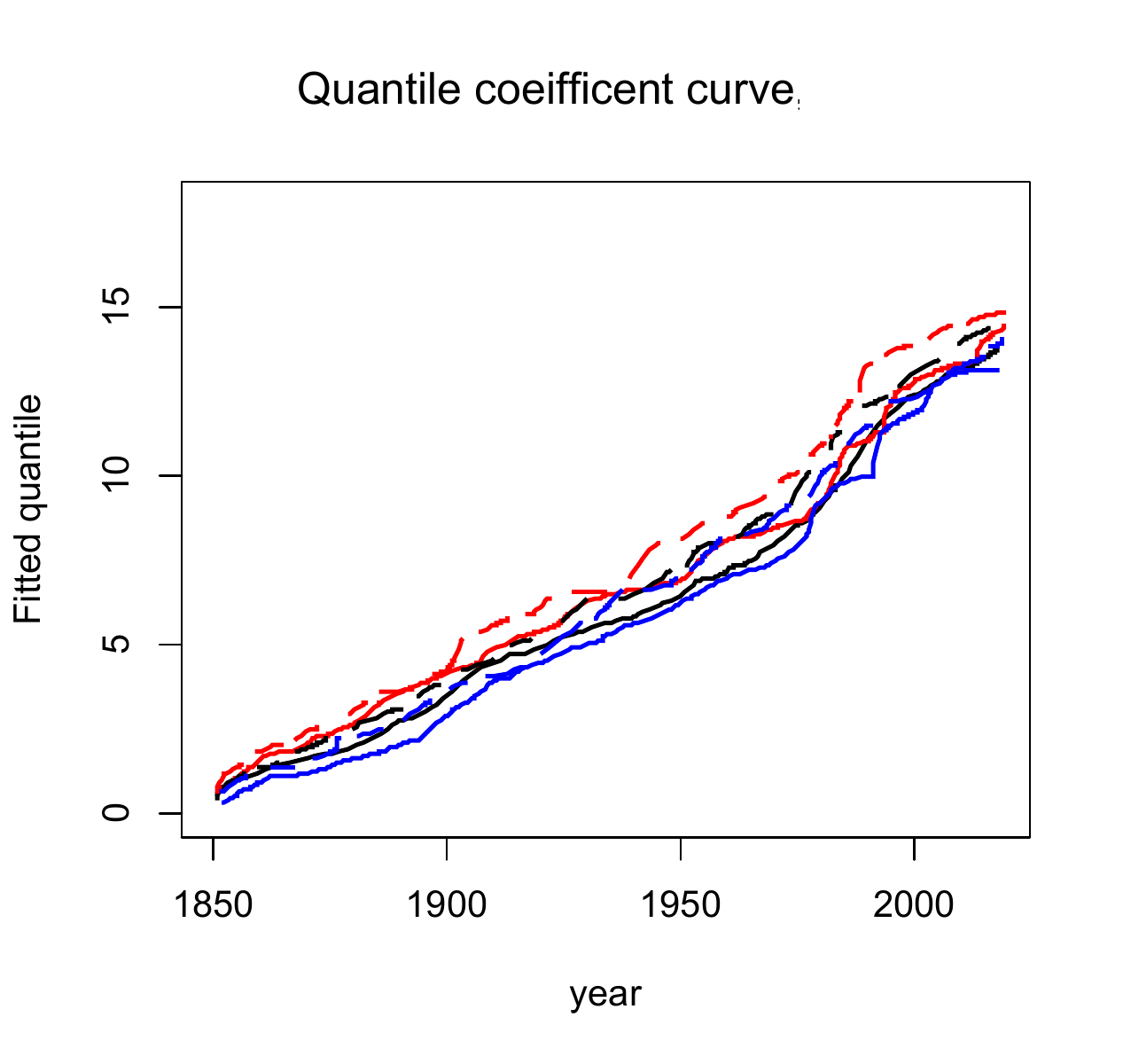}
	\caption{\it  
		\label{Fig4}
		The plots of fitted quantile coefficients curves associated with $x_{1}$, i.e., gas (left diagram), $x_{2}$, i.e., liquid (middle diagram) and $x_{3}$, i.e., solid (right diagram). In each diagram, the line curve represents for the northern hemisphere, and the dotted curve represents for the southern hemisphere. The red curve {\color{black}corresponds to} $\tau = 0.8$, the black curve {\color{black}corresponds to} $\tau = 0.5$, and the blue curve {\color{black}corresponds to} $\tau = 0.2$.	} 
\end{figure}

We then discuss a few more observations on this data. The diagrams in Figure \ref{Fig3} indicate that for both northern and southern hemispheres, $y$, $x_1$, $x_2$ and $x_3$ increase over time, which is a well-known feature in climate science. Moreover,  we observe from Figure \ref{Fig4} that the fitted quantile coefficient curves associated with $x_1$, $x_2$ and $x_{3}$ are monotonically increasing over time for a given quantile (in Figure \ref{Fig4}). We now investigate the performance of the test based on $T_{n_1, n_2}$ (here $n_1 = n_2 = n = 169$) to check whether this data favours $H_{0}$ (see \eqref{hypo1}) or not when ${\bf c} = (c_{1, 1},  c_{2, 1}, c_{3, 1}, c_{1, 2}, c_{2, 2}, c_{3, 2})$ equals  $(1, 0, 0, 1, 0, 0)$ and $(0, 0, 1, 0, 0, 1)$,  and the test is carried out following the procedure described in Section \ref{BBT}. For $\tau = (0.8, 0.5, 0.2)$ and ${\bf c} = (1, 0, 0, 1, 0, 0)$, we obtain $\hat{d} = (0.082, 0.093, 0.087)$, and the $p$-values of the SIT test and the SCB test are $(0.279, 0.347, 0.311)$ and $(0.275, 0.296, 0.288)$, respectively. Next, when ${\bf c} = (0, 0, 1, 0, 0, 1)$ and $\tau = (0.8, 0.5, 0.2)$, we obtain $\hat{d} = (0.103, 0.111, 0.108)$, and the $p$-values of the SIT test and the SCB test are $(0.517, 0.574, 0.528)$ and $(0.479, 0.513, 0.494)$, respectively. 
These $p$-values of both the tests SIT and SCB indicate that this data set favours the null hypothesis for ${\bf c} = (1, 0, 0, 1, 0, 0)$ and ${\bf c}= (0, 0, 1, 0, 0, 1)$, which is consistent with the nature of the curves drawn on the left and the right diagrams of Figure \ref{Fig4}. {\color{black} The same phenomena i.e., the large $p$-values, is observed for the SIT and the SCB tests based on Theorem \ref{core-dependent} as well.} 

However, for $\tau = (0.8, 0.5, 0.2)$ and ${\bf c} = (c_{1, 1}, c_{2, 1}, c_{3, 1}, c_{1, 2}, c_{2, 2}, c_{3, 2}) = (0, 1, 0, 0, 1, 0)$, the $p$-values of the SIT test and the SCB tests are $(0.068, 0.087, 0.071)$ and $(0.056, 0.081, 0.069)$, respectively, which indicates a rejection of the null hypothesis at $9\%$ level of significance. {\color{black} In fact, the SIT and the SCB tests based on Theorem \ref{core-dependent} also obtain small $p$-values.} These small $p$-values obtained in the last case is consistent with the feature of the curves illustrated in the middle diagram in Figure \ref{Fig4},  which clearly indicates that there is no any constant shift between the quantile coefficient curves of emitting solid carbon for the northern and the southern hemispheres. This fact leads to  relatively small $p$-values.

{\bf Acknowledgment} Weichi Wu (corresponding author) is funded by NSFC (No.11901337) Young program, and Subhra Sankar Dhar is funded by SERB project MATRICS (MTR/2019/000039), Government of India. The authors are grateful to the referees and the Associate Editor for their constructive comments on an earlier version of this paper. The authors would also like to thank Dr.\ Yeonwoo Rho for sharing their code to implement the algorithm proposed in their paper on unit root test. 


\renewcommand{\thelemma}{\Alph{section}.\arabic{lemma}}
\renewcommand{\theproposition}{\Alph{section}.\arabic{proposition}}
\renewcommand{\theequation}{\Alph{section}.\arabic{equation}}
\renewcommand{\thetheorem}{\Alph{section}.\arabic{theorem}}
\renewcommand{\thecorol}{\Alph{section}.\arabic{corol}}
\renewcommand{\thesection}{\Alph{section}}  

\setcounter{equation}{0}

\setcounter{section}{0}\ \\

 \textbf{
 	\begin{center}
 	Supplemental material for ``Comparing time varying regression quantiles under shift invariance"
 	\end{center} }
\begin{abstract}
	This supplementary material contains a sketch of the proofs as well as all the detailed proofs for Theorem \ref{asymptotic}, Theorem \ref{SCB} and {\color{black}Theorem \ref{core-dependent}} in the main article, proof of proposition \ref{prop-monotone} and a remark \ref{remark1} on the parameter $\eta$. {\color{black}Moreover, the simulation results for mutually dependent data sets are also available here. }
\end{abstract}

\setcounter{equation}{1}
\section{Sketch of the Proofs of Theorems  \ref{asymptotic} and \ref{SCB}}\label{proof}
 The properties of the estimators $\hat a$ and $\hat b$ follow from the  stochastic expansion of  the deviation of the local linear quantile estimator $\hat m_s(t)-m_s(t)$, (cf. Porposition 1 of \cite{wu2017nonparametric}) as well as the monotonicity of functions $\mu_s$, $s=1$ and 2. The proof of Theorem \ref{asymptotic} {\color{black}has} two steps. The first step is to expand the deviation of the test statistics under null and local alternatives, and approximate it by certain Gaussian processes using the extended  argument of  proof of  Theorem 4.1 of \cite{dette2019detecting}. Notice that in our case, we consider two samples as well as the local linear quantile regression. The second step is {\color{black}to use} Theorem 2.1 of \cite{de1987central} to figure out the asymptotic behavior of a quadratic Gaussian integrals via tedious calculations. Based on Step 1 of the proof of Theorem \ref{asymptotic} and using the approximation formula in Proposition 1 of \cite{sun1994simultaneous}, we derive the proof of Theorem \ref{SCB}, which obtains the simultaneous confidence band.  Intensive calculations are provided in our proof to determine the parameters in the approximation formula of \cite{sun1994simultaneous}. 

\section{Proofs of Theorem \ref{asymptotic} and \ref{SCB}.}

\subsection{Proof of Theorem \ref{asymptotic}}

{\it Proof.} Write \begin{align}\label{tildeTn}
\tilde T_{n_1,n_2}=\int\limits_{\mathbb R} (\hat g_1(t)-\hat g_2(t))^2w(t)dt\text{, where} ~~ w(t)=\mf 1(a+\eta\leq t\leq  b-\eta).
\end{align}
In the following, we shall prove that, under conditions of Theorem \ref{asymptotic},
\begin{align}
n_1b_{n,1}^{5/2}\tilde T_{n_1,n_2}-b_{n,1}^{-1/2}(\check B_1+\gamma_0\gamma_1^3\check B_2))-\int\limits_{\mathbb R} \kappa^2(t)w(t)dt\Rightarrow N(0,V_T),\label{tp1}\\
n_1b_{n,1}^{5/2}(\tilde T_{n_1,n_2}-T_{n_1,n_2})=o_p(1).\label{tp2}
\end{align}\bigskip

\noindent{\bf Proof of \eqref{tp1}:}\\

Define $ g_{s }(t) = \frac{1}{Nh_s}\sum\limits_{i = 1}^{N}H\Big (\frac{{m}_{s}(\frac{i}{N}) - t}{h_s}\Big )$. Then we have the following decomposition:
\begin{align}
\tilde T_{n_1,n_2}=\int\limits_{\mathbb R} (J_1(t)-J_2(t)+J_3(t))^2w(t)dt,\\
J_s(t)=\hat g_s(t)-g_s(t), s = 1~\mbox{and}~2, J_3(t)=g_1(t)-g_2(t). 
\end{align}
Using the similar argument of page 471 of \cite{dette2006simple}, we have
\begin{align}\label{3.20}
J_3(t)=(m_1^{-1})'(t)-(m_2^{-1})'(t)+O\left(h_1+h_2+\frac{1}{Nh_1}+\frac{1}{Nh_2}\right)\notag\\
=\rho_n\kappa(t)+O\left(h_1+h_2+\frac{1}{Nh_1}+\frac{1}{Nh_2}\right).
\end{align}
Next, by Taylor series expansion, we have that for $s=1$ and 2, the following decomposition holds: 
\begin{align}
J_s(t)=J_{s,1}(t)+J_{s,2}(t),
\end{align}
where \begin{align*}
&J_{s,1}(t)=\frac{1}{Nh^2_s}\sum_{i=1}^N
H'\left(\frac{m_s(i/N)-t}{h_s}\right)(\hat m_s(i/N)-m_s(i/N)),\\
&
J_{s,2}(t)=\frac{1}{2Nh^3_s}\sum_{i=1}^{N}
H''\left(\frac{m_s(i/N)-t+\nu^*_s(\hat m_s(i/N)-m_s(i/N))}{h_s}\right)(\hat m_s(i/N)-m_s(i/N))^2
\end{align*}
for some $\nu^*_s\in [-1,1]$ ($s=1$ and 2). Notice that $\frac{\log^{4/3}n_s}{\sqrt {(n_sb_{n,s})}h_s}=o(1)$, and thus the number of non-zero
summands in $J_{s,2}(t)$  is of order $O\left(\frac{\log^{4/3}n_s}{\sqrt {(n_sb_{n,s})}}\right)$ with probability $1$. Using Propositions \ref{propGaussian} and \ref{propMs}, with the same  arguments  in the proof of  Theorem 4.1 in  the online  supplement of \cite{dette2019detecting} for obtaining bound for $\Delta_{2,N}$ in their paper, we obtain that for $s=1$ and 2, and uniformly for $t\in [a+\eta, b-\eta]$,
\begin{align}\label{Mach-10}
J_{s,2} (t)=O_p\left(\frac{\log ^4 n_s}{(n_sb_{n,s})^{3/2}h_s^2}\right).
\end{align}
Therefore, by applying the assertion of Proposition \ref{propGaussian} to $J_{s,1}(t)$, equations \eqref{3.20}--\eqref{Mach-10}, on a possibly richer probability space, there exists a sequence of i.i.d.\ standard normal random variables $(V_{s,i})_{i\in \mathbb Z}$,  $s=1$ and 2 such that $\tilde T_{n_1,n_2}$ can be written as
\begin{align}\label{B10}
\tilde T_{n_1,n_2}=\int\limits_{\mathbb R}[(m_1^{-1})'(t)-(m_2^{-1})'(t)+Z(t)+R(t)]^2w(t)dt\notag
\\=\int\limits_{\mathbb R}[\rho_n\kappa(t)+Z(t)+R(t)]^2w(t)dt,
\end{align}
where 
\begin{align}\label{Z(t)}
&Z(t)=\sum_{s=1}^2 \frac{(-1)^{s-1}}{n_sb_{n,s}Nh_s^2}\sum_{i=1}^N\sum_{j=1}^{n_s}M_{\mf c_s}(i/N)H'\left(\frac{m_s(i/N)-t}{h_s}\right)\bar K_{b_{n,s}}(j/n_s-i/N)V_{j,s},\\
&~~\mbox{and}~~\sup_{t\in [a+\eta,b-\eta]}|R(t)|=O_p(\Omega_n).\label{Rt}
\end{align}

As \begin{align}
n_1b_{n,1}^{5/2}\rho_n^2\int\limits_{\mathbb R} \kappa^2(t)w(t)dt=\int\limits_{\mathbb R} \kappa^2(t)w(t)dt,
\end{align}to prove \eqref{tp1}, it is sufficient to show that
\begin{align}
n_1b_{n,1}^{5/2}\int\limits_{\mathbb R} Z^2(t)w(t)dt -b_{n,1}^{-1/2}(\check B_1+\gamma_0\gamma_1^3\check B_2))-\int\limits_{\mathbb R} \kappa^2(t)w(t)dt\Rightarrow N(0,V_T),\label{tp1_1}\\
n_1b_{n,1}^{5/2}\left(\int\limits_{\mathbb R} R(t)(R(t)+\rho_n \kappa(t)+Z(t))w(t)dt+\int\limits_{\mathbb R} \rho_n\kappa(t)Z(t)w(t)dt\right)=o_p(1).\label{tp1_2}\end{align}
\smallskip
{\bf Proof of \eqref{tp1_1}}:\\
We decompose $Z(t)$ by $Z(t):=Z_1(t)-Z_2(t)$. Here for $s=1$ and 2, we have 
\begin{align}
&Z_s(t)=\sum_{j=1}^{n_s}W_s(m_s,j,t)V_{j,s},\label{5.33-3-26}\mbox{,~where}\\
&W_s(m_s,j,t)=\frac{1}{n_sb_{n,s}Nh_s^2}\sum_{i=1}^NM_{\mf c_s}(i/N)H'\left(\frac{m_s(i/N)-t}{h_s}\right)\bar K_{b_{n,s}}(j/n_s-i/N).\label{5.34-3-26} 
\end{align}
As a result, we have
\begin{align}
\int\limits_{\mathbb R} (Z_1(t)-Z_2(t))^2w(t)dt=A_1+A_2-2A_{12},\label{5.35-2020-26}~\mbox{where}\\
A_s=\int\limits_{\mathbb R} \left(\sum_{j=1}^{n_s}W_s(m_s,j,t)V_{j,s}\right)^2w(t)dt,~~s=1~\mbox{and}~2,\\
A_{12}=\int\limits_{\mathbb R}  \left(\sum_{j=1}^{n_1}W_1(m_1,j,t)V_{j,1}\right)\left(\sum_{j=1}^{n_2}W_2(m_2,j,t)V_{j,2}\right)w(t)dt.
\end{align}

We first prove the results for $A_1$, and the result for $A_2$ can be evaluated in a similar way. Notice that $A_1=A_{1,a}+A_{1,b}$, where
\begin{align}
A_{1,a}=\sum_{j=1}^{n_1}\left[\int\limits_{\mathbb R}W^2_1(m_1,j,t)w(t)dt\right] V^2_{j,1},\\
A_{1,b}=2\sum_{1\leq j_1<j_2\leq n_1}\left[\int\limits_{\mathbb R} W_1(m_1,j_1,t)W_1(m_1,j_2,t)w(t)dt\right] V_{j_1,1}V_{j_2,1}.
\end{align}

Now, using the bandwidth condition $h_s=o(b_{n,s})$, calculating the Riemann sum with widths of summands $1/N$ and change of variable $y=(m_s(u)-t)/h_s$, a few steps algebraic calculations show that 
\begin{align}
W_s(m_s,j,t)&=\frac{1}{n_sb_{n,s}h_s}\int\limits_{\mathbb R} M_{\mf c_s}(m_s^{-1}(t+h_sy))H'(y)\times\notag
\\&\bar K_{b_{n,s}}(j/n_s-m_s^{-1}(t+h_sy))(m_s^{-1})'(t+h_sy)dy+O(R(j,s,n,t)),\label{5.40-3-26}
\end{align}
where
\begin{align}
R(j,s,n,t)=\frac{1}{n_sb_{n,s}h_s^2N}\mf 1\left(\left|\frac{j/n_s-m_s^{-1}(t)}{b_{n,s}+Mh_s}\right|\leq 1\right)
\end{align}
for a sufficiently large constant $M$. Since $H$ is chosen to be symmetric, we have $\int\limits_{\mathbb R} H'(x)dx=0$. Therefore, by Taylor series expansion, for $t$ with $w(t)\neq 0$, it follows that for $s=1$ and 2, the leading term of $W_s(m_s,j,t)$ can be written as $\tilde W_s(m_s,j,t)(1+O(b_{n,s}+\frac{h_s}{b_{n,s}}))$, 
where
\begin{align}\label{leading}
\tilde W_s(m_s,j,t)&=\frac{-1}{n_sb^2_{n,s}}M_{\mf c_s}(m_s^{-1}(t))\bar K'\left(\frac{j/n_s-m_s^{-1}(t)}{b_{n,s}}\right)((m_s^{-1})'(t))^2\int\limits_{\mathbb R} H'(y)ydy\notag
\\&=\frac{-1}{n_sb^2_{n,s}}\check g_s(t)\bar K'\left(\frac{j/n_s-m_s^{-1}(t)}{b_{n,s}}\right).
\end{align}

Next, by using \eqref{leading}, we have 
\begin{align}\label{bias1}
\E(A_{1,a})&=\sum_{j=1}^{n_1}\int\limits_{\mathbb R}\tilde W_1^2(m_1,j,t)w(t)dt(1+o(1))\notag
\\&=\frac{1}{n_1b^3_{n,1}}\left(\int\limits_{\mathbb R} H'(y)dy\right)^2\int\limits_{\mathbb R} \bar K^{'2}(x)dx \int\limits_{\mathbb R} M_{\mf c_1}^2(u)(m_1'(u))^{-3}w(m_1(u))du(1+o(1))\notag\\
&=\frac{1}{n_1b^3_{n,1}} \check B_1(1+o(1)).
\end{align}
On the other hand, the similar calculations show that
\begin{align}
\Var(A_{1,a})=\sum_{j=1}^{n_1}\left[\int\limits_{\mathbb R}W_1^2(m_1,j,t)w(t)dt\right]^2=O\left(\frac{n_1b^2_{n,1}}{n_1^4b_{n,1}^8}\right)=O\left(\frac{1}{n_1^3b^6_{n,1}}\right).\label{VarA1a}
\end{align}

Then for $A_{1,b}$, we have that due to symmetry of the $\int\limits_{\mathbb R} W_1(m_1,j_1,t)W_1(m_1,j_2,t)w(t)dt$ in $j_1$ and $j_2$,
\begin{align}
\Var(A_{1,b})/2&=\sum_{j_1=1}^{n_s}\sum_{j_2=1,j_2\neq j_1}^{n_s}\left[\int\limits_{\mathbb R} W_1(m_1,j_1,t)W_1(m_1,j_2,t)w(t)dt\right]^2\notag
\\&=\sum_{j_1=1}^{n_s}\sum_{j_2=1,j_2\neq j_1}^{n_s}\left[\int\limits_{\mathbb R} \tilde W_1(m_1,j_1,t)\tilde W_1(m_1,j_2,t)w(t)dt \right]^2(1+o(1))\notag
\\&=\frac{1}{n_s^2b_{n,s}^{8}}\int\limits_{\mathbb R}\int\limits_{\mathbb R}\left[\int\limits_{\mathbb R}\check g_1^2(t)\bar K'\left(\frac{u-m_1^{-1}(t)}{b_{n,1}}\right)\bar K'\left(\frac{v-m_1^{-1}(t)}{b_{n,1}}\right)w(t)dt\right]^2dudv\notag \\&\times(1+o(1)).
\end{align}
Now, by changing variable, i.e., letting $\frac{u-m_1^{-1}(t)}{b_{n,1}}=x$, and using the fact that $\tilde g_1(m_1(u))=\check g_1^2(m_1(u))w(m_1(u))m_1'(u)$, we have
\begin{align}
&\Var(A_{1,b})/2\notag
\\&=\frac{1}{n^2_1b^6_{n,1}}\int\limits_{\mathbb R}\int\limits_{\mathbb R}\left[\int\limits_{\mathbb R} \check g_1^2(m_1(u))\bar K'(x)\bar K'\left(x+\frac{v-u}{b_{n,1}}\right)w(m(u))m'(u)dx\right]^2dudv(1+o(1))\notag\\
&=\frac{1}{n_1^2b_{n,1}^6}\int\limits_{\mathbb R}\int\limits_{\mathbb R}\tilde g^2_1(m_1(u))\left[\int\limits_{\mathbb R} \bar K'(x)\bar K'\left(x+\frac{v-u}{b_{n,1}}\right)dx\right]^2dudv(1+o(1))\notag\\
&=\frac{1}{n_1^2b^{5}_{n,1}}\int\limits_{\mathbb R}\int\limits_{\mathbb R}\tilde g^2_1(m_1(u)) (\bar K'\star \bar K'(y))^2dudy(1+o(1))=\frac{1}{n_1^2b^{5}_{n,1}}\check V_1(1+o(1)).\label{VarA1b}
\end{align}
Combining \eqref{bias1} \eqref{VarA1a} and \eqref{VarA1b}, it follows that
\begin{align}\label{A1result}
\E (A_{1,a})=\frac{1}{n_1b^3_{n,1}} \check B_1(1+o(1)),\\ \Var (A_{1,a})=O\left(\frac{1}{n_1^3b^6_{n,1}}\right),\\  \Var(A_{1,b})\notag=\frac{2}{n_1^2b^{5}_{n,1}}\check V_1(1+o(1)).
\end{align}
Similarly, 
\begin{align}\label{A2result}
\E (A_{2,a})=\frac{1}{n_2b^3_{n,2}} \check B_2(1+o(1)), \Var (A_{2,a})=O\left(\frac{1}{n_2^3b^6_{n,2}}\right), \Var(A_{2,b})=\frac{2}{n_2^2b^{5}_{n,2}}\check V_2(1+o(1)).
\end{align}
On the other hand, for $A_{12}$, we have
\begin{align}
&\Var(A_{12})=\sum_{j_1=1}^{n_1}\sum_{j_2=1}^{n_2}\left[\int\limits_{\mathbb R} W_1(m_1,j_1,t)W_2(m_2,j_2,t)w(t)dt\right]^2\notag
\\&=\frac{1}{n_1n_2b_{n,1}^4b^4_{n,2}}\int\limits_{\mathbb R} \int\limits_{\mathbb R} \left[\int\limits_{\mathbb R} \check g_1(t)\check g_2(t)\bar K'\left(\frac{u-m_1^{-1}(t)}{b_{n,1}}\right)\bar K'\left(\frac{v-m_2^{-1}(t)}{b_{n,2}}\right)w(t)dt\right]^2dudv(1+o(1)),\\
\end{align}
 By change of variable using $x=(u-m_1^{-1}(t))/b_{n,1}$,
we have that  
\begin{align}
&\int\limits_{\mathbb R} \int\limits_{\mathbb R} \left[\int\limits_{\mathbb R} \check g_1(t)\check g_2(t)\bar K'\left(\frac{u-m_1^{-1}(t)}{b_{n,1}}\right)\bar K'\left(\frac{v-m_2^{-1}(t)}{b_{n,2}}\right)w(t)dt\right]^2dudv\notag\\
&=b^2_{n,1}\int\limits_{\mathbb R} \int\limits_{\mathbb R} \tilde g^2_{1,2}(m_1(u))\left[\int\limits_{\mathbb R} \bar K'(x)\bar K'\left(\frac{m_{21}'(u)b_{n,1}x+v-m_{21}(u)}{b_{n,2}}\right)dx\right]^2dudv(1+o(1))\notag\\
&=b^2_{n,1}b_{n,2}\int\limits_{\mathbb R} \int\limits_{\mathbb R} \tilde g^2_{1,2}(m_1(u))\left[\int\limits_{\mathbb R} \bar K'(x)\bar K'\left(\frac{m_{21}'(u)b_{n,1}x}{b_{n,2}}+y\right)dx\right]^2dudy(1+o(1)).
\end{align}
Therefore, we have that
\begin{align}\label{A12-result}
I=\frac{\check V_{12}(\gamma_1)}{n_1n_2b_{n,1}^2b_{n,2}^3}(1+o(1))
\end{align} 
Notice that $A_{1,b}, A_{2,b}, A_{12}$ are mutually uncorrelated. Using this fact, bandwidth conditions and \eqref{5.35-2020-26}, \eqref{A1result}, \eqref{A2result} and \eqref{A12-result}, equation \eqref{tp1_1} follows from Theorem 2.1 of  \cite{de1987central} and tedious calculations. \smallskip

{\bf Proof of \eqref{tp1_2}:} By \eqref{5.33-3-26}, \eqref{5.34-3-26}, \eqref{5.40-3-26}-\eqref{leading}, we obtain for $s=1,2$
\begin{align}\label{MaxZ}
\sup_{t\in [0,1]}|Z_s(t)|=\frac{\sqrt{\log n_s}}{\sqrt{n_sb_{n,s}}b_{n,s}},
\end{align}
which together with bandwidth conditions yields \eqref{tp1_2}. Hence \eqref{tp1} holds.\smallskip

{\bf Proof of Theorem \ref{asymptotic} (conclusion part):} By the proof of \eqref{tp1} and the bandwidth conditions, it suffices to show that
\begin{align}\label{finalerror}
n_1b^{5/2}_{n,1}\int\limits_{\mathbb R} (Z(t))^2(\hat w(t)-w(t))dt=o_p(1).
\end{align}
Meanwhile, assertion in \eqref{finalerror} follows from \eqref{MaxZ}, Proposition \ref{propbound} and bandwidth conditions. Now, by \eqref{tp1} and \eqref{tp2}, Theorem \ref{asymptotic} follows. \hfill $\Box$

\subsection{ Proof of Theorem \ref{SCB}}
{\it Proof.} We prove the theorem in two steps.

Step 1:
Recall $Z(t)$ defined in \eqref{Z(t)} in the proof of Theorem \ref{asymptotic} . We first evaluate 
$\E(Z^2(t))$, $\E(Z'^2(t))$ and $\E(Z(t)Z'(t))$. 
Since $h_s=o(b_{n,s})$, uniformly for $t\in [a+\eta, b-\eta]$, we have 
\begin{align*}
&\E(Z^2(t))\\&=\sum_{s=1}^2 \frac{1}{(n_sb_{n,s}Nh_s^2)^2}\sum_{j=1}^{n_s}\left(\sum_{i=1}^NM_{\mf c_s}(i/N)H'\left(\frac{m_s(i/N)-t}{h_s}\right)\bar K_{b_{n,s}}(j/n_s-i/N)\right)^2\\
&=\sum_{s=1}^2 \frac{1}{(n_sb_{n,s}h_s^2)^2}\sum_{j=1}^{n_s}\left(\int\limits_{\mathbb R} M_{\mf c_s}(u)H'\left(\frac{m_s(u)-t}{h_s}\right)\bar K_{b_{n,s}}(j/n_s-u)du+O\left(\frac{\mf 1(|j/n_s-t|\leq \bar M b_{n,s})}{N}\right)\right)^2\\
&=\sum_{s=1}^2 \frac{1}{n_s(b_{n,s}h_s^2)^2}\int\limits_{\mathbb R}\left(\int\limits_{\mathbb R} M_{\mf c_s}(u)H'\left(\frac{m_s(u)-t}{h_s}\right)\bar K_{b_{n,s}}(v-u)du\right)^2dv+O\left(\frac{1}{n_s^2b^2_{n,s}h_s^2}+\frac{1}{n_sNb^2_{n,s}h_s^2}\right)\\
&=\sum_{s=1}^2 \frac{M^2_{\mf c_s}(m_s^{-1}(t))}{n_s(b_{n,s}h_s)^2(m_s'(m_s^{-1}(t)))^2}\int\limits_{\mathbb R}\left(\int\limits_{\mathbb R} H'(x)\bar K_{b_{n,s}}(v-m_s^{-1}(t+xh_s))dx\right)^2dv+O\left(\frac{1}{n_sb_{n,s}h_s}\right)
\\&=\sum_{s=1}^2 \frac{M^2_{\mf c_s}(m_s^{-1}(t))(m_s^{-1})'^2(t)}{n_sb^3_{n,s}(m_s'(m_s^{-1}(t)))^2}\int\limits_{\mathbb R} \bar K'^2(y)dy\left(\int\limits_{\mathbb R} H'(x)xdx\right)^2\left(1+O\left(\frac{h_s}{b_{n,s}}\right)\right)+O\left(\frac{1}{n_sb_{n,s}h_s}\right).
\end{align*}
Next, we have
\begin{align*}
&\E(Z'^2(t))\\&=\sum_{s=1}^2 \frac{1}{(n_sb_{n,s}Nh_s^3)^2}\sum_{j=1}^{n_s}\left(\sum_{i=1}^NM_{\mf c_s}(i/N)H''\left(\frac{m_s(i/N)-t}{h_s}\right)\bar K_{b_{n,s}}(j/n_s-i/N)\right)^2\\
&=\sum_{s=1}^2 \frac{1}{(n_sb_{n,s}h_s^3)^2}\sum_{j=1}^{n_s}\left(\int\limits_{\mathbb R} M_{\mf c_s}(u)H''\left(\frac{m_s(u)-t}{h_s}\right)\bar K_{b_{n,s}}(j/n_s-u)du\left(1+O\left(\frac{1}{Nh_s}\right)\right)\right)^2\\
&=\sum_{s=1}^2 \frac{1}{n_s(b_{n,s}h_s^3)^2}\int\limits_{\mathbb R}\left(\int\limits_{\mathbb R} M_{\mf c_s}(u)H''\left(\frac{m_s(u)-t}{h_s}\right)\bar K_{b_{n,s}}(v-u)du\right)^2dv\left(1+O\left(\frac{1}{Nh_s}\right)\right)\\\notag&~~~~~\quad\quad+O\left(\frac{1}{n^2_sb^2_{n,s}h_s^4}\right)\\
&=\sum_{s=1}^2 \frac{M^2_{\mf c_s}(m_s^{-1}(t))}{n_s(b_{n,s}h^2_s)^2(m_s'(m_s^{-1}(t)))^2}\int\limits_{\mathbb R}\left(\int\limits_{\mathbb R} H''(x)\bar K_{b_{n,s}}(v-m_s^{-1}(t+xh_s))dx\right)^2dv\left(1+O\left(\frac{1}{Nh_s}\right)\right)\notag\\&~~~~~\quad\quad+O(\frac{1}{n_sb_{n,s}h_s^3})\notag
\\&=\sum_{s=1}^2 \frac{M^2_{\mf c_s}(m_s^{-1}(t))}{n_sb_{n,s}h_s^4(m_s'(m_s^{-1}(t)))^2}\int\limits_{\mathbb R} \bar K^2(y)dy\left(\int\limits_{\mathbb R} H''(x)dx\right)^2\left(1+O\left(\frac{1}{Nh_s}+\frac{h_s}{b_{n,s}}\right)\right)+O\left(\frac{1}{n_sb_{n,s}h_s^3}\right).
\end{align*}
Finally, by the symmetry of $K$ and $H$, we have
\begin{align*}
&\E(Z(t)Z'(t))\\&=\sum_{s=1}^2 \frac{1}{(n_sb_{n,s}Nh_s^2)^2h_s}\sum_{j=1}^{n_s}\left(\sum_{i=1}^NM_{\mf c_s}(i/N)H'\left(\frac{m_s(i/N)-t}{h_s}\right)\bar K_{b_{n,s}}(j/n_s-i/N)\right)
\\&\times \left(\sum_{i=1}^NM_{\mf c_s}(i/N)H''\left(\frac{m_s(i/N)-t}{h_s}\right)\bar K_{b_{n,s}}(j/n_s-i/N)\right)\\
&=\sum_{s=1}^2 \frac{1}{(n_sb_{n,s}h_s^2)^2h_s}\sum_{j=1}^{n_s}\left(\int\limits_{\mathbb R} M_{\mf c_s}(u)H'\left(\frac{m_s(u)-t}{h_s}\right)\bar K_{b_{n,s}}(j/n_s-u)du+O\left(\frac{\mf 1(|j/n_s-t|\leq \bar M b_{n,s})}{N}\right)\right)\\&
\left(\int\limits_{\mathbb R} M_{\mf c_s}(u)H''\left(\frac{m_s(u)-t}{h_s}\right)\bar K_{b_{n,s}}(j/n_s-u)du+O\left(\frac{\mf 1(|j/n_s-t|\leq \bar M b_{n,s})}{N}\right)\right)\\
&=\sum_{s=1}^2 \frac{1}{n_s(b_{n,s}h_s^2)^2h_s}\int\limits_{\mathbb R}\left(\int\limits_{\mathbb R} M_{\mf c_s}(u)H'\left(\frac{m_s(u)-t}{h_s}\right)\bar K_{b_{n,s}}(v-u)du\right)\\&
\left(\int\limits_{\mathbb R} M_{\mf c_s}(u)H''\left(\frac{m_s(u)-t}{h_s}\right)\bar K_{b_{n,s}}(v-u)du\right)dv+O\left(\frac{1}{n_s^2b^2_{n,s}h_s^3}+\frac{1}{n_sNb_{n,s}h_s^4}\right)\\
&=\sum_{s=1}^{2}\frac{M^2_{\mf c_s}(m_s^{-1}(t))}{n_s(b_{n,s}h_s)^2h_sm_s'(m_s^{-1}(t))}\int\limits_{\mathbb R}\left(\int\limits_{\mathbb R} H'(x)\bar K_{b_{n,s}}(v-m_s^{-1}(t+xh_s))dx\right)\\&\left(\int\limits_{\mathbb R} H''(x)\bar K_{b_{n,s}}(v-m_s^{-1}(t+xh_s))dx\right)dv+O\left(\frac{1}{nb_{n,s}h_s^2}\right)
\\&=O\left(\frac{h_s+(h_s/b_{n,s})^2}{n_sb_{n,s}h_s^3}\right).\label{ZZ'(t)}
\end{align*}
Step 2: We use Proposition 1 of \cite{sun1994simultaneous} to evaluate the maximum deviation of $Z(t)$.
For any two $p$-dimensional vectors $\mf u=(u_1,...u_p)^\top$ and $\mf v=(v_1,....,v_p)^\top$, write $<\mf u,\mf v>=\sum\limits_{i=1}^pu_iv_i$, and $\|\mf u\|_E=<\mf u, \mf u>^{1/2}$.
Define
\begin{align}
T_{j,s}(t)=w_s(m_s,j,t), s=1~\mbox{and}~2, 1\leq j\leq n_s,\\
\mf T(x)=\Big(T_{1,1}(x),....,T_{n_1,1}(x), T_{1,2}(x),...,T_{n_2,2}(x)\Big)^\top,
\end{align}
and $\mf V=(V_1,....V_{n_1}, V_{n_1+1},...,V_{n_1+n_2})^\top$, where $\{V_i\}_{i=1}^{n_1+n_2}$ is a i.i.d.\ sequence of standard normal random variables. Then, for any $0<a<b<1$ and $\eta > 0$, $\displaystyle\sup_{t\in[a,b]}|Z(t)|$ has the same distribution as $\displaystyle\sup_{t\in[a+\eta,b-\eta]}|<\mf T(t), \mf V>|$.
Therefore, by Proposition 1 of  \cite{sun1994simultaneous}, we have that
\begin{align}\label{final-1}
\lim_{c\rightarrow \infty}\p\left(\sup_{t\in [a,b]}|<\frac{\mf T(t)}{\|\mf T(t)\|_E}, \mf V>|\geq c\right)=\frac{\kappa_0(a+\eta,b-\eta)}{\pi}\exp\left(-\frac{c^2}{2}\right)+2(1-\phi(c))+o(\exp(-c^2/2)),
\end{align}
where $\kappa_0(a+\eta,b-\eta)=\int\limits_{a+\eta}^{b-\eta}\Big\|\frac{\partial }{\partial x}\big(\frac{\mf T(x)}{\|\mf T(x)\|_E}\big)\Big\|_Edx$.
Notice that \begin{align}
\|\mf T(x)\|_E=\|Z(x)\|,~\left|\left|\frac{\partial \mf T(x)}{\partial x}\right|\right|_E=\left|\left|\frac{\partial}{\partial x}Z(x)\right|\right|,\label{Sep1-2020}\\
\frac{\partial }{\partial x}\|\mf T(x)\|_E=\frac{<\mf T(x), \frac{\partial }{\partial x}\mf T(x)>}{\|\mf T(x)\|_E}=\frac{\E (Z(t)Z'(t))}{\|Z(t)\|},~\mbox{and}~ \\
\Big\|\frac{\partial }{\partial x}\left(\frac{\mf T(x)}{\|\mf T(x)\|_E}\right)\Big\|_E=\left|\left|\frac{\frac{\partial}{\partial x}\mf T(x)\|\mf T(x)\|_E-\frac{\partial}{\partial x}\|\mf T(x)\|_E\mf T(x)}{\|T(x)\|_E^2}\right|\right|_E.\label{Sep2-2020}
\end{align}
By using \eqref{Sep1-2020}--\eqref{Sep2-2020} and the results of 
$\E(Z^2(t))$, $\E(Z'^2(t))$ and $\E(Z(t)Z'(t))$ from Step 1, we have that
\begin{align}\label{final-0}
\Big\|\frac{\partial }{\partial x}\left(\frac{\mf T(x)}{\|\mf T(x)\|_E}\right)\Big\|_E=\frac{b_{n,1}}{h_1^2} \sqrt{\frac{K_2(t)}{K_1(t)}}(1+o(1)).
\end{align}
Furthermore, by\eqref{eqab}, we have
\begin{align}
\lim_{n_1\rightarrow \infty}\p\left(\max(|\hat a-m_1(0)|, |\hat b-m_1(1-m_2^{-1}(m_1(0)))|)\leq \eta/2\right)=1. 
\end{align} 
Therefore, \begin{align}\label{final-2}
&\p\left(\sup_{t\in [\hat m_1(0)+3\eta/2,\hat m_1(1-m_2^{-1}(m_1(0)))-3\eta/2]}|<\frac{\mf T(t)}{\|\mf T(t)\|_E},\mf V>|\geq c\right)\notag\\
&\leq \p\left(\sup_{t\in [\hat a+\eta,\hat b-\eta]}|<\frac{\mf T(t)}{\|\mf T(t)\|_E}, \mf V>|\geq c\right)\notag\\
&\leq 
\p\left(\sup_{t\in [\hat m_1(0)+\eta/2,\hat m_1(1-m_2^{-1}(m_1(0)))-\eta/2]}|<\frac{\mf T(t)}{\|\mf T(t)\|_E},\mf V>|\geq c\right).
\end{align}
As a consequence, by \eqref{final-0}, \eqref{final-1} and \eqref{final-2}, and the fact that $\eta=o(1)$,
the theorem follows  when choosing $c=\sqrt{-2\log(\pi \alpha/\kappa_0)}$. 
\section{Proof of Proposition \ref{prop-monotone}}
In this subsection we prove Proposition \ref{prop-monotone} in the main article, which enables us to test the null hypothesis $H_{0}$ in \eqref{hypo1} via investigating $(m_1^{-1}(u))' - (m_2^{-1}(u))'$.

\noindent{\bf Proof of Proposition \ref{prop-monotone}:} 
We extend the proof of Lemma 2.1 in \cite{dette2021identifying}. 
We shall show \eqref{hypo1} is equivalent to 
\begin{align}\label{eq-m-2}
(m_2^{-1}(u)-m_1^{-1}(u))'=0
\end{align}
for $m_1(0)< u < m_1(1-d)$.

If $m_1(t)=m_2(t+d)$ for $0<t< 1-d$ for some unknown $d$, and $m_1(t)$ and $m_2(t)$ are monotonically increasing for $0< t< 1-d$, one then can write
$u=m_1(t)=m_2(t+d)$ for $m_1(0)< u < m_1(1-d)$, which implies that
$t=m^{-1}(u)$ and $t+d=m^{-1}_2(u)$. Therefore, 
$$(m_2^{-1}(u)-m_1^{-1}(u))=d$$ 
for $m_1(0)< u < m_1(1-d)$. Since $d$ is a constant, we have proven that \eqref{hypo1} implies \eqref{eq-m-2}.

On the other hand, by \eqref{eq-m-2}, one can see that for any $t\in (m_1(0), m_1(1-d))$,
\begin{align}
\int\limits_{m_1(0)}^t(m_1^{-1}(u))'du=\int\limits_{m_1(0)}^t(m_2^{-1}(u))'du.
\end{align}
As a result, we have $m_1^{-1}(t)=m_2^{-1}(t)-m_2^{-1}(m_1(0))$, and $d=m_2^{-1}(m_1(0))$.
Therefore, by rearranging the equation and taking $m_2'$ on both sides of it, one can conclude that 
\begin{align}
t=m_2(m_1^{-1}(t)+m_2^{-1}(m_1(0)))=m_1(m_1^{-1}(t))
\end{align}
for $0<m_1^{-1}(t)<1-d$, which finishes the proof by setting $u=m_1^{-1}(t)$. \hfill $\Box$
\section{Technical propositions for proving Theorems \ref{asymptotic}
	and \ref{SCB}. }
In the section, we discuss three propositions that are needed for the proof of Theorems \ref{asymptotic}
and \ref{SCB}. We first have the following result related to uniform approximation of $\hat m_s(t)-m_s(t)$ through a Gaussian process. The proof of this proposition is based on a Bahadur Representation of $\hat m_s(t)-m_s(t)$. 
\begin{proposition}\label{propGaussian}
	Assume conditions (A1)-(A8) and (B3). Then on a possibly richer probability space, there exist an i.i.d. sequence of standard normal random variables   $(V_{i,1})_{i\in \mathbb Z}$ and  $(V_{i,2})_{i\in \mathbb Z}$, such that for $s=1$ and 2,
	\begin{align}
		\sup_{t\in \mathcal T_{n,s} }\left|\hat m_s(t)-m_s(t)-\frac{ M_{\mf c_s}(t)\sum_{i=1}^{n_s}V_{i,s}\bar K_{b_{n,s}}(i/n_s-t)}{nb_{n,s}}\right|=O_p(\Theta_{n,s}),
	\end{align}
	where $\Theta_{n,s}$ is defined in the main article above condition (B1).
\end{proposition}

\noindent {\it Proof.} The proposition follows immediately from equation (59) of the supplemental material of  \cite{wu2017nonparametric}. \hfill \hfill $\Box$ 

Therefore we have the following proposition regarding the magnitude of $|\hat m_s(t)-m_s(t)|$.
\begin{proposition}\label{propMs}
	Along with the conditions of Proposition \ref{propGaussian}, assume that $\pi_{n,s}=o(\sqrt{\log n_s})$,  $n_sb_{n,s}^2\log ^2 n_s\rightarrow \infty$ for $s=1$ and 2. Then, we have that for $s=1$ and 2,
	\begin{align}
	\sup_{t\in \mathcal T_{n,s}}|\hat m_s(t)-m_s(t)|=O_p\left(\frac{\sqrt{\log n_s}}{\sqrt{n_sb_{n,s}}}\right).
	\end{align}
\end{proposition}
{\it Proof.} It follows from Lemma 1 of \cite{zhou2010simultaneous} that
\begin{align}
\sup_{t\in \mathcal T_{n,s}}\left|\frac{\sum_{i=1}^{n_s}V_{i,s}\bar K_{b_{n,s}}(i/n_s-t)}{n_sb_{n,s}}\right|=O_p\left(\frac{\sqrt{\log n}}{\sqrt{n_sb_s}}\right).
\end{align} 
Then the assertion of this proposition follows from  Proposition \ref{propGaussian}. \hfill $\Box$
The following proposition provides the convergence rate of  $\hat a$ and $\hat b$ under the null and the local alternatives. 
\begin{proposition}\label{propbound}
	Under the conditions of Proposition \ref{propGaussian}, assuming (B1), (B2) and $\eta=o(1)$,
	$\eta^{-1}=O(\log (n_1+n_2))$. Then if  $(m_1^{-1})^{'}(t)-(m_2^{-1})^{'}(t)=\rho_{n_1,n_2}\kappa(t)$ for some non-zero bounded function $\kappa(t)$ and $\rho_{n_1,n_2}=o(\eta)$,  
	we have that 
	\begin{align}\label{eqab}
		i) \max(|\hat a-m_1(0)|, |\hat b-m_1(1-m_2^{-1}(m_1(0)))|)=O_p\left(\sum_{s=1}^2\frac{\sqrt{\log n_s}}{\sqrt{n_sb_{n,s}}b_{n,s}}+\rho_{n_1,n_2}\right). 
	\end{align} Notice that under null, the LHS of \eqref{eqab} will be reduced to	$\max(|\hat a-a|, |\hat b-b|)$. Moreover, 
	\begin{align}
		ii)\lim_{n_1\rightarrow \infty, n_2\rightarrow \infty}\p(m^{-1}_s(t)\in (b_n,1-b_n),~s=1,2, ~\mbox{for all}~ t\in(\hat a+c_1\eta, \hat b-c_2\eta))=1.
	\end{align}
	for any given positive constant $c_1, c_2>0$.
\end{proposition}
\begin{remark}\label{remark1} Note that i) in 
	Proposition \ref{propbound} shows that $\hat w(t)$ in \eqref{Tn1n2} is consistent under null hypothesis and local alternatives. Further, observe that  ii) in Proposition \ref{propbound} shows that by introducing $\eta$, we avoid bandwidth conditions since ii)  excludes regions where $m^{-1}_s(t)$ ($s=1$ and 2) are close to $0$ and $1$. 
\end{remark}

\noindent{\bf Proof of Proposition \ref{propbound}:} Notice that $(m_2)^{-1}(t)-(m_1')^{-1}(t)=\rho_{n_1,n_2}\kappa(t)$ implies that 
\begin{align}\label{pro-integral}
	m_2^{-1}(u)-m_1^{-1}(u)=m_2^{-1}(m_1(0))+\int\limits_{m_1(0)}^u\rho_{n_1,n_2}\kappa(t)dt.
\end{align}
Proof of assertion i): By proposition \ref{propMs} and \eqref{pro-integral}, it is sufficient to show that for $s=1$ and 2, 
\begin{align}
	\sup_{u\in[a,b]}|\hat g_s(u)-g_s(u)|=O_p\left(\frac{\sqrt{\log n_s}}{\sqrt{(n_sb_{n,s})}b_{n,s}}\right),
\end{align}
where $g_{s}(t) = \frac{1}{Nh_s}\sum\limits_{i = 1}^{N}H\Big (\frac{\hat{m}_{s}(\frac{i}{N}) - t}{h_s}\Big ).$ After carefully inspecting the proof of Theorem \ref{asymptotic},
one can find that $\sup_{t\in [a,b]}|\hat g_s(t)-g_s(t)|=O_p\left(\sup_{t\in [a,b]}Z_s(t)\right)$, where $Z_s(t)$ is defined in \eqref{5.33-3-26} in the proof of Theorem \ref{asymptotic}. Then i) of the proposition follows from 
\eqref{MaxZ} in the proof of Theorem \ref{asymptotic}. The assertion in (ii) follows from condition (B2), \eqref{pro-integral} (with $u=\hat m_1\left(1-\hat d\right)$ especially), strict monotonicity of $m_s$ $s=1$ and 2, the mean value theorem and the fact that $b_{n_1}=o(\eta),
b_{n_2}=o(\eta)$ and $\rho_{n_1,n_2}=o(\eta)$, which can be explained as follows. Note that using (i), with probability tending to 1, we have
\begin{align}
	m_1(0)<\hat a_1+c_1\eta.
\end{align}
Now, since $b_{n_1}=o(\eta)$, and $m_1^{-1}$ is differentiable, using mean value theorem with probability tending to 1, we have
\begin{align}
	m_1(b_{n_1})<\hat a+c_1\eta, 
\end{align} and hence, $\displaystyle\lim_{{n\rightarrow \infty}}\p(b_{n_1}<m_1^{-1}(\hat a+c_1\eta))=1$. Arguing in a similar way, one can establish that  $\displaystyle\lim_{{n\rightarrow \infty}}\p(1-b_{n_1}>m_1^{-1}( \hat b-c_2\eta))=1$. Then (ii) holds since the function $m^{-1}(\cdot)$ is monotone. \hfill $\Box$
{\color{black}\section{Proof of results of Section \ref{mutual-dependent}}
In this section, we assume that $n_1\geq n_2$, $1\leq n_1/n_2\leq M<\infty$ for some large constant $M$. Here we  stick to the scenario that  two series are both collected in the same period, i.e., $\tilde{\mf y}_{i,1}$ is realized at time $\frac{i}{n_1}$ for $1\leq i\leq n_1$, and   $\tilde{\mf y}_{i,2}$ is realized at time $\frac{\lf in_1/n_2\rf}{n_1}$ for $1\leq i\leq n_2$.
\begin{proposition}\label{dependentGaussian}
	Under the conditions of Proposition \ref{propGaussian} with (A7) replaced by (A7'), then on a possibly richer probability space, there exist an i.i.d. sequence of normal  $N\Big(0,\begin{pmatrix}
	1 &0\\0&1
	\end{pmatrix}\Big)$ random vectors $\mf V_i=(V_{i,1},V_{i,2})^\top$  such that
	
		\begin{align}
	\sup_{t\in \mathcal T_{n,1}\cap \mathcal T_{n,2} }\left|\hat m_1(t)-m_1(t)-\frac{ \sum_{i=1}^{n_1}(M^c_{1,1}(t)V_{i,1}+M^c_{1,2}(t)V_{i,2})\bar K_{b_{n,1}}(i/n_1-t)}{nb_{n,1}}\right|=O_p(\Theta_{n,1}),
	\\	\sup_{t\in \mathcal T_{n,1}\cap \mathcal T_{n,2} }\left|\hat m_2(t)-m_2(t)-\frac{ \sum_{i=1}^{n_2}(M^c_{1,2}(t)V_{i',1}+M^c_{2,2}(t)V_{i',2})\bar K_{b_{n,2}}(i'/n_1-t)}{nb_{n,2}}\right|=O_p(\Theta_{n,2}),
	\end{align}
	where $i'=\lf in_1/n_2\rf$.
\end{proposition}
{\it Proof.}
By Lemma 7 of \cite{wu2017nonparametric}, we have 
\begin{align}\label{E1}
\max_{s=1,2}\sup_{t\in \mathcal T_{n,s}}|\tilde {\bs \theta}_{\tau,s,b_{n,s}}(t)-\bs \theta_{\tau,s}(t)-\bs \Sigma_s^{-1}(t)\frac{\sum_{i=1}^{n_s}\psi(e_{i,s})\mf x_{i,s}\bar K_{b_{n,s}}(i/n_s-t)}{nb_{n,s}}|=O_p(\max_{s=1,2}(n_sb_{n,s})^{-1/2}\pi_{n,s}),
\end{align}
where $\pi_{n,s}$ is defined in Proposition \ref{propGaussian}.
By Proposition 3 in \cite{wu2017nonparametric}, for any $s\times p$ ($s\leq p$) matrix $\mf C$,  we have
 on a possibly richer probability space, there exists an i.i.d. sequence of normal  $N\Big(0,\begin{pmatrix}
1 &0\\0&1
\end{pmatrix}\Big)$ random vectors $\mf V_i=(V_{i,1},V_{i,2})^\top$ such that 
\begin{align}\label{E61}
\max_{1\leq j\leq n_1}|\sum_{i=1}^j\mf U(\frac{i}{n_1},\FF_i,\GG_i)-\sum_{i=1}^j\bs \nu_{\mf C}(i/n_1)\mf V_i|=O_p(n_1^{1/4}\log n_1)
\end{align}
with $\bs \nu_{\mf C}(t)=(\mf C\mf V^2(t)\mf C^\top)^{1/2}.$
Now, define the index set $A=\{\lf\frac{i}{n_2}n_1\rf, 1\leq i\leq n_2\}$ and observe that for any bandwidth $b_n$, we have 
\begin{align}
\sum_{i=1}^{n_2}\psi(e_{i,2})\mf x_{i,2}K_{b_{n}}(\lf in_1/n_2\rf/n_1-t)=
\sum_{i=1}^{n_1}\psi(e_{i,1})\mf x_{i,1}K_{b_{n}}(i/n_1-t)\mf 1(i\in A).
\end{align}
Let us now consider \begin{align}
\Delta_1=\sup_{t\in \mathcal T_{n,1}}|\tilde {\bs \theta}_{\tau,1,b_{n,1}}(t)-\bs \theta_{\tau,1}(t)-\bs \Sigma_1^{-1}(t)\frac{\sum_{i=1}^{n_1}\psi(e_{i,1})\mf x_{i,1}\bar K_{b_{n,1}}(i/n_1-t)}{nb_{n,1}}|,\\
\Delta_2=\sup_{t\in \mathcal T_{n,2}}|\tilde {\bs \theta}_{\tau,2,b_{n,2}}(t)-\bs \theta_{\tau,2}(t)-\bs \Sigma_2^{-1}(t)\frac{\sum_{i=1}^{n_1}\psi(e_{i,1})\mf x_{i,1}\bar K_{b_{n,2}}(i/n_1-t)\mf 1(i\in A)}{nb_{n,2}}|
),
\end{align}
then \eqref{E1} can be written as  \begin{align}
\max(\Delta_1,\Delta_2)=O_p(\max_{s=1,2}(n_sb_{n,s})^{-1/2}\pi_{n,s}).
\end{align}
Use the expression, an application of summation by part formula together with \eqref{E61}, we have 
\begin{align}\label{E68}
\sup_{t\in\mathcal T_{n_1}\cap \mathcal T_{n,2}}\Big|\begin{pmatrix}
\hat m_1(t)-m_1(t)\\
\hat m_2(t)-m_2(t)
\end{pmatrix}-\sum_{i=1}^{n_1}
\mf K(i,t)\mf M_{\mf c}(t) \mf V_i|=O_p(\max_{s=1,2}\Theta_{n,s}),\end{align}
where \begin{align}
\mf K(i,t)=\begin{pmatrix}
\bar K_{b_{n,1}}(i/n_1-t)/(nb_{n,1})&\\
& \bar K_{b_{n,2}}(i/n_1-t)/(nb_{n,2})\mf 1(i\in A)
\end{pmatrix}.
\end{align}
Straightforward calculation shows that \eqref{E68} implies both the two assertions of the proposition. \hfill $\Box$ 

\begin{proposition}\label{propE2}
Let $\rho_n=o(1)$, and assume for $t\in [a,b]$, $(m_1^{-1}-m_2^{-1})'(t)=\rho_n\kappa(t)$ for some bounded function $\kappa(t)$, where $a=m_1(0)\wedge m_2(0)$ and $b=m_1(1)\vee m_2(1)$. Then for any $u\in [m_1(0)\wedge m_2(0),m_1(1)\vee m_1(1)]$,  we have 
\begin{align}
m_1^{-1}(u)-m_2^{-1}(u)=-m_2^{-1}(m_1(0))+O(\rho_n)= m_1^{-1}(m_2(0))+O(\rho_n).
\end{align}
In particular,
\begin{align}
t-m_2^{-1}(m_1(t))=-m_2^{-1}(m_1(0))+O(\rho_n)= m_1^{-1}(m_2(0))+O(\rho_n),\\m_1^{-1}(m_2(t))-t=-m_2^{-1}(m_1(0))+O(\rho_n)= m_1^{-1}(m_2(0))+O(\rho_n).
\end{align}
\end{proposition}
{\it Proof.} Observe that
\begin{align}
m_1^{-1}(t)-m_2^{-1}(t)-(0-m_2^{-1}(m_1(0)))=\int_{m_1(0)}^t (m_1^{-1}(s)-m_2^{-1}(s))'ds,\\
m_1^{-1}(t)-m_2^{-1}(t)-(m_1^{-1}(m_2(0))-0)=\int_{m_2(0)}^t (m_1^{-1}(s)-m_2^{-1}(s))'ds
\end{align}
Notice that Under $H_0$ $m_1(0)=m_2(d)$, so $m_2^{-1}(m_1(0))=d$, and by the fact that $(m_1^{-1}-m_2^{-1})'(t)=\rho_n\kappa(t)$, the proposition follows. $\Box$

To prove the assertion in Theorem \ref{core-dependent}, we shall study the asymptotic behavior of $T_{n_1,n_2}$ and $\hat m^{-1}_{1}-\hat m^{-2}(t)$ under the case that the two time series are correlated. The results are presented and proved in Theorem \ref{themE1} and \ref{themE2} below, respectively. 

Let us first consider the following variables: 
\begin{align}
\check g_{uv}(t)=M_{uv}^c(m_u^{-1}(t))((m_u^{-1})'(t))^2\int_RH'(y)ydy,\\
\end{align}
Define the following quantities with  $r,a,b,c$ functions of $u$, $a^*,b^* \in \mathbb R$,
\begin{align}
\tilde E_1=\int \bar K'^2(u)du\int \check{g}_{11}^2(m_1(x))w(m_1(x))m_1'(x)dx,\\
\tilde C_1=\int (\bar K'\star \bar K'(z))^2dz\int (\check{g}_{11}^2(m_1(x))w(m_1(x))m_1'(x))^2dx\\
\tilde E_2=\int \bar K'^2(u)du\int \check{g}_{21}^2(m_2(x))w(m_2(x))m_2'(x)dx,\\
\tilde C_2=\int (\bar K'\star \bar K'(z))^2dz\int (\check{g}_{21}^2(m_2(x))w(m_2(x))m_2'(x))^2dx\\
\tilde E_3(a^*,b^*)=\int w(t)\check g_{11}(t)\check g_{21}(t)\int \bar K'(z)\bar K'\left(a^*+zb^*\right)dzdt, \\
\tilde C_{3,1}(r)=\int\int [\int \bar K'(x)\bar K'(y+r(u)x)dx]^2\notag[\check g_{11}(m_1(u))\check g_{21}(m_1(u))w(m_1(u))m_1'(u)]^2dudy,\\
\tilde C_{3,2}(a,b,c)=\frac{1}{n_1^2b_{n,1}b_{n,2}^4}\int \int(\check g_{11}(m_1(u))\check g_{22}(m_1(u))w(m_1(u))m_1'(u))^2\times\notag\\\Big[\int \bar K'(x)\bar K'(a(u)+b(u)y+c(u)x)dx\int \bar K'(y+x)\bar K'(a(u)+c(u)x)dx\Big]dudy\\
D_{1,2}(a,b,c)=\int\int \check g_{21}^2(m_1(u))\check g^2_{11}(m_1(u))[m_1'(u)w(m_1(u))]^2\Big[\int \bar K'(x)\bar K'(y+x)dx\Big]\notag\\
\Big[\int \bar K'(a(u)+b(u)x)\bar K'(a(u)+b(u)x+c(u)y))dx\Big]dudy\\
D_{1,3}(a,b,c)=\int\int \check g_{11}^3(m_1(u))\check g_{21}(m_1(u))[m_1'(u)w(m_1(u))]^2\Big[\int \bar K'(x)\bar K'(y+x)dx\Big]\notag\\
~~~~~~~~~~~\Big[\int \bar K'(x)\bar K'(a(u)+b(u)x+c(u)y)dx\Big]dudy\\
D_{2,3}(a,b,c)=\int\int \check g_{21}^3(m_2(u))\check g_{11}(m_2(u))[m_2'(u)w(m_2(u))]^2\Big[\int \bar K'(x)\bar K'(y+x)dx\Big]\notag\\
~~~~~~~~~~~\Big[\int \bar K'(x)\bar K'(a(u)+b(u)x+c(u)y))dx\Big]dudy
\end{align}
Further define for $a,b,c$ functions of $u$, $z\in \mathbb R$,
\begin{align*}
&\tilde I (z,a,b,c)=\int\int dudy[w(m_1(u))m_1'(u)]^2I^2(z,u,a,b,c, y)dudy, \text{where}\\
&I(z,u,a,b,c,y)=\int \Big(z\check g_{11}(m_1(u))\bar K'(x)-\check g_{21}'(m_1(u))\bar K'(a(u)+b(u)x)\Big)\notag\\&\times\Big(\check g_{21}(m_1(u))\bar K'(a(u+b(u)x+c(u)y))-z\check g_{12}(m_1(u))\bar K'(y+x)\Big)dx.\\&
\tilde {II}(z, a,b)=\int\int z^2\check g_{12}^2(m_1(u))[w(m_1(u))m_1'(u)]^2II^2(z,u,a,b,c,y)dudy, \text{where}\\
&II(z,u,a,b,y)=\int\Big(z\check g_{11}(m_1(u))\bar K'(x)-\check g_{21}(m_1(u))\bar K'\big(a+bx\big)\Big)\bar K'(y+x)dx.\notag\\&
\tilde {III}(z,a,b)=\int\int c^2_{2,1,n}\check g_{11}^2(m_1(u))[w(m_1(u))m_1'(u)]^2III^2(z,u,a,b,y)dudy, \text{where}\\
&III(z,u,a,b,y)=\int\Big(-z\check g_{12}(m_1(u))\bar K'(x)+\check g_{22}(m_1(u))\bar K'\big(a+bx\big)\Big)\bar K'(y+x)dx.\\
&\tilde{IV}=\int\int [\check g_{11}(m_1(u))\check g_{12}(m_1(u))w(m_1(u))m_1'(u)]^2[\bar K'\star \bar K'(y)]^2dydu.
\end{align*}


\begin{theorem}\label{themE1}
Assume the conditions stated in (A1)-(A6), (A7'), (A8) and (B1), (B2), (B3), and that as $n\rightarrow \infty$, $n_1/n_2\rightarrow n_{1,2}\in (0,\infty)$, $b_{n_1}/b_{n_2} \rightarrow b_{1,2}\in (0,\infty)$, $n_{1,2}=1/n_{2,1}$, $b_{1,2}=1/b_{2,1}$, $\eta^{-1}=O(\log(n_1+n_2))$,  $\eta=o(1)$.
Further, let $(m_1^{-1})'(t)-(m_2^{-1})'(t)=\rho_n\kappa(t)$ for some bounded function $\kappa(t)$, and $\rho_n:=\rho_{n_1,n_2}=[(n_1b^{5/2}_{n_1})^{-1/2}(n_2b^{5/2}_{n_2})^{-1/2}]^{-1/2}$. Assuming $\frac{\rho_{n_1,n_2}}{b_{n,1}}=o(1).$	
Let us now define \begin{align}
\tilde B=n_{2,1}b_{2,1}^{3/2}\tilde E_1+n_{1,2}b_{1,2}^{3/2}\tilde E_2-2n_{2,1}^{1/2}b_{1,2}^{1/2}\tilde E_3(0,b_{1,2}), 
\end{align}
and suppose that the supscript $1-2$ mean the exchange of indices $1$ and $2$. For example, 
let \begin{align}
\tilde B^{1-2}=n_{1,2}b_{1,2}^{3/2}\tilde E_1^{1-2}+n_{2,1}b_{2,1}^{3/2}\tilde E_2^{1-2}-2n_{1,2}^{1/2}b_{2,1}^{1/2}\tilde E^{1-2}_3(0,b_{2,1}),
\end{align}
where, for instance,
\begin{align}
\tilde E_2^{1-2}=\int \bar K'^2(u)du\int \check{g}_{12}^2(m_1(x))w(m_1(x))m_1'(x)dx.
\end{align}
Further, define
\begin{align}
\tilde V=2\tilde C_1n_{2,1}b_{2,1}^{5/2}+2\tilde C_2n_{1,2}b_{1,2}^{5/2}+4n_{2,1}b_{1,2}^{3/2}\tilde C_{32}(0,b_{1,2},b_{1,2})+4n_{2,1}b_{1,2}^{3/2}D_{12}(0,b_{1,2},b_{1,2})\notag\\
-8n_{2,1}b_{2,1}^{1/2}D_{13}(0,b_{1,2},b_{1,2})-8b_{2,1}^{1/2}D_{2,3}(0,b_{2,1},b_{2,1}),\\
\tilde W=n_{1,2}b_{1,2}^{11/2}\tilde I(n_{2,1}b_{2,1}^2,0,b_{1,2},b_{1,2})
-(n_{1,2}^2-n_{1,2})b_{1,2}^{11/2}\tilde {II}(n_{2,1}b_{2,1}^2,0,b_{1,2})+(n_{1,2}^2-\notag\\n_{1,2})b_{1,2}^{11/2}\tilde{III}(n_{2,1}b_{2,1}^2,0,b_{1,2})-n_{2,1}(1-n_{2,1})^2b_{2,1}^{5,2}\tilde{IV}.\end{align}
Now, if $m_2^{-1}(m_1(0))= 0$, we have 
 \begin{align}
 (n_1n_2)^{1/2}(b_{n_1}^5b_{n_2}^5)^{1/4}T_{n_1n_2}-(b_{n,1}b_{n,2})^{-1/4}(\tilde B+\tilde B^{1-2})\Rightarrow N(0,\tilde V+\tilde V^{1-2}+4\tilde W).
 \end{align}
Now, if $|m_2^{-1}(m_1(0))|> 0$, we have 
  \begin{align}
 (n_1n_2)^{1/2}(b_{n_1}^5b_{n_2}^5)^{1/4}T_{n_1n_2}-(b_{n,1}b_{n,2})^{-1/4}(\underline{\tilde B}+\underline{\tilde B}^{1-2})\Rightarrow N(0,\underline{\tilde V}+\underline{\tilde V}^{1-2}+4\underline{\tilde W}),
 \end{align}
 where 
 \begin{align}
 \underline{\tilde B}=n_{2,1}b_{2,1}^{3/2}\tilde E_1+n_{1,2}b_{1,2}^{3/2}\tilde E_2,\notag\\
 \underline{\tilde V}=2\tilde C_1n_{2,1}b_{2,1}^{5/2}+2\tilde C_2n_{1,2}b_{1,2}^{5/2}
 -8n_{2,1}b_{2,1}^{1/2}D_{13}(0,b_{1,2},b_{1,2})-8b_{2,1}^{1/2}D_{2,3}(0,b_{2,1},b_{2,1}),\\
 \underline{\tilde W}=n_{1,2}b_{1,2}^{11/2}\tilde I(n_{2,1}b_{2,1}^2,\infty,b_{1,2},b_{1,2})
 -(n_{1,2}^2-n_{1,2})b_{1,2}^{11/2}\tilde {II}(n_{2,1}b_{2,1}^2,\infty,b_{1,2})+(n_{1,2}^2-\notag\\n_{1,2})b_{1,2}^{11/2}\tilde{III}(n_{2,1}b_{2,1}^2,\infty,b_{1,2})-n_{2,1}(1-n_{2,1})^2b_{2,1}^{5,2}\tilde{IV}.\end{align}

\end{theorem}

{\it Proof.} Notice that Propositions \ref{prop-monotone} and \ref{propbound} still hold. Therefore, following the  proof of Theorem \ref{asymptotic}, we shall see that 
\begin{align}
(n_1n_2b_{n,1}^{5/2}b_{n,2}^{5/2})^{1/2}(\tilde T_{n_1,n_2}-T_{n_1,n_2})=o_p(1).
\end{align}
Here $\tilde T_{n_1,n_2}$ is defined in \eqref{tildeTn}, and 
 applying Proposition \ref{dependentGaussian} instead of Proposition \ref{propGaussian}, we shall see that the decomposition of $\tilde T_{n_1,n_2}$ in  \eqref{B10} can be written as 
\begin{align}
\tilde T_{n_1,n_2}=\int\limits_{\mathbb R}[\rho_n\kappa(t)+Z(t)+R(t)]^2w(t)dt,
\end{align}
where $Z(t)$ is defined in \eqref{Z(t)} replaced by 
\begin{align}\label{E98}
Z(t)=Z_{11}(t)+Z_{12}(t)-Z_{22}(t)-Z_{21}(t).
\end{align}
Here
\begin{align*}
Z_{11}(t)=\sum_{j=1}^{n_1}W_{11}(j,t)V_{j,1},~W_{11}(j,t)=\sum_{i=1}^N\frac{M_{11}^c(i/N)}{n_1b_{n,1}Nh_1^2}H'(\frac{m_1(i/N)-t}{h_1})\bar K_{b_{n,1}}(j/n_1-i/N),\\
Z_{12}(t)=\sum_{j=1}^{n_1}W_{12}(j,t)V_{j,2},~W_{12}(j,t)=\sum_{i=1}^N\frac{M_{12}^c(i/N)}{n_1b_{n,1}Nh_1^2}H'(\frac{m_1(i/N)-t}{h_1})\bar K_{b_{n,1}}(j/n_1-i/N),\\
Z_{21}(t)=\sum_{j=1}^{n_2}W_{21}(j,t)V_{\lf \frac{n_1j}{n_2}\rf,1},~W_{21}(j,t)=\sum_{i=1}^N\frac{M_{12}^c(i/N)}{n_2b_{n,2}Nh_2^2}H'(\frac{m_2(i/N)-t}{h_2})\bar K_{b_{n,2}}(j/n_2-i/N),\\
Z_{22}(t)=\sum_{j=1}^{n_2}W_{22}(j,t)V_{\lf \frac{n_1j}{n_2}\rf,2},~W_{22}(j,t)=\sum_{i=1}^N\frac{M_{22}^c(i/N)}{n_2b_{n,2}Nh_2^2}H'(\frac{m_2(i/N)-t}{h_2})\bar K_{b_{n,2}}(j/n_2-i/N).
\end{align*}
For the remaining term $R(t)$, it satisfies 
\eqref{Rt}.
Again following the  proof of Theorem \ref{asymptotic}, we shall see 
that
\begin{align}
(n_1n_2b_{n,1}^{5/2}b_{n,2}^{5/2})^{1/2}\left(\int\limits_{\mathbb R} R(t)(R(t)+\rho_n \kappa(t)+Z(t))w(t)dt+\int\limits_{\mathbb R} \rho_n\kappa(t)Z(t)w(t)dt\right)=o_p(1).\end{align}
In the following, we shall show that 
\begin{align}\label{E100}
(n_1b_{n,1}^{5/2}n_2b_{n,2}^{5/2})^{1/2}\int\limits_{\mathbb R} Z^2(t)w(t)dt -(b_{n,1}b_{n,2})^{-1/4}(\tilde B+\tilde B^{1-2})\Rightarrow N(0,\tilde V+\tilde V^{1-2}+4\tilde W)
\end{align}
for $m_2^{-1}(m_1(0))=0$, and 
\begin{align}\label{E101}
(n_1b_{n,1}^{5/2}n_2b_{n,2}^{5/2})^{1/2}\int\limits_{\mathbb R} Z^2(t)w(t)dt -(b_{n,1}b_{n,2})^{-1/4}(\underline{\tilde B}+\underline{\tilde B}^{1-2})\Rightarrow N(0,\underline{\tilde V}+\underline{\tilde V}^{1-2}+4\underline{\tilde W}).
\end{align}
for $|m_2^{-1}(m_1(0))|>0$.

We now calculate the asymptotic variance and bias.
 To ease the notation, write 
\begin{align}
W(j_1,j_2)=\int W_{11}(j_1,t)W_{11}(j_2,t)w(t)dt,\\
W'(j_1,j_2)=\int W_{11}(j_1,t)W_{21}(j_2,t)w(t)dt,\\
W_{21}'(j_1,j_2)=\int W_{21}(j_1,t)W_{21}(j_2,t)w(t)dt.
\end{align}
Similarly to the proof of Theorem \ref{asymptotic},
 $W_{uv}(j,t)$ can be written as $\tilde W_{uv}(j,t)(1+O(b_{n,u}+\frac{h_u}{b_{n,u}})))$ for $u,v=1,2$, where
\begin{align}\label{leadingW}
\tilde W_{uv}(j,t)&=\frac{-1}{n_ub^2_{n,u}}M^c_{uv}(m_u^{-1}(t))\bar K'\left(\frac{j/n_u-m_u^{-1}(t)}{b_{n,u}}\right)((m_u^{-1})'(t))^2\int\limits_{\mathbb R} H'(y)ydy\notag
\\&=\frac{-1}{n_ub^2_{n,u}}\check g_{uv}(t)\bar K'\left(\frac{j/n_u-m_u^{-1}(t)}{b_{n,u}}\right),
\end{align}	
since for notational simplicity we have defined $M^c_{12}(t)=M^c_{21}(t)$.

Define $\tilde Z_1(t)=Z_{11}(t)-Z_{21}(t)$ and $\tilde Z_2(t)=Z_{22}(t)-Z_{12}(t)$, so
\begin{align}\label{tildeZtdecompose}
Z(t)=\tilde Z_1(t)-\tilde Z_2(t)
\end{align}
Then \begin{align}
\int \tilde Z^2_1(t)w(t)dt=&\int (\sum_{j=1}^{n_1}W_{11}(j,t)V_{j,1})^2w(t)dt+\int (\sum_{j=1}^{n_2}W_{21}(j,t)V_{\lf n_1j/n_2\rf,1})^2w(t)dt\notag\\
-&2\int \sum_{j_1=1}^{n_1}W_{11}(j_1,t)V_{j,1}\sum_{j_2=1}^{n_2}W_{21}(j_2,t)V_{\lf n_1j_2/n_2\rf,1}w(t)dt:=
A+B-2C,
\end{align}
where $A$, $B$ and $C$ are defined in an obvious manner.
Following the proof of Theorem \ref{asymptotic}, we shall see that
\begin{align}
\E(A)=&\E(\int (\sum_{j=1}^{n_1}W_{11}(j,t)V_{j,1})^2w(t)dt)
=\frac{1}{n_1b_{n_1}^3}\tilde E_1(1+o(1)),\\
Var(A)=&Var(\int (\sum_{j=1}^{n_1}W_{11}(j,t)V_{j,1})^2w(t)dt)
=\frac{2}{n_1^2b_{n_1}^5}\tilde C_1(1+o(1)),\\
\E(C)=&\E(\int \sum_{j_1=1}^{n_1}W_{11}(j,t)V_{j,1}\sum_{j=1}^{n_2}W_{21}(j,t)V_{\lf n_1j/n_2\rf,1}w(t)dt)\notag\\&=
\frac{1}{n_1b_{n_1}b_{n_2}^2}\tilde E_3(\frac{m_1^{-1}(t)-m_2^{-1}(t)}{b_{n,2}},\frac{b_{n,1}}{b_{n,2}})(1+o(1)).
\end{align}
To calculate $Var(C)=Var(\int \sum_{j_1=1}^{n_1}W_{11}(j,t)V_{j,1}\sum_{j=1}^{n_2}W_{21}(j,t)V_{\lf n_1j/n_2\rf,1}w(t)dt)$, notice that
\begin{align}
Var(C)=&Var(\int \sum_{j_1=1}^{n_1}W_{11}(j,t)V_{j,1}\sum_{j=1}^{n_2}W_{21}(j,t)V_{\lf n_1j/n_2\rf,1}w(t)dt)\notag\\&=Cov(\sum_{j_1=1}^{n_1}\sum_{j_2=1}^{n_2}W_{21}'(j_1,j_2)V_{j_1,1}V_{\lf \frac{n_1j_2}{n_2}\rf},\sum_{j_3=1}^{n_1}\sum_{j_4=1}^{n_2}W_{21}'(j_3,j_4)V_{j_3,1}V_{\lf \frac{n_1j_4}{n_2}\rf})
\end{align}
When $j_1=j_3, j_2=j_4$ and $j_1\neq \lf n_1j/n_2\rf$, we have 
\begin{align}
&\E\sum_{j_1=1}^{n_1}\sum_{j_2=1}^{n_2}\sum_{j_3=1}^{n_1}\sum_{j_4=1}^{n_2}W_{21}'(j_1,j_2)V_{j_1,1}V_{\lf \frac{n_1j_2}{n_2}\rf}W_{21}'(j_3,j_4)V_{j_3,1}V_{\lf \frac{n_1j_4}{n_2}\rf}\mf 1(j_1=j_3, j_2=j_4, j_1\neq \lf \frac{n_1j_2}{n_2}\rf )\notag\\
&=\frac{1}{n_1b^4_{n,1}n_2b_{n,2}^4}\int\int[\int \check g_{11}(t)\check g_{22}(t)\bar K'(\frac{u-m_1^{-1}(t)}{b_{n,1}})\bar K'(\frac{v-m_2^{-1}(t)}{b_{n,2}})w(t)dt]^2dudv(1+o(1))\notag
\\&=
\frac{1}{n_1n_2b_{n,1}^2b_{n,2}^3}\int\int [\int \bar K'(x)\bar K'(y+m_{21}'(u)\frac{b_{n,1}x}{b_{n,2}})dx]^2\notag\\&~~~~~~~~~~~~~~~~\times[\check g_{11}(m_1(u))\check g_{21}(m_1(u))w(m_1(u))m_1'(u)]^2dudy(1+o(1)).\label{E82}
\end{align}
Similarly, since
\begin{align}
W_{21}'(\lf \frac{n_1j_4}{n_2} \rf, j_2 )=\frac{1}{n_1n_2b_{n,1}^2b_{n,2}^2}\int \check g_{11}(t)\check g_{21}(t)\bar K'(\frac{j_4/n_2-m_1^{-1}(t)}{b_{n,1}})\notag\\\bar K'(\frac{j_2/n_2-m_2^{-1}(t)}{b_{n,2}})w(t)dt(1+O(\sum_{s=1}^2(b_{n,s}+\frac{h_s}{b_{n,s}})))
\end{align}
and similarly expression holds for $W_{21}'(\lf \frac{n_1j_2}{n_2} \rf, j_4 )$, 
we have
\begin{align}
&\E\sum_{j_1=1}^{n_1}\sum_{j_2=1}^{n_2}\sum_{j_3=1}^{n_1}\sum_{j_4=1}^{n_2}W_{21}'(j_1,j_2)V_{j_1,1}V_{\lf \frac{n_1j_2}{n_2}\rf}W_{21}'(j_3,j_4)V_{j_3,1}V_{\lf \frac{n_1j_4}{n_2}\rf}\mf 1(j_1=\lf\frac{n_1j_4}{n_2}\rf, j_3=\lf\frac{n_1j_2}{n_2}\rf, j_1\neq \lf \frac{n_1j_2}{n_2}\rf )\notag\\
&=\frac{1}{n^2_1b^4_{n,1}b_{n,2}^4}\int\int\Big[\int \check g_{11}(t)\check g_{22}(t)\bar K'(\frac{u-m_1^{-1}(t)}{b_{n,1}})\bar K'(\frac{v-m_2^{-1}(t)}{b_{n,2}})w(t)dt
\notag\\&\quad \quad \quad\times\int \check g_{11}(t)\check g_{22}(t)\bar K'(\frac{v-m_1^{-1}(t)}{b_{n,1}})\bar K'(\frac{u-m_2^{-1}(t)}{b_{n,2}})w(t)dt\Big]dudv(1+o(1))\notag
\\&=\frac{1}{n_1^2b_{n,1}b_{n,2}^4}\int \int(\check g_{11}(m_1(u))\check g_{22}(m_1(u))w(m_1(u))m_1'(u))^2\notag\\&~~~~~~~~~\times\Big[\int \bar K'(x)\bar K'(\frac{u-m_{21}(u)}{b_{n,2}}+\frac{b_{n_1}}{b_{n_2}}y+m_{21}'(u)\frac{b_{n,1}}{b_{n,2}}x)dx\notag\\
&~~~~~~~~~\times\int \bar K'(y+x)\bar K'(\frac{u-m_{21}(u)}{b_{n,2}}+m_{21}'(u)\frac{b_{n,1}}{b_{n,2}}x)dx\Big] dudy.\label{E84}
\end{align}
Also, it can be verified that
\begin{align}
\E\sum_{j_1=1}^{n_1}\sum_{j_2=1}^{n_2}&\sum_{j_3=1}^{n_1}\sum_{j_4=1}^{n_2}W_{21}'(j_1,j_2)V_{j_1,1}V_{\lf \frac{n_1j_2}{n_2}\rf}W_{21}'(j_3,j_4)V_{j_3,1}V_{\lf \frac{n_1j_4}{n_2}\rf}\mf 1(j_1=\lf\frac{n_1j_4}{n_2}\rf, j_3=\lf\frac{n_1j_2}{n_2}\rf= j_1 )\notag
\\&=o(\frac{1}{n_1^2b_{n,1}^2b_{n,2}^3}),\label{E85}
\end{align}
Combining \eqref{E82}, \eqref{E84} and \eqref{E85}, we have 
\begin{align}
&Var(C)=\notag\\&(\frac{1}{n_1n_2b_{n,1}^2b_{n,2}^3}\tilde C_{3,1}(m_{21}'(u)\frac{b_{n,1}}{b_{n,2}})+\frac{1}{n_1^2b_{n,1}b_{n,2}^4}\tilde C_{3,2}(\frac{u-m_{21}(u)}{b_{n,2}},\frac{b_{n_1}}{b_{n_2}}, m_{21}'(u)\frac{b_{n,1}}{b_{n,2}})
)(1+o(1))
\end{align}

Using similar argument, we obtain
\begin{align}
&Cov(A,C)=Cov(\sum_{j_1=1}^{n_1}\sum_{j_2=1}^{n_2}\int W_{11}(j_1,t)W_{11}(j_2,t)w(t)dtV_{j_1,1}V_{j_2,1}, \notag\\&~~~~~~~~~~~~~~~~\sum_{j_3=1}^{n_1}\sum_{j_4=1}^{n_2}\int W_{11}(j_3,t)W_{21}(j_4,t)w(t)dtV_{j_3}V_{\lf n_1j_4/n_2\rf, 1})
\notag\\&=\frac{2}{n_1^2b_{n,1}^3b_{n,2}^2}D_{1,3}(\frac{u-m_{21}(u)}{b_{n,2}}, m_{21}'(u)\frac{b_{n,1}}{b_{n,2}},\frac{b_{n,1}}{b_{n,2}})(1+o(1))
\end{align}
and
\begin{align}
&Cov(B,C)=\frac{2}{n_1n_2b_{n,2}^3b_{n,1}^2}D_{2,3}(\frac{u-m_{12}(u)}{b_{n,1}}, m_{12}'(u)\frac{b_{n,2}}{b_{n,1}},\frac{b_{n,2}}{b_{n,1}})(1+o(1))
\\
&Cov(A,B)=\frac{2}{n_1^2b_{n,1}b_{n,2}^4}
D_{1,2}(\frac{u-m_{21}(u)}{b_{n,2}}, m_{21}'(u)\frac{b_{n,1}}{b_{n,2}},\frac{b_{n,1}}{b_{n,2}})(1+o(1))
\\&\E(B)=\frac{1}{n_2b_{n,2}^3}\tilde E_2
,Var(B)=\frac{2}{n_2^2b_{n,2}^5}\tilde C_2
\end{align}

Therefore, 
\begin{align}\label{E120}
\E \int& \tilde Z^2_1(t)w(t)dt=\Big\{\frac{\tilde E_{1}}{n_1b_{n,1}^3}+\frac{\tilde E_{2}}{n_2b_{n,2}^3}-\frac{2}{n_1b_{n_1}b_{n_2}^2}\tilde E_3(\frac{m_1^{-1}(t)-m_2^{-1}(t)}{b_{n,2}},\frac{b_{n,1}}{b_{n,2}})\}(1+o(1)),\\
&Var \int\tilde Z^2_1(t)w(t)dt=\notag\\& \Big\{\frac{2\tilde C_1}{n_1^2b_{n,1}^5}+\frac{2\tilde C_2}{n^2_2b_{n,2}^5}+\notag\\
&(\frac{4}{n_1n_2b_{n,1}^2b_{n,2}^3}\tilde C_{3,1}(m_{21}'(u)\frac{b_{n,1}}{b_{n,2}})+\frac{4}{n_1^2b_{n,1}b_{n,2}^4}\tilde C_{3,2}(\frac{u-m_{21}(u)}{b_{n,2}},\frac{b_{n_1}}{b_{n_2}}, m_{21}'(u)\frac{b_{n,1}}{b_{n,2}})
)\notag\\
&+\frac{4}{n_1^2b_{n,1}b_{n,2}^4}
D_{1,2}(\frac{u-m_{21}(u)}{b_{n,2}}, m_{21}'(u)\frac{b_{n,1}}{b_{n,2}},\frac{b_{n,1}}{b_{n,2}})-\frac{8}{n_1^2b_{n,1}^3b_{n,2}^2}D_{1,3}(\frac{u-m_{21}(u)}{b_{n,2}}, m_{21}'(u)\frac{b_{n,1}}{b_{n,2}},\frac{b_{n,1}}{b_{n,2}})\notag\\&-\frac{8}{n_1n_2b_{n,2}^3b_{n,1}^2}D_{2,3}(\frac{u-m_{12}(u)}{b_{n,1}}, m_{12}'(u)\frac{b_{n,2}}{b_{n,1}},\frac{b_{n,2}}{b_{n,1}})\Big\}(1+o(1))\label{E121}.
\end{align}
It is easy to see that $\E\int \tilde Z_2^2(t)w(t)dt$ and  $Var\int \tilde Z_2^2(t)w(t)dt$ has the same form as that of $\E\int \tilde Z_1^2(t)w(t)dt$ and  $Var\int \tilde Z_1^2(t)w(t)dt$ with indices $1$ and $2$ exchanged.
On the other hand, notice that 
$$\int Z^2(t)w(t)dt=\int (\tilde Z^2_1(t)+\tilde Z^2_2(t)-2\tilde Z_1(t)\tilde Z_2(t))w(t)dt.$$
It is obvious that $\E\int\tilde Z_1(t)\tilde Z_2(t)w(t)dt=0$, $Cov(\int\tilde Z_1^2(t)w(t)dt,\int\tilde Z_2^2(t)w(t)dt)=0$, $$Cov(\int\tilde Z_1^2(t)w(t)dt,\int\tilde Z_1(t)\tilde Z_2(t)w(t)dt)=0,$$ and  $$Cov(\int \tilde Z_2^2(t)w(t)dt,\int\tilde Z_1(t)\tilde Z_2(t)w(t)dt)=0.$$ It remains to calculate $Var(\int \tilde Z_1(t)\tilde Z_2(t)w(t)dt)$. Without loss generality consider $n_1/n_2=k\in \mathbb Z$ where $k\geq 1$.
Then, we have
\begin{align*}
\tilde Z_1(t)=\sum_{j=1}^{n_1}[W_{11}(j,t)-W_{21}(j/k)\mf 1(j/k\in \mathbb Z)]V_{j,1}:=\sum_{j=1}^{n_1}\tilde W_{1}(j,t)V_{j,1},\\
\tilde Z_2(t)=\sum_{j_1=1}^{n_1}[W_{22}(j/k,t)\mf 1(j/k\in \mathbb Z)-W_{12}(j,t)]V_{j,2}:=\sum_{j=1}^{n_1}\tilde W_{2}(j,t)V_{j,2},
\end{align*}
where $\tilde W_{1}(j,t)$ and $\tilde W_2(j,t)$ are defined in an obvious way.  Further, define
\begin{align}
\tilde W_{11}(j,t)=W_{11}(j,t)-W_{21}(j/k,t),\\\tilde  W_{22}(j,t)=W_{22}(j/k,t)-W_{12}(j,t).
\end{align}Let us now define  
\begin{align}
V=\int \tilde Z_1(t)\tilde Z_2(t)w(t)dt=\sum_{j_1=1}^{n_1}\sum_{j_2=1}^{n_1}\int \tilde W_1(j_1,t)\tilde W_2(j_2,t)w(t)dtV_{j_1,1}V_{j_2,1}.
\end{align}
Hence,  
\begin{align*}
Var(V)&=\sum_{j_1=1}^{n_1}\sum_{j_2=1}^{n_1}[\int \tilde W_1(j_1,t)\tilde W_2(j_2,t)w(t)dt]^2\notag\\
&=\sum_{j_1=1}^{n_1}\sum_{j_2=1}^{n_1}\big[\int\big(\tilde W_{11}(j_1,t)\mf 1(j_1/k\in \mathbb Z)+W_{11}(j_1,t)\mf 1(j_1/k\not\in \mathbb Z)\big)\notag\\&~~~~~~~~\big(\tilde W_{22}(j_2,t)\mf 1(j_2/k\in \mathbb Z)-W_{12}(j_2,t)\mf 1(j_1/k\not\in \mathbb Z))\big)w(t)dt\big]^2\\&:=I-II+III-IV,
\end{align*}
where \begin{align}
I=\sum_{j_1=1}^{n_1}\sum_{j_2=1}^{n_1}\Big[\int \tilde W_{11}(j_1,t)\tilde W_{22}(j_2,t)\mf 1(j_1/k\in \mathbb Z)\mf 1(j_2/k\in \mathbb Z)w(t)dt\Big]^2,\\
II=\sum_{j_1=1}^{n_1}\sum_{j_2=1}^{n_1}\Big[\int \tilde W_{11}(j_1,t) W_{12}(j_2,t)\mf 1(j_1/k\in \mathbb Z)\mf 1(j_2/k\not\in \mathbb Z)w(t)dt\Big]^2,\\
III=\sum_{j_1=1}^{n_1}\sum_{j_2=1}^{n_1}\Big[\int  W_{11}(j_1,t)\tilde W_{22}(j_2,t)\mf 1(j_1/k \not \in \mathbb Z)\mf 1(j_2/k\in \mathbb Z)w(t)dt\Big]^2,\\
IV=\sum_{j_1=1}^{n_1}\sum_{j_2=1}^{n_1}\Big[\int  W_{11}(j_1,t) W_{12}(j_2,t)\mf 1(j_1/k \not \in \mathbb Z)\mf 1(j_2/k\not \in \mathbb Z)w(t)dt\Big]^2.
\end{align}
To sake of notational complexity, write $c_{2,1,n}=\frac{n_2b_{n,2}^2}{n_1b_{n,1}^2}$, $m_{12}(u)=(m_1)^{-1}(m_2(u))$. Then via tedious calculus, we see that
\begin{align*}\
I=\frac{b_{n,1}^3}{n_2^2b_{n,2}^8}\tilde I(c_{2,1,n},\frac{u-m_{21}(u)}{b_{n,2}},\frac{b_{n,1}}{b_{n,2}}m_{21}'(u),\frac{b_{n,1}}{b_{n,2}})(1+o(1))
\end{align*}

For $II$, by Riemann sum approximation, we see that
\begin{align*}
II=\frac{(k-1)n_1^2}{k^2(n_2b_{n,2}^2)^4}\int\int \Big\{c_{2,1,n}\check g_{11}(t)\bar K'(\frac{u-m_1^{-1}(t)}{b_{n,1}})-\check g_{21}(t)\bar K'(\frac{u-m_2^{-1}(t)}{b_{n,2}}))\notag \\\times c_{2,1,n}\check g_{12}(t)\bar K'(\frac{v-m_1^{-1}(t)}{b_{n,1}})w(t)dt\Big\}^2dudv(1+o(1)),
\end{align*}
and therefore,
\begin{align*}
II=\frac{(n_1-n_2)b_{n,1}^3}{n_2^3b_{n,2}^8}\tilde {II}(c_{2,1,n},\frac{u-m_{21}(u)}{b_{n,2}},\frac{b_{n,1}}{b_{n,2}}m_{21}'(u))(1+o(1))
\end{align*}

Using the same  argument, we have
\begin{align*}
III=\frac{(n_1-n_2)b_{n,1}^3}{n_2^3b_{n,2}^8}\tilde{III}(c_{2,1,n},\frac{u-m_{21}(u)}{b_{n,2}},\frac{b_{n,1}}{b_{n,2}}m_{21}'(u))(1+o(1))\end{align*}

Finally, 
\begin{align*}
IV=\Big(\frac{n_1-n_2}{n_1}\Big)^2\frac{1}{n_1^2b_{n,1}^5}\tilde{IV}(1+o(1))\end{align*}
In summary, $Var(V)$ equals with  \begin{align}\label{E129}
&=\Big\{\frac{b_{n,1}^3}{n_2^2b_{n,2}^8}\tilde I(c_{2,1,n},\frac{u-m_{21}(u)}{b_{n,2}},\frac{b_{n,1}}{b_{n,2}}m_{21}'(u),\frac{b_{n,1}}{b_{n,2}})-\frac{(n_1-n_2)b_{n,1}^3}{n_2^3b_{n,2}^8}\tilde {II}(c_{2,1,n},\frac{u-m_{21}(u)}{b_{n,2}},\frac{b_{n,1}}{b_{n,2}}m_{21}'(u))\notag\\+&\frac{(n_1-n_2)b_{n,1}^3}{n_2^3b_{n,2}^8}\tilde{III}(c_{2,1,n},\frac{u-m_{21}(u)}{b_{n,2}},\frac{b_{n,1}}{b_{n,2}}m_{21}'(u))-\Big(\frac{n_1-n_2}{n_1}\Big)^2\frac{1}{n_1^2b_{n,1}^5}\tilde{IV}\Big\}(1+o(1))
\end{align}

Therefore, the asymptotic bias and vairance of \eqref{E100} and \eqref{E101} follow from  \eqref{E120},\eqref{E121}, the term obtained by \eqref{E120},\eqref{E121} via swapping indices $1$ and $2$, \eqref{E129} and Proposition \ref{propE2}.  The asymptotic normality follows tedious calculation via Theorem 2.1 of  \cite{de1987central} and the independence of centered $\int \tilde Z_1^2(t)w(t)dt$,  $\int \tilde Z_2^2(t)w(t)dt$ and $\int \tilde Z_1(t)\tilde Z_2(t)w(t)dt$. Therefore, \eqref{E100} and \eqref{E101} hold and the proof of  theorem follows. \hfill $\Box$

\begin{theorem}\label{themE2}
Assume the conditions stated in (A1)-(A6), (A7'), (A8) and (B1), (B2), (B3), and suppose that as $n\rightarrow \infty$, $n_1/n_2\rightarrow n_{1,2}\in (0,\infty)$, $b_{n_1}/b_{n_2} \rightarrow b_{1,2}\in (0,\infty)$, $h_2/h_1\rightarrow h_{1,2}\in (0,\infty)$, $n_{1,2}=1/n_{2,1}$, $b_{1,2}=1/b_{2,1}$,  $h_{1,2}=1/h_{2,1}$, $\eta^{-1}=O(\log(n_1+n_2))$,  $\eta=o(1)$.
 For $u=1,2$ and $v=1,2$, let 
	\begin{align}
\check g^\circ_{uv}(t)	=M^c_{uv}(m_u^{-1}(t))((m_u^{-1})'(t))\int\limits_{\mathbb R} H''(y)ydy.
	\end{align} 
	Define 
	\begin{align}
	\bar K_1^\circ(t)=\Big((n_{2,1}b_{2,1}^3)^{1/2}\check g_{11}^{\circ^2}(t)+(n_{1,2}b_{1,2})^{1/2}\check g_{21}^{\circ^2}(t)\Big)^2\int \bar K'^2(x)dx\notag\\-2\check g_{11}^{\circ^2}(t)\check g_{21}^{\circ^2}(t)n_{2,1}^{1/2}b_{1,2}^{1/2}\int \bar K'(x)\bar K'(b_{1,2}x)dx\mf 1(m_{21}(0)=0),\\
		\bar K_2^\circ(t)=\Big((n_{2,1}b_{2,1}h_{2,1}^4)^{1/2}\check g_{11}^{\circ^2}(t)+(n_{1,2}b_{1,2}h_{1,2}^4)^{1/2}\check g_{21}^{\circ^2}(t)\Big)^2\int \bar K^2(x)dx\notag\\-2\check g_{11}^{\circ^2}(t)\check g_{21}^{\circ^2}(t)n_{2,1}^{1/2}b_{1,2}^{1/2}\int \bar K(x)\bar K(b_{1,2}x)dx\mf 1(m_{21}(0)=0),
	\end{align}
	and consequently,
	 \begin{align}
	K^\circ_1(t)=\bar K_1^\circ(t)+(\bar K_1^\circ)^{1-2}(t),\\
	K^\circ_2(t)=	\bar K_2^\circ(t)+(	\bar K_2^\circ)^{1-2}(t).
	\end{align} Now, if  $(m_1^{-1})^{'}(t)-(m_2^{-1})^{'}(t)=\rho_{n_1,n_2}\kappa(t)$ for some non-zero bounded function $\kappa(t)$ and $\rho_{n_1,n_2}=o(\eta\wedge b_{n,1})$, then as $\min(n_1, n_2)\rightarrow\infty$, we have
	\begin{align}
	\p\left(\sup_{t\in \mathcal I_{\hat a, \hat b}}\frac{(n_1b_{n,1}^3n_2b_{n,2}^3)^{1/4}|(\hat m_1^{-1})'(t)-(\hat m_2^{-1})'(t)-((m_1^{-1})'(t)-(m_2^{-1})'(t))|}{K^{\circ^{1/2}}_1(t)}\geq \sqrt{-2\log(\pi\alpha/\kappa_0)} \right)\rightarrow \alpha,
	\end{align}
	where $\mathcal I_{\hat a, \hat b}=(\hat a+\eta, \hat b-\eta)$ and
	\begin{align}
	\kappa_0=\frac{(b_{n,1}b_{n,2})^{1/2}}{h_1h_2}\int\limits_{m_1(0)}^{m_1(1-m_2^{-1}(m_1(0)))} \left(\frac{K_2(t)}{K_1(t)}\right)^{1/2}dt.
	\end{align}
\end{theorem}

{\it Proof.} Without loss of generality, consider $n_1\geq n_2$ and that $n_1/n_2=k$ for some integer $k$. Following the proof of Theorem \ref{SCB}, it suffices to evaluate  $\E (Z^2(t))$, $\E (Z^{'2}(t))$ and $\E( Z(t)Z'(t))$, where now $Z(t)$ is defined in \eqref{E98}. Once the above quantities are obtained, the theorem will follow from an application of  Proposition 1 of \cite{sun1994simultaneous} and the same argument as given in the proof of Theorem \ref{SCB}.

Recall the decomposition of $Z(t)$ in \eqref{tildeZtdecompose}.  
\begin{align}
\E Z^2(t)=\E \tilde Z^2_1(t)+\E \tilde Z^2_2(t), 
\end{align}
and recall the definition of $W_{uv}(j,t)$, $u,v\in\{1,2\}^2$ in the proof of Theorem \ref{E1}.  
\begin{align}
\E \tilde Z_{1}^2(t)&=\sum_{j=1}^{n_1}
[W^2_{11}(j,t)-W_{21}(j/k,t)\mf 1(j/k\in \mathbb Z)]^2\notag\\&=\sum_{j=1}^{n_1}W_{11}(j,t)+\sum_{j=1}^{n_1}W_{21}^2(j/k,t)\mf 1(j/k\in \mathbb Z)-2\sum_{j=1}^{n_1}W_{11}(j,t)W_{21}(j/k,t)\mf 1(j/k\in \mathbb Z).
\end{align}
By the similar but simpler argument than the proof of Theorem \ref{themE1}, we have
\begin{align}
\sum_{j=1}^{n_1}W_{11}^2(j,t)=\frac{1}{n_1b_{n,1}^3}\check g^2_{11}(t)\int \bar K'^2(x)dx(1+o(1)),\\
\sum_{j=1}^{n_1}W_{21}^2(j/k,t)\mf 1(j/k\in \mathbb Z)=\frac{1}{n_2b_{n,2}^3}\check g_{21}^2(t)\int \bar K'^2(x)dx(1+o(1)),\\
\sum_{j=1}^{n_1}W_{11}(j,t)W_{21}(j/k,t)\mf 1(j/k\in \mathbb Z)=\frac{\check g_{11}(t)\check g_{21}(t)}{n_1b_{n,1}b_{n,2}^2}\int \bar K'(x)\notag\\ \times\bar K'(\frac{m_1^{-1}(t)-m_2^{-1}(t)+xb_{n,1}}{b_{n,2}})dx(1+o(1)).
\end{align}
Therefore,  
\begin{align}\label{v1}
\E(\tilde Z_1^2(t))=\Big(\frac{1}{n_1b_{n,1}^3}\check g^2_{11}(t)\int \bar K'^2(x)dx+\frac{1}{n_2b_{n,2}^3}\check g_{21}^2(t)\int \bar K'^2(x)dx\notag\\-2\frac{\check g_{11}(t)\check g_{21}(t)}{n_1b_{n,1}b_{n,2}^2}\int \bar K'(x)\bar K'(\frac{m_1^{-1}(t)-m_2^{-1}(t)+xb_{n,1}}{b_{n,2}})dx\Big)(1+o(1)),\\\label{v2}
\E(\tilde Z_2^2(t))=(\E(\tilde Z_1^2(t)))^{1-2}.
\end{align}
To calculate $\E(Z'(t))=\E(\tilde Z'^2_1(t))+\E(\tilde Z'^2_2(t))$,
similarly to the proof of Theorem \ref{asymptotic},
$\frac{\partial }{\partial t}W_{uv}(j,t)$ can be written as $\tilde W'_{uv}(j,t)(1+o(1))$ for $u,v=1,2$, where
\begin{align}
\tilde W'_{uv}(j,t)&=\frac{-1}{n_ub_{n,u}h_u^2}M^c_{uv}(m_u^{-1}(t))\bar K\left(\frac{j/n_u-m_u^{-1}(t)}{b_{n,u}}\right)((m_u^{-1})'(t))\int\limits_{\mathbb R} H''(y)ydy\notag
\\&:=\frac{-1}{n_ub_{n,u}h_u^2}\check g^\circ_{uv}(t)\bar K\left(\frac{j/n_u-m_u^{-1}(t)}{b_{n,u}}\right).
\end{align}
Here $\check g_{uv}^\circ(t)$ is defined in a straightforward way.	
Using this fact, it follows that
\begin{align}
\sum_{j=1}^{n_1}W'^2_{11}(j,t)=\frac{(\check g_{11}^\circ(t))^2}{n_1b_{n,1}h_1^4}\int \bar K^2(x)dx(1+o(1)),\\ \sum_{j=1}^{n_1}W'^2_{21}(j/k,t)\mf 1(j/k\in\mathbb Z)=\frac{(\check g_{21}^\circ(t))^2}{n_2b_{n,2}h_2^4}\int \bar K^2(x)dx(1+o(1)),\\\sum_{j=1}^{n_1}W_{11}'(j,t)W'_{21}(j/k,t)\mf 1(j/k\in \mathbb Z)=\frac{\check g^\circ_{11}(t)\check g^\circ_{21}(t)}{n_1b_{n,2}h_1^2h_2^2}\int \bar K(x)\notag\\\times \bar K(\frac{m_1^{-1}(t)-m_2^{-1}(t)+xb_{n,1}}{b_{n,2}})dx(1+o(1)).
\end{align}
Hence
\begin{align}\label{v3}
\E(\tilde Z'^2_1(t))=\Big(\frac{(\check g_{11}^\circ(t))^2}{n_1b_{n,1}h_1^4}\int \bar K^2(x)dx+\frac{(\check g_{21}^\circ(t))^2}{n_2b_{n,2}h_2^4}\int \bar K^2(x)dx\notag\\-2\frac{\check g^\circ_{11}(t)\check g^\circ_{21}(t)}{n_1b_{n,2}h_1^2h_2^2}\int \bar K(x) \bar K(\frac{m_1^{-1}(t)-m_2^{-1}(t)+xb_{n,1}}{b_{n,2}})dx\Big),\\
\E(\tilde Z'^2_2(t))=(\E(\tilde Z'^2_1(t)))^{1-2}\label{v4}
\end{align}
Following the proof of Theorem \ref{SCB}, we have
\begin{align}\label{v5}
\E(\tilde Z(t)\tilde Z'(t))=O(\frac{1}{n_1b_{n,1}^2h_1^2}\frac{h_1}{b_{n,1}})=o(\frac{1}{n_1b_{n,1}^2h_1^2}).
\end{align}
Therefore, the theorem follows from \eqref{v1},
 \eqref{v2},
  \eqref{v3},
   \eqref{v4},
    \eqref{v5} and Proposition \ref{propE2}.
    \hfill $\Box$ 
    
\noindent \textbf{Proof of Theorem \ref{core-dependent}}. The theorem directly follows from the proof of Theorem \ref{themE1} and Theorem \ref{themE2}. Notice that under the null hypothesis, the asymptotic results of  Theorem \ref{themE1} and Theorem \ref{themE2} will differ between the scenarios of $d=m_{21}(0)=0$ and  $d=m_{21}(0)\neq 0$. $\hfill$ $\Box$}

{\color{black}\section{Results of Simulation Study : Dependent Data}\label{Sec_F} 
In Section 5.2 in the main manuscript, we briefly discuss the results of simulation studies for dependent data sets. The detailed  results are reported here. The results in Tables \ref{tabs1}, \ref{tabs2}, \ref{tabs3} and \ref{tabs4} indicate that the estimated sizes of the SIT test and the SCB test (described in Section 5 in the main article) are not deviated more than 1\% from the estimated sizes when the data sets are independent, and the estimated powers of the SIT test and the SCB test are not deviated more than 6\% from the estimated powers when the data sets are independent.

{\color{black}
\begin{table}[h!]
	
	\begin{center}		
		
		\begin{tabular}{ccc}\hline
			model & $n = 100$ & $n = 500$\\ \hline 
			Example 1 ($\alpha = 5\%$ , $\tau = 0.5$, $b_{n, s} = n_{s}^{-1/4}$)  & {$(0.061, 0.058, 0.063)$} & {$(0.050, 0.055, 0.049)$}\\ \hline
			Example 1 ($\alpha = 5\%$, $\tau = 0.5$, $b_{n, s} = n_{s}^{-1/5}$)  & {$(0.055, 0.058, 0.059)$} & {$(0.052, 0.054, 0.050)$}\\ \hline
			Example 1 ($\alpha = 5\%$, $\tau = 0.7$, $b_{n, s} = n_{s}^{-1/4}$)  & {$(0.056, 0.061, 0.060)$}  & {$(0.052, 0.051, 0.055)$} \\ \hline
			Example 1 ($\alpha = 5\%$, $\tau = 0.7$, $b_{n, s} = n_{s}^{-1/5}$)  & {$(0.057, 0.062, 0.060)$} & {$(0.051, 0.053, 0.052)$}
			 \\ \hline
			Example 1 ($\alpha = 5\%$, $\tau = 0.8$, $b_{n, s} = n_{s}^{-1/4}$)  & {$(0.053, 0.052, 0.058)$} & 
			 {$(0.055, 0.052, 0.051)$} \\ \hline
			Example 1 ($\alpha = 5\%$, $\tau = 0.8$, $b_{n, s} = n_{s}^{-1/5}$)  & {$(0.055, 0.056, 0.052)$} & 
			 {$(0.053, 0.051, 0.051)$} \\ \hline
			Example 2 ($\alpha = 5\%$, $\tau = 0.5$, $b_{n, s} = n_{s}^{-1/4}$)  & {$(0.057, 0.058, 0.055)$} &  {$(0.050, 0.053, 0.056)$}\\ \hline
			Example 2 ($\alpha = 5\%$, $\tau = 0.5$, $b_{n, s} = n_{s}^{-1/4}$)  & {$(0.051, 0.055, 0.054)$} &  {$(0.052, 0.054, 0.050)$}\\ \hline
			Example 2 ($\alpha = 5\%$, $\tau = 0.7$, $b_{n, s} = n_{s}^{-1/4}$)  & {$(0.054, 0.055, 0.058)$} &  {$(0.052, 0.054, 0.050)$}\\ \hline
			Example 2 ($\alpha = 5\%$, $\tau = 0.7$, $b_{n, s} = n_{s}^{-1/5}$)  & {$(0.053, 0.061, 0.059)$} &  {$(0.054, 0.049, 0.051)$}\\ \hline
			Example 2 ($\alpha = 5\%$, $\tau = 0.8$, $b_{n, s} = n_{s}^{-1/4}$)  & {$(0.053, 0.062, 0.058)$} &  {$(0.054, 0.050, 0.052)$}\\ \hline
			Example 2 ($\alpha = 5\%$, $\tau = 0.8$, $b_{n, s} = n_{s}^{-1/5}$)  & {$(0.051, 0.056, 0.056)$} &  {$(0.052, 0.050, 0.050)$}\\ \hline
		\end{tabular}
	\end{center}
	\caption{\it The estimated size of the SIT test for different sample sizes $n_1= n_2=n$. The levels of significance (denoted as $\alpha$) is $5\%$ for dependent data. In each cell, from the left, the first, the second and the third values are corresponding to $w_{n, s} = n_{s}^{-1/3}$, $n_{s}^{-1/5}$ and $n_{s}^{-1/7}$, respectively.}
	
	\label{tabs1}
	
\end{table}}

\begin{table}[h!]
	
	\begin{center}		
		
		\begin{tabular}{ccc}\hline
			model & $n = 100$ & $n = 500$\\ \hline 
			Example 1 ($\alpha = 5\%$ , $\tau = 0.5$, $b_{n, s} = n_{s}^{-1/4}$)  & {$(0.062, 0.055, 0.064)$} & {$(0.053, 0.054, 0.051)$}\\ \hline
			Example 1 ($\alpha = 5\%$, $\tau = 0.5$, $b_{n, s} = n_{s}^{-1/5}$)  & {$(0.062, 0.059, 0.062)$} & {$(0.053, 0.051, 0.050)$}\\ \hline
			Example 1 ($\alpha = 5\%$, $\tau = 0.7$, $b_{n, s} = n_{s}^{-1/4}$)  & {$(0.063, 0.061, 0.064)$}  & {$(0.052, 0.050, 0.055)$} \\ \hline
			Example 1 ($\alpha = 5\%$, $\tau = 0.7$, $b_{n, s} = n_{s}^{-1/5}$)  & {$(0.052, 0.061, 0.063)$} & {$(0.050, 0.053, 0.055)$}
			 \\ \hline
			Example 1 ($\alpha = 5\%$, $\tau = 0.8$, $b_{n, s} = n_{s}^{-1/4}$)  & {$(0.057, 0.056, 0.058)$} & 
			 {$(0.051, 0.050, 0.051)$} \\ \hline
			Example 1 ($\alpha = 5\%$, $\tau = 0.8$, $b_{n, s} = n_{s}^{-1/5}$)  & {$(0.058, 0.055, 0.054)$} & 
			 {$(0.056, 0.055, 0.055)$} \\ \hline
			Example 2 ($\alpha = 5\%$, $\tau = 0.5$, $b_{n, s} = n_{s}^{-1/4}$)  & {$(0.050, 0.056, 0.054)$} &  {$(0.050, 0.051, 0.054)$}\\ \hline
			Example 2 ($\alpha = 5\%$, $\tau = 0.5$, $b_{n, s} = n_{s}^{-1/4}$)  & {$(0.052, 0.055, 0.057)$} &  {$(0.050, 0.052, 0.056)$}\\ \hline
			Example 2 ($\alpha = 5\%$, $\tau = 0.7$, $b_{n, s} = n_{s}^{-1/4}$)  & {$(0.057, 0.059, 0.059)$} &  {$(0.054, 0.056, 0.054)$}\\ \hline
			Example 2 ($\alpha = 5\%$, $\tau = 0.7$, $b_{n, s} = n_{s}^{-1/5}$)  & {$(0.055, 0.057, 0.058)$} &  {$(0.056, 0.052, 0.054)$}\\ \hline
			Example 2 ($\alpha = 5\%$, $\tau = 0.8$, $b_{n, s} = n_{s}^{-1/4}$)  & {$(0.055, 0.062, 0.059)$} &  {$(0.055, 0.052, 0.058)$}\\ \hline
			Example 2 ($\alpha = 5\%$, $\tau = 0.8$, $b_{n, s} = n_{s}^{-1/5}$)  & {$(0.055, 0.058, 0.060)$} &  {$(0.052, 0.055, 0.059)$}\\ \hline
		\end{tabular}
	\end{center}
	\caption{\it The estimated size of the SCB test for different sample sizes $n_1= n_2=n$ for dependent data. The levels of significance (denoted as $\alpha$) are $5\%$ and $10\%$. In each cell, from the left, the first, the second and the third values are corresponding to $w_{n, s} = n_{s}^{-1/3}$, $n_{s}^{-1/5}$ and $n_{s}^{-1/7}$, respectively.}
	
	\label{tabs2}
	
\end{table}

\begin{table}[h!]
	
	\begin{center}		
		
		\begin{tabular}{ccc}\hline
			model & $n = 100$ & $n = 500$\\ \hline 
			Example 3 ($\alpha = 5\%$ , $\tau = 0.5$, $b_{n, s} = n_{s}^{-1/4}$)  & {$(0.477, 0.454, 0.491)$} & {$(0.749, 0.777, 0.763)$}\\ \hline
			Example 3 ($\alpha = 5\%$, $\tau = 0.5$, $b_{n, s} = n_{s}^{-1/5}$)  & {$(0.461, 0.475, 0.502)$} & {$(0.781, 0.799, 0.795)$}\\ \hline
			Example 3 ($\alpha = 5\%$, $\tau = 0.7$, $b_{n, s} = n_{s}^{-1/4}$)  & {$(0.505, 0.517, 0.522)$}  & {$(0.774, 0.841, 0.833)$} \\ \hline
			Example 3 ($\alpha = 5\%$, $\tau = 0.7$, $b_{n, s} = n_{s}^{-1/5}$)  & {$(0.479, 0.581, 0.543)$} & {$(0.865, 0.803, 0.888)$}
			 \\ \hline
			Example 3 ($\alpha = 5\%$, $\tau = 0.8$, $b_{n, s} = n_{s}^{-1/4}$)  & {$(0.475, 0.496, 0.517)$} & 
			 {$(0.872, 0.831, 0.900)$} \\ \hline
			Example 3 ($\alpha = 5\%$, $\tau = 0.8$, $b_{n, s} = n_{s}^{-1/5}$)  & {$(0.534, 0.540, 0.522)$} & 
			 {$(0.888, 0.849, 0.903)$} \\ \hline
			Example 4 ($\alpha = 5\%$, $\tau = 0.5$, $b_{n, s} = n_{s}^{-1/4}$)  & {$(0.477, 0.486, 0.532)$} &  {$(0.817, 0.833, 0.825)$}\\ \hline
			Example 4 ($\alpha = 5\%$, $\tau = 0.5$, $b_{n, s} = n_{s}^{-1/5}$)  & {$(0.512, 0.471, 0.510)$} &  {$(0.833, 0.855, 0.895)$}\\ \hline
			Example 4 ($\alpha = 5\%$, $\tau = 0.7$, $b_{n, s} = n_{s}^{-1/4}$)  & {$(0.534, 0.555, 0.527)$} &  {$(0.870, 0.879, 0.904)$}\\ \hline
			Example 4 ($\alpha = 5\%$, $\tau = 0.7$, $b_{n, s} = n_{s}^{-1/5}$)  & {$(0.543, 0.555, 0.565)$} &  {$(0.875, 0.899, 0.909)$}\\ \hline
			Example 4 ($\alpha = 5\%$, $\tau = 0.8$, $b_{n, s} = n_{s}^{-1/4}$)  & {$(0.471, 0.492, 0.505)$} &  {$(0.843, 0.868, 0.881)$}\\ \hline
			Example 4 ($\alpha = 5\%$, $\tau = 0.8$, $b_{n, s} = n_{s}^{-1/5}$)  & {$(0.468, 0.471, 0.499)$} &  {$(0.902, 0.908, 0.922)$}\\ \hline
		\end{tabular}
	\end{center}
	\caption{\it The estimated power of the SIT test for different sample sizes $n_1= n_2=n$ for dependent data. The levels of significance (denoted as $\alpha$) is $5\%$. In each cell, from the left, the first, the second and the third values are corresponding to $w_{n, s} = n_{s}^{-1/3}$, $n_{s}^{-1/5}$ and $n_{s}^{-1/7}$, respectively.}
	
	\label{tabs3}
	
\end{table}

\begin{table}[h!]
	
	\begin{center}		
		
		\begin{tabular}{ccc}\hline
			model & $n = 100$ & $n = 500$\\ \hline 
			Example 3 ($\alpha = 5\%$ , $\tau = 0.5$, $b_{n, s} = n_{s}^{-1/4}$)  & {$(0.471, 0.425, 0.475)$} & {$(0.722, 0.757, 0.743)$}\\ \hline
			Example 3 ($\alpha = 5\%$, $\tau = 0.5$, $b_{n, s} = n_{s}^{-1/5}$)  & {$(0.444, 0.468, 0.498)$} & {$(0.743, 0.788, 0.802)$}\\ \hline
			Example 3 ($\alpha = 5\%$, $\tau = 0.7$, $b_{n, s} = n_{s}^{-1/4}$)  & {$(0.484, 0.488, 0.491)$}  & {$(0.786, 0.822, 0.813)$} \\ \hline
			Example 3 ($\alpha = 5\%$, $\tau = 0.7$, $b_{n, s} = n_{s}^{-1/5}$)  & {$(0.511, 0.527, 0.535)$} & {$(0.903, 0.841, 0.865)$}
			 \\ \hline
			Example 3 ($\alpha = 5\%$, $\tau = 0.8$, $b_{n, s} = n_{s}^{-1/4}$)  & {$(0.489, 0.472, 0.511)$} & 
			 {$(0.844, 0.845, 0.890)$} \\ \hline
			Example 3 ($\alpha = 5\%$, $\tau = 0.8$, $b_{n, s} = n_{s}^{-1/5}$)  & {$(0.496, 0.502, 0.498)$} & 
			 {$(0.855, 0.872, 0.894)$} \\ \hline
			Example 4 ($\alpha = 5\%$, $\tau = 0.5$, $b_{n, s} = n_{s}^{-1/4}$)  & {$(0.492, 0.475, 0.557)$} &  {$(0.842, 0.859, 0.871)$}\\ \hline
			Example 4 ($\alpha = 5\%$, $\tau = 0.5$, $b_{n, s} = n_{s}^{-1/5}$)  & {$(0.500, 0.484, 0.499)$} &  {$(0.844, 0.866, 0.884)$}\\ \hline
			Example 4 ($\alpha = 5\%$, $\tau = 0.7$, $b_{n, s} = n_{s}^{-1/4}$)  & {$(0.498, 0.506, 0.525)$} &  {$(0.845, 0.872, 0.877)$}\\ \hline
			Example 4 ($\alpha = 5\%$, $\tau = 0.7$, $b_{n, s} = n_{s}^{-1/5}$)  & {$(0.504, 0.528, 0.544)$} &  {$(0.848, 0.866, 0.887)$}\\ \hline
			Example 4 ($\alpha = 5\%$, $\tau = 0.8$, $b_{n, s} = n_{s}^{-1/4}$)  & {$(0.442, 0.459, 0.472)$} &  {$(0.810, 0.835, 0.844)$}\\ \hline
			Example 4 ($\alpha = 5\%$, $\tau = 0.8$, $b_{n, s} = n_{s}^{-1/5}$)  & {$(0.455, 0.468, 0.479)$} &  {$(0.829, 0.856, 0.883)$}\\ \hline
		\end{tabular}
	\end{center}
	\caption{\it The estimated power of the SCB test for different sample sizes $n_1= n_2=n$ for dependent data. The levels of significance (denoted as $\alpha$) is $5\%$. In each cell, from the left, the first, the second and the third values are corresponding to $w_{n, s} = n_{s}^{-1/3}$, $n_{s}^{-1/5}$ and $n_{s}^{-1/7}$, respectively.}
	
	\label{tabs4}
	
\end{table}}

\bibliography{lit}

\begin{thebibliography}{}

\bibitem[Chaudhuri, 1991]{chaudhuri1991nonparametric}
Chaudhuri, P. (1991).
\newblock Nonparametric estimates of regression quantiles and their local
  bahadur representation.
\newblock {\em The Annals of Statistics}, 19(2):760--777.

\bibitem[Collier and Dalalyan, 2015]{colldala2015}
Collier, O. and Dalalyan, A.~S. (2015).
\newblock Curve registration by nonparametric goodness-of-fit testing.
\newblock {\em Journal of Statistical Planning and Inference}, 162:20--42.

\bibitem[Craven and Wahba, 1978]{craven1978smoothing}
Craven, P. and Wahba, G. (1978).
\newblock Smoothing noisy data with spline functions.
\newblock {\em Numerische mathematik}, 31(4):377--403.

\bibitem[Dahlhaus, 1997]{dahlhaus1997fitting}
Dahlhaus, R. (1997).
\newblock Fitting time series models to nonstationary processes.
\newblock {\em The Annals of Statistics}, 25(1):1--37.

\bibitem[Dahlhaus et~al., 2019]{dahlhaus2019towards}
Dahlhaus, R., Richter, S., and Wu, W.~B. (2019).
\newblock Towards a general theory for nonlinear locally stationary processes.
\newblock {\em Bernoulli}, 25(2):1013--1044.

\bibitem[de~Jong, 1987]{de1987central}
de~Jong, P. (1987).
\newblock A central limit theorem for generalized quadratic forms.
\newblock {\em Probability Theory and Related Fields}, 75(2):261--277.

\bibitem[Dette et~al., 2021]{dette2021identifying}
Dette, H., Dhar, S.~S., and Wu, W. (2021).
\newblock Identifying shifts between two regression curves.
\newblock {\em Annals of the Institute of Statistical Mathematics},
  73:855--889.

\bibitem[Dette et~al., 2006]{dette2006simple}
Dette, H., Neumeyer, N., and Pilz, K.~F. (2006).
\newblock A simple nonparametric estimator of a strictly monotone regression
  function.
\newblock {\em Bernoulli}, 12(3):469--490.

\bibitem[Dette and Volgushev, 2008]{dette2008non}
Dette, H. and Volgushev, S. (2008).
\newblock Non-crossing non-parametric estimates of quantile curves.
\newblock {\em Journal of the Royal Statistical Society: Series B (Statistical
  Methodology)}, 70(3):609--627.

\bibitem[Dette et~al., 2011]{dette2011comparing}
Dette, H., Wagener, J., and Volgushev, S. (2011).
\newblock Comparing conditional quantile curves.
\newblock {\em Scandinavian Journal of Statistics}, 38(1):63--88.

\bibitem[Dette and Wu, 2019]{dette2019detecting}
Dette, H. and Wu, W. (2019).
\newblock Detecting relevant changes in the mean of nonstationary processes—a
  mass excess approach.
\newblock {\em The Annals of Statistics}, 47(6):3578--3608.

\bibitem[Dette and Wu, 2020]{dette2020prediction}
Dette, H. and Wu, W. (2020).
\newblock Prediction in locally stationary time series.
\newblock {\em arXiv preprint arXiv:2001.00419, to appear Journal of Business
  \& Economic Statistics}.

\bibitem[Efron, 1991]{efron1991}
Efron, B. (1991).
\newblock Regression percentiles using asymmetric squared error loss.
\newblock {\em Statistica Sinica}, 1:93--125.

\bibitem[Gamboa et~al., 2007]{gamloumaz2007}
Gamboa, F., Loubes, J., and Maza, E. (2007).
\newblock Semi-parametric estimation of shifts.
\newblock {\em Electronic Journal of Statistics}, 1:616--640.

\bibitem[Gutenbrunner and Jureckova, 1992]{gutenbrunner1992}
Gutenbrunner, G. and Jureckova, J. (1992).
\newblock Regression rank scores and regression quantiles.
\newblock {\em The Annals of Statistics}, 20:305--330.

\bibitem[He and Zhu, 2003]{He2003}
He, X. and Zhu, L. (2003).
\newblock A lack-of-fit test for quantile regression.
\newblock {\em Journal of the American Statistical Association}, 98:1013--1022.

\bibitem[Horowitz and Spokoiny, 2002]{Horowitz2002}
Horowitz, J.~L. and Spokoiny, V.~G. (2002).
\newblock An adaptive, rate-optimal test of linearity for median regression
  models.
\newblock {\em Journal of the American Statistical Association}, 97:822--835.

\bibitem[Kim, 2007]{kim2007quantile}
Kim, M.-O. (2007).
\newblock Quantile regression with varying coefficients.
\newblock {\em The Annals of Statistics}, pages 92--108.

\bibitem[Koenker and Bassett, 1978]{koenker1978}
Koenker, R. and Bassett, G. (1978).
\newblock Regression quantiles.
\newblock {\em Econometrica}, 46:33--50.

\bibitem[Koenker and Bassett, 1982]{koenker1982}
Koenker, R. and Bassett, G. (1982).
\newblock Robust tests for heteroscedasticity based on regression quantiles.
\newblock {\em Econometrica}, 50:43--61.

\bibitem[Koenker et~al., 1994]{koenker1994}
Koenker, R., Ng, P., and Portnoy, S. (1994).
\newblock Quantile smoothing splines.
\newblock {\em Biometrika}, 81:673--680.

\bibitem[Munk and Dette, 1998]{munk1998}
Munk, A. and Dette, H. (1998).
\newblock Nonparametric comparison of several regression functions: exact and
  asymptotic theory.
\newblock {\em The Annals of Statistics}, 26(6):2339--2368.

\bibitem[Qu, 2008]{qu2008testing}
Qu, Z. (2008).
\newblock Testing for structural change in regression quantiles.
\newblock {\em Journal of Econometrics}, 146(1):170--184.

\bibitem[Qu and Yoon, 2015]{qu2015nonparametric}
Qu, Z. and Yoon, J. (2015).
\newblock Nonparametric estimation and inference on conditional quantile
  processes.
\newblock {\em Journal of Econometrics}, 185(1):1--19.

\bibitem[Raupach et~al., 2014]{Raupach2014}
Raupach, M.~R., Davis, S.~J., Peters, G.~P., Andrew, R.~M., Canadell, J.~G.,
  Ciais, P., Friedlingstein, P., Jotzo, F., van Vuuren, D.~P., and Le~Quéré,
  C. (2014).
\newblock Sharing a quota on cumulative carbon emissions.
\newblock {\em Nature Climate Change}, 4:873--879.

\bibitem[Ruppert and Carroll, 1980]{ruppert1980}
Ruppert, D. and Carroll, R. (1980).
\newblock Trimmed least square estimation in the linear models.
\newblock {\em Journal of the American Statistical Association}, 75:828--838.

\bibitem[Schucany and Sommers, 1977]{schucany1977improvement}
Schucany, W. and Sommers, J.~P. (1977).
\newblock Improvement of kernel type density estimators.
\newblock {\em Journal of the American Statistical Association},
  72(358):420--423.

\bibitem[Sun and Loader, 1994]{sun1994simultaneous}
Sun, J. and Loader, C.~R. (1994).
\newblock Simultaneous confidence bands for linear regression and smoothing.
\newblock {\em The Annals of Statistics}, 22(3):1328--1345.

\bibitem[Takeuchi et~al., 2006]{Takeuchi2006}
Takeuchi, I., Le, Q., Sears, Q.~V., and Smola, A.~J. (2006).
\newblock Nonparametric quantile estimation.
\newblock {\em Journal of Machine Learning Research}, 7:1231--1264.

\bibitem[Vimond, 2010]{vimond2010}
Vimond, M. (2010).
\newblock Efficient estimation for a subclass of shape invariant models.
\newblock {\em The Annals of Statistics}, 38:1885--1912.

\bibitem[Wu and Zhou, 2017]{wu2017nonparametric}
Wu, W. and Zhou, Z. (2017).
\newblock Nonparametric inference for time-varying coefficient quantile
  regression.
\newblock {\em Journal of Business \& Economic Statistics}, 35(1):98--109.

\bibitem[Wu and Zhou, 2018a]{wu2018gradient}
Wu, W. and Zhou, Z. (2018a).
\newblock Gradient-based structural change detection for nonstationary time
  series m-estimation.
\newblock {\em The Annals of Statistics}, 46(3):1197--1224.

\bibitem[Wu and Zhou, 2018b]{wu2018simultaneous}
Wu, W. and Zhou, Z. (2018b).
\newblock Simultaneous quantile inference for non-stationary long-memory time
  series.
\newblock {\em Bernoulli}, 24(4A):2991--3012.

\bibitem[Wu and Zhao, 2007]{wu2007inference}
Wu, W.~B. and Zhao, Z. (2007).
\newblock Inference of trends in time series.
\newblock {\em Journal of the Royal Statistical Society: Series B (Statistical
  Methodology)}, 69(3):391--410.

\bibitem[Yu and Jones, 1998]{yu1998local}
Yu, K. and Jones, M. (1998).
\newblock Local linear quantile regression.
\newblock {\em Journal of the American statistical Association},
  93(441):228--237.

\bibitem[Zhao and Wu, 2008]{zhao2008confidence}
Zhao, Z. and Wu, W.~B. (2008).
\newblock Confidence bands in nonparametric time series regression.
\newblock {\em The Annals of Statistics}, 36(4):1854--1878.

\bibitem[Zheng, 1998]{zheng1998}
Zheng, J.~X. (1998).
\newblock A consistent nonparametric test of parametric regression models under
  conditional quantile restrictions.
\newblock {\em Econometric Theory}, 14:123--138.

\bibitem[Zhou, 2010]{zhou2010nonparametric}
Zhou, Z. (2010).
\newblock Nonparametric inference of quantile curves for nonstationary time
  series.
\newblock {\em The Annals of Statistics}, 38(4):2187--2217.

\bibitem[Zhou and Wu, 2009]{zhou2009local}
Zhou, Z. and Wu, W.~B. (2009).
\newblock Local linear quantile estimation for nonstationary time series.
\newblock {\em The Annals of Statistics}, 37(5B):2696--2729.

\bibitem[Zhou and Wu, 2010]{zhou2010simultaneous}
Zhou, Z. and Wu, W.~B. (2010).
\newblock Simultaneous inference of linear models with time varying
  coefficients.
\newblock {\em Journal of the Royal Statistical Society: Series B (Statistical
  Methodology)}, 72(4):513--531.

\end{thebibliography}

\end{document}